\documentclass[prx,twocolumn,english,superscriptaddress,floatfix,longbibliography,nofootinbib]{revtex4-2}
\usepackage[table]{xcolor}

\newcommand\brpm{\mathbin{\vcenter{\hbox{\oalign{$\scriptstyle({+})$\cr
					\noalign{\kern-.3ex}
					\hfil$\scriptscriptstyle-$\hfil\cr}}}}}
\usepackage{booktabs}
\usepackage{pgfplotstable}

\usepackage{csvsimple}

\usepackage{tikz}
\usepackage{tikzit}

\tikzstyle{gate}=[shape=rectangle, text height=1.5ex, text depth=0.25ex, yshift=0.5mm, fill=white, draw=black, minimum height=5mm, yshift=-0.5mm, minimum width=5mm, font={\small}, tikzit category=circuit]
\tikzstyle{big gate}=[shape=rectangle, text height=1.5ex, text depth=0.25ex, yshift=0.5mm, fill=white, draw=black, minimum height=10mm, yshift=-0.5mm, minimum width=5mm, font={\small}, tikzit category=circuit]
\tikzstyle{Z dot}=[inner sep=0mm, minimum size=2mm, shape=circle, draw=black, fill=zxgreen, tikzit fill={rgb,255: red,221; green,255; blue,221}, tikzit category=zx]
\tikzstyle{Z bold dot}=[inner sep=0mm, minimum size=2mm, shape=circle, draw=black, fill=zxgreen, tikzit fill={rgb,255: red,221; green,255; blue,221}, line width=1.2pt, tikzit category=zx]
\tikzstyle{Z phase dot}=[minimum size=5mm, font={\footnotesize\boldmath}, shape=rectangle, rounded corners=2mm, inner sep=0.2mm, outer sep=-2mm, scale=0.8, tikzit shape=circle, draw=black, fill=zxgreen, tikzit fill={rgb,255: red,221; green,255; blue,221}, tikzit draw=blue, tikzit category=zx]
\tikzstyle{Z tiny dot}=[inner sep=0mm, minimum size=1mm, shape=circle, draw=black, fill=zxgreen, tikzit fill={rgb,255: red,221; green,255; blue,221}]
\tikzstyle{X dot}=[Z dot, shape=circle, draw=black, fill=zxred, tikzit fill={rgb,255: red,255; green,136; blue,136}, tikzit category=zx]
\tikzstyle{X dot pi}=[Z dot, shape=circle, draw=black, fill=zxred, tikzit fill={rgb,255: red,255; green,136; blue,136}, tikzit category=zx, label={center:$\pi$}]
\tikzstyle{X bold dot}=[inner sep=0mm, minimum size=2mm, shape=circle, draw=black, fill=zxred, tikzit fill={rgb,255: red,255; green,136; blue,136}, line width=1.2pt, tikzit category=zx]
\tikzstyle{X phase dot}=[Z phase dot, tikzit shape=circle, tikzit draw=blue, fill=zxred, tikzit fill={rgb,255: red,255; green,136; blue,136}, font={\footnotesize\boldmath}, tikzit category=zx]
\tikzstyle{X tiny dot}=[inner sep=0mm, minimum size=1mm, shape=circle, draw=black, fill=zxred, tikzit fill={rgb,255: red,255; green,136; blue,136}]
\tikzstyle{hadamard}=[fill=yellow, draw=black, shape=rectangle, inner sep=0.6mm, minimum height=1.5mm, minimum width=1.5mm, tikzit category=zx]
\tikzstyle{paulibox}=[fill={rgb,255: red,221; green,221; blue,255}, draw=black, shape=rectangle, inner sep=0.6mm, minimum height=5mm, minimum width=5mm, font={\footnotesize}, text height=1.5ex, text depth=0.25ex, tikzit category=zx]
\tikzstyle{vertex}=[inner sep=0.2mm, minimum size=1mm, shape=circle, draw=black, fill=black, tikzit category=misc]
\tikzstyle{vertex set}=[inner sep=0.2mm, minimum size=1mm, shape=circle, draw=black, fill=white, font={\footnotesize\boldmath}, tikzit category=misc]
\tikzstyle{small black dot}=[fill=black, draw=black, shape=circle, inner sep=0pt, minimum width=1.2mm, tikzit category=circuit]
\tikzstyle{cnot ctrl}=[fill=black, draw=black, shape=circle, inner sep=0pt, minimum width=1.2mm, tikzit category=circuit]
\tikzstyle{cnot targ}=[fill=white, draw=white, shape=circle, tikzit category=circuit, label={center:$\oplus$}, inner sep=0pt, minimum width=2.1mm, tikzit fill={rgb,255: red,102; green,204; blue,255}, tikzit draw=black]
\tikzstyle{ket}=[fill=white, draw=black, shape=regular polygon, regular polygon sides=3, regular polygon rotate=-30, scale=0.7, inner sep=1pt, tikzit category=circuit, tikzit shape=rectangle, tikzit fill=green]
\tikzstyle{bra}=[fill=white, draw=black, shape=regular polygon, regular polygon sides=3, regular polygon rotate=30, scale=0.7, inner sep=1pt, tikzit category=circuit, tikzit shape=rectangle, tikzit fill=red]
\tikzstyle{scalar}=[shape=rectangle, text height=1.5ex, text depth=0.25ex, yshift=0.5mm, fill=white, draw=black, minimum height=5mm, yshift=-0.5mm, minimum width=5mm, font={\small}]
\tikzstyle{clabel}=[fill=white, draw=none, shape=rectangle, tikzit fill={rgb,255: red,56; green,255; blue,242}, font={\footnotesize}, inner sep=1pt, tikzit category=labels]
\tikzstyle{empty diagram}=[draw={gray!40!white}, dashed, shape=rectangle, minimum width=1cm, minimum height=1cm, tikzit category=misc]
\tikzstyle{amap}=[fill=white, draw=black, shape=NEbox, tikzit category=asymmetric, tikzit fill=yellow, tikzit shape=rectangle]
\tikzstyle{amap conj}=[fill=white, draw=black, shape=NWbox, tikzit category=asymmetric, tikzit fill=green, tikzit shape=rectangle]
\tikzstyle{amap adj}=[fill=white, draw=black, shape=SEbox, tikzit category=asymmetric, tikzit fill=red, tikzit shape=rectangle]
\tikzstyle{amap trans}=[fill=white, draw=black, shape=SWbox, tikzit category=asymmetric, tikzit fill=orange, tikzit shape=rectangle]
\tikzstyle{astate}=[fill=white, draw=black, shape=NEtriangle, tikzit category=asymmetric, tikzit shape=circle, tikzit fill=yellow]
\tikzstyle{astate conj}=[fill=white, draw=black, shape=NWtriangle, tikzit category=asymmetric, tikzit shape=circle, tikzit fill=green]
\tikzstyle{astate adj}=[fill=white, draw=black, shape=SEtriangle, tikzit category=asymmetric, tikzit shape=circle, tikzit fill=red]
\tikzstyle{astate trans}=[fill=white, draw=black, shape=SWtriangle, tikzit category=asymmetric, tikzit shape=circle, tikzit fill=orange]
\tikzstyle{box}=[shape=rectangle, text height=1.5ex, text depth=0.25ex, yshift=0.5mm, fill=white, draw=black, minimum height=5mm, yshift=-0.5mm, minimum width=5mm, font={\small}]
\tikzstyle{medium box}=[shape=rectangle, text height=1.5ex, text depth=0.25ex, yshift=0.5mm, fill=white, draw=black, minimum height=10mm, yshift=-0.5mm, minimum width=5mm, font={\small}]

\tikzstyle{simple}=[-]
\tikzstyle{hadamard edge}=[-, dashed, dash pattern=on 2pt off 0.5pt, thick, draw={rgb,255: red,68; green,136; blue,255}]
\tikzstyle{box edge}=[-, dashed, dash pattern=on 2pt off 0.5pt, thick, draw={rgb,255: red,203; green,192; blue,225}, fill=none]
\tikzstyle{blue box edge}=[-, dashed, dash pattern=on 2pt off 0.5pt, thick, draw={rgb,255: red,203; green,192; blue,225}, fill={rgb,255: red,230; green,219; blue,255}]
\tikzstyle{orange box edge}=[-, dashed, dash pattern=on 2pt off 0.5pt, thick, draw={rgb,255: red,203; green,192; blue,225}, fill={rgb,255: red,255; green,185; blue,164}]
\tikzstyle{brace edge}=[-, tikzit draw=blue, decorate, decoration={brace,amplitude=1mm,raise=-1mm}]
\tikzstyle{diredge}=[->]
\tikzstyle{double edge}=[-, double, shorten <=-1mm, shorten >=-1mm, double distance=2pt]
\tikzstyle{gray edge}=[-, {gray!60!white}]
\tikzstyle{pointer edge}=[->, very thick, gray]
\tikzstyle{boldedge}=[-, line width=1.2pt, shorten <=-0.17mm, shorten >=-0.17mm]
\tikzstyle{bidir edge}=[<->, very thick, draw={rgb,255: red,191; green,191; blue,191}]
\tikzstyle{surface X}=[-, tikzit fill=red, fill=zxred]
\tikzstyle{surface Z}=[-, tikzit fill=green, fill=zxgreen]

\input{quantum.tikzdefs}
\definecolor{mygreen}{rgb}{0.328125,0.6796875,0.1953125}
\definecolor{myblue}{rgb}{0.12156862745098039, 0.4666666666666667, 0.7058823529411765}
\usepackage{hyperref}
\hypersetup{
     colorlinks=true,
     linkcolor=myblue,
     filecolor=magenta,      
     urlcolor=myblue,
     pdftitle={General-state encoding circuits for universal fault-tolerant quantum computation},
     citecolor=myblue
}

\usepackage{amsmath}
\usepackage{mathtools}
\usepackage{amssymb}
\usepackage{nicefrac}
\usepackage{siunitx}
\usepackage{bbold}
\usepackage{braket}
\usepackage{hhline}
\usepackage{lipsum}
\usepackage{multirow}
\usepackage{hhline}

\usetikzlibrary{quantikz}
\usepackage{yquant}
\usepackage{rotating}
\usepackage{adjustbox}

\usepackage{amsfonts}
\usepackage{stmaryrd}
\usepackage[nameinlink]{cleveref}

\tikzset{noisy/.style={starburst,fill=red,draw=black,line width=1pt,inner xsep=1pt,inner ysep=1pt}}
\crefname{Circuit}{quantum circuit}{quantum circuit} %
\usepackage{newfloat}
\DeclareFloatingEnvironment[fileext=qua, listname={List of Quantum Circuits}, placement=ht] {Circuit}

\usepackage{enumitem}

\usepackage{subcaption}
\usepackage{ragged2e}
\DeclareCaptionJustification{justified}{\justifying}

\newcommand{\code}[1]{\llbracket #1 \rrbracket} %

\usepackage{tom}
\crefname{algocf}{Algorithm}{Algorithms}
\Crefname{algocf}{Algorithm}{Algorithms}

\newcommand{\SubEncoderFig}[2]{%
  \resizebox{!}{#1}{\input{#2.tikz}}%
}

\newcommand{\ExtractedFig}[2]{%
  \resizebox{!}{#1}{\input{#2.tikz}}%
}

\newcommand{\Tableau}[3][24pt]{%
  \begingroup
  \setlength{\arraycolsep}{2.2pt}%
  \renewcommand{\arraystretch}{0.95}%
  \resizebox{!}{#1}{$
    \left(
    \begin{array}{#2}
      #3
    \end{array}
    \right)
  $}%
  \endgroup
}

\begin{document}

\title{General-state encoding circuits for universal fault-tolerant quantum computation}
\title{Synthesis and Optimization of General-state Encoding Circuits}
\title{Methods for Synthesizing and Optimizing General Quantum Encoding Circuits}
\title{Synthesis and Optimization of General-state Encoding Circuits for Universal Fault-tolerant Quantum Computation}
\title{Synthesis and Optimization of Generalized State-Encoding Circuits for Fault-Tolerant Quantum Computation}
\title{Synthesis and Optimization of Encoding Circuits for Fault-Tolerant Quantum Computation}

\author{Tom Peham}
\email{tom.peham@tum.de}
\affiliation{Chair for Design Automation,
Technical University of Munich, 80333 Munich,
Germany}

\author{Matthew Steinberg}
\affiliation{\color{black} Global Technology Applied Research, JPMorgan Chase, New York, NY 10017 USA}
\affiliation{QuTech, Delft University of Technology, 2628 CJ Delft, The Netherlands}
\affiliation{Quantum and Computer Engineering Department, Delft University of Technology, 2628 CD Delft, The Netherlands}

\author{Robert Wille}
\email{robert.wille@tum.de}
\affiliation{Chair for Design Automation,
Technical University of Munich, 80333 Munich,
Germany}

\author{Sascha Heußen}
\email{s.heussen@neqxt.org}
\affiliation{\href{https://www.neqxt.org/}{\color{black} neQxt, 50670 Cologne, Germany}}

\begin{abstract}
  Preparing arbitrary logical states is a central primitive for universal fault-tolerant quantum computation and the cost of encoded-state preparation contributes directly to the overall resource overhead.
  This makes the synthesis of efficient general-state encoding circuits an important problem, particularly with respect to two-qubit gate count and circuit depth.
  Yet the synthesis of such encoders has been studied less extensively than general Clifford circuit synthesis or the preparation of specific logical Pauli-eigenstates.
  In this work, we develop methods for synthesizing efficient encoders for arbitrary stabilizer codes.
  We formulate encoder synthesis as a search over stabilizer tableaus and introduce greedy and rollout-based algorithms that exploit the freedom among stabilizer-equivalent realizations of the same encoding isometry.
  For code families with a modular structure, such as generalized concatenated and holographic codes, we show how large encoders can be assembled from optimized local constituent encoders, and we use SMT-based exact synthesis to obtain optimal local circuits for small instances.
  We further evaluate the proposed methods on a broad set of stabilizer codes, including holographic and quantum low-density parity-check (qLDPC) codes, and compare them against recent encoder-synthesis methods and existing constructions from the literature, obtaining improvements of up to $43\%$ in two-qubit gate count and up to $70\%$ in depth.
  Our results support the optimization of encoded-state preparation in several fault-tolerant quantum-computing schemes, and all methods are openly available as part of the Munich Quantum Toolkit~\cite{wille2024mqthandbook}.
\end{abstract}

\maketitle

\section{Introduction}

Due to the inherent noise present in any quantum system, utility-scale quantum computing (QC) is widely believed to require quantum error correction to actively correct errors occurring during a computation. The concept of \emph{quantum error correction} (QEC) codes, which encode logical qubits redundantly into entangled quantum states, has developed into a critical research field towards the realization of practically useful quantum computers~\cite{lidar2013quantum,devitt2013quantum,google2024quantumerrorcorrection,pogorelov2025experimentalfaulttolerant,ryananderson2024highfidelityfaulttolerantteleportationlogical,bluvstein2024logicalquantumprocessor}.
A fundamental step in any fault-tolerant~(FT) architecture is the preparation of encoded states, necessitating the synthesis of circuits that map input states into the logical code space.

Such circuits are of particular importance in resource-intensive workflows such as magic state distillation, which repeatedly consume encoded resource states~\cite{bravyi2005universalquantumcomputation}. In some constructions, decoding (or unencoding) circuits---the adjoints of the corresponding encoders---also appear explicitly as part of the protocol~\cite{beverland2021costofuniversality,rodriguez2024experimentaldemonstrationlogicalmagic}. Encoding circuits are also relevant for logical magic state preparation and cultivation protocols, wherein encoded magic states are prepared and then verified by measuring logical operators and stabilizers with postselection. In such settings, the structure of the preparation and measurement circuits directly affects both the resource cost and the success characteristics of the protocol~\cite{chamberland2019faulttolerantmagic,gidney2024magicstatecultivationgrowing,sahay2026foldtransversalsurfacecodecultivation}.
Finally, optimized encoders are also attractive in modular or distributed architectures, where the physical qubits of a logical block may be spread across multiple modules and nonlocal operations are constrained by inter-module communication and entanglement generation~\cite{tham2025distributedfaulttolerantquantummemories,clayton2025distributedquantumerrorcorrection}.

These applications show that encoded-state preparation can contribute directly to the overall resource overhead of FT quantum computation.
Optimizing encoding circuits with low two-qubit gate count and low depth is therefore an important problem despite the fact that such encoding circuits cannot, in general, be fault-tolerant for arbitrary input states, since errors on the input qubits are directly mapped to logical errors in the code space.

Encoding circuits for stabilizer codes can be implemented using only gates from the Clifford gate set.
The synthesis and optimization of Clifford and CNOT circuits have been extensively studied.
Existing approaches range from normal-form constructions~\cite{gottesman2026survivingquantumcomputer,gidney2021stimfaststabilizer,lu2024universalgraphrepresentationstabilizer} to heuristic and optimal synthesis methods~\cite{webster2025heuristicoptimalsynthesis,peham2023depthoptimalsynthesis,bravyi2022qubitoptimalclifford,bravyi2021cliffordcircuit,patel2008optimalsynthesis}, but they are primarily designed for unitary Clifford operators.
Encoding circuits, by contrast, realize \emph{isometries} rather than full unitaries.
A generic approach is therefore to extend the target isometry to a unitary and then apply a unitary-synthesis method.
Crucially, this perspective misses a key feature of the encoding problem; namely, that an encoder is not specified uniquely as a full unitary, since only its action on the logical subspace matters. As a result, generic unitary methods fail to exploit the resulting optimization freedom and can produce unnecessarily large circuits.

In this work, we develop methods for synthesizing and optimizing encoding circuits for two complementary settings.
For arbitrary stabilizer codes, we formulate encoder synthesis as a search over stabilizer tableaus and develop greedy and rollout-based methods that optimize two-qubit gate count and circuit depth.
For code families whose encoders admit a modular decomposition, including generalized concatenated~\cite{yamasaki2024time,yoshida2025concatenate,knill1996concatenated} and holographic codes~\cite{fan2024lego_hqec,jahn2021holographic,pastawski2015holographic,harris2018calderbank,harris2020decoding,steinberg2025far,steinberg2025universal}, we reduce the problem to the optimization of local constituent encoders and use SMT-based exact synthesis to solve these local instances optimally.
Across a broad benchmark set, these methods yield competitive and often improved circuits compared to recent encoder-synthesis methods and existing constructions from the literature, with improvements of up to $43\%$ in two-qubit gate count and up to $70\%$ in depth.

The main contributions of this work are as follows:
\begin{itemize}
\item We develop a greedy synthesis method for arbitrary stabilizer encoders that exploits the degrees of freedom of the stabilizer group to reduce two-qubit gate count and circuit depth.

\item We propose a rollout-based extension that further improves synthesis quality by evaluating candidate gate applications through greedy completion of the remaining synthesis task.

\item We show how modular structure in generalized concatenated and holographic code constructions can be exploited to synthesize large encoding circuits from optimized local constituent encoders, and we develop SMT-based exact-synthesis techniques for the resulting small local instances.

\item We evaluate the proposed methods on a range of stabilizer codes and benchmark them against recent encoder-synthesis methods, including Rustiq~\cite{goubaultdebrugiere2025graphstatebased}, obtaining competitive and often improved circuits. For example, our methods yield a $49$-CNOT general-state encoder for the $\code{23,1,7}$ Golay code, an $82$-gate depth-$13$ encoder for the $\code{30,6,5}$ symplectic double code~\cite{kanomata2025fault}, and a logical $\ket{0}^{\otimes 12}$ preparation circuit for the $\code{144,12,12}$ bivariate bicycle (BB) code~\cite{bravyi2024highthreshold} requiring $330$ CNOT gates at a depth of $17$.
\end{itemize}

All developed methods are available as part of the Munich Quantum Toolkit~\cite{wille2024mqthandbook}.

The remainder of this work is structured as follows.
\Cref{sec:background} reviews stabilizer tableaus, defines the encoding circuit synthesis problem, and summarizes the relevant literature on Clifford circuit synthesis.
\Cref{sec:encod-circ-synth} presents the state-space search formulation and the rollout-based synthesis method for arbitrary stabilizer and CSS codes.
\Cref{sec:encod-circ-conc} shows how encoding circuits for generalized concatenated codes can be optimized via their local constituent encoders and introduces an optimal SMT-based synthesis method for the resulting local instances.
\Cref{sec:circuits} presents encoding circuits obtained using the proposed methods, and \Cref{sec:conclusion} concludes this work.

\section{Background}
\label{sec:background}

To keep this work as self-contained as possible, we provide the necessary background on stabilizer tableaus and Clifford encoding isometries in the following.

\subsection{Symplectic Representations of Stabilizer Codes}
\label{sec:symp}

Given a (multi-qubit) Pauli operator $\calP$ on $n$ qubits, we denote the \emph{symplectic} representation of $\calP$ as $\overline{\calP} \in \F_2^{2n}$.
To keep the notation lightweight, we use the same symbols for Pauli operators (or sets of Pauli operators) and for their symplectic representations whenever the intended meaning is clear from context.
Similarly, the symplectic representation of a sequence of $n$-qubit Pauli operators \(\calO = \calP_1, \ldots, \calP_m\) is defined as

\[\overline{\calO} =
  \begin{pmatrix}
    \overline{\calP}_1\\
    \vdots\\
    \overline{\calP}_m\\
  \end{pmatrix} \in \F_2^{m \times 2n}~.
\]

We refer to such a symplectic representation as the \emph{stabilizer tableau} of \(\calO\).
It is useful to think of a stabilizer tableau \(T\) in block matrix representation
\begin{equation}
  \label{eq:stab-tableau}
  T =
  \begin{pmatrix}
    T_X & T_Z \\
  \end{pmatrix}~,
\end{equation}
where \(T_X\) is the submatrix denoting the \(X\)-parts of the Pauli operators in \(\calO\) and \(T_Z\) denotes the \(Z\)-parts of the Pauli operators in \(\calO\).
We denote the \(i\)-th column of \(T_X\) (\(T_Z\)) as \(T_X[i]\) (\(T_Z[i]\)).

\begin{example}
  Consider the stabilizer generators of the two-qubit Bell state \(S = \{XX, ZZ\}\).
  The tableau representation of \(S\) is then
\[
  \begin{pmatrix}
    1 & 1 & 0 & 0 \\
    0 & 0 & 1 & 1
  \end{pmatrix},~\]
with $T_X=
\begin{pmatrix}
  1 & 1 \\ 0 & 0
\end{pmatrix}
$ and $T_Z=
\begin{pmatrix}
  0 & 0 \\ 1 & 1
\end{pmatrix}
$.
\end{example}

Stabilizer codes can be represented using stabilizer tableaus.
Let $C$ be an $\code{n,k,d}$ stabilizer code with stabilizer group $\mathcal{S}$ with independent stabilizer generators $\{s_1,\dots,s_{n-k}\}$.
The logical \(X\) operators \mbox{\(\calX = \{\calX_1, \ldots, \calX_k\}\)} and logical \(Z\) operators \mbox{\(\calZ= \{\calZ_1, \ldots, \calZ_k\}\)} are chosen such that \(\calX_i\) anticommutes with \(\calZ_i\) and \(\calX_i\) commutes with \(\calZ_j\) if \(i\neq j\), i.e., $\calX_i, \calZ_i$ denote the logical operators of the $i$-th logical qubit.
The symplectic representation of the code is then given by

\[\overline{\calC} =
  \begin{pmatrix}
    \overline{\calX}\\
    \overline{\calZ}\\
    \overline{S}    \\
  \end{pmatrix}
  \in \F_2^{(n+k)\times 2n}~.
\]

\begin{example}
  The five-qubit code is defined by the stabilizer generators \(\{XZZXI, IXZZX, XIXZZ, ZXIXZ\}\).
  A choice of logical basis is \(\calX = ZZZZZ\), \(\calZ = XXXXX\).
  The symplectic representation of this code is then

  \[
    \begin{pmatrix}
      0 & 0 & 0 & 0 & 0 & 1 & 1 & 1 & 1 & 1 \\
      1 & 1 & 1 & 1 & 1 & 0 & 0 & 0 & 0 & 0 \\\midrule
      1 & 0 & 0 & 1 & 0 & 0 & 1 & 1 & 0 & 0 \\
      0 & 1 & 0 & 0 & 1 & 0 & 0 & 1 & 1 & 0 \\
      1 & 0 & 1 & 0 & 0 & 0 & 0 & 0 & 1 & 1 \\
      0 & 1 & 0 & 1 & 0 & 1 & 0 & 0 & 0 & 1
    \end{pmatrix}~.
  \]
\end{example}

\subsection{Clifford Encoding Isometries}
\label{sec:enc-iso}

An encoding isometry from \(k\) qubits to \(n\) qubits is an operator \(\calE \in \C^{2^n \times 2^k}\) such that \(\calE^\dagger \calE = I_{2^k}\).
An encoder \(\calE\) is a Clifford isometry if it maps Pauli operators to Pauli operators under conjugation, and Clifford isometries can be interpreted as defining an \(\code{n,k}\) stabilizer code.
The logicals of the respective code are obtained from propagating Pauli \(X\) and \(Z\) through the isometry, i.e., \(\calX_i = \calE X_i \calE^\dagger\) and \(\calZ_i = \calE Z_i\calE^\dagger\).
The stabilizers of the code are then all Pauli operators \(S\) such that \(S\calE = \calE\), or equivalently, \(\calE^\dagger S \calE = I_{2^k}\).

An \emph{encoding circuit}, then, is a concrete quantum circuit implementing a specific encoding isometry.
Due to reversibility, such a quantum circuit must act on \(n\) input qubits to produce \(n\) output qubits.
In addition to the \(k\) input qubits encoded by the isometry, \((n-k)\) extra ancillary qubits initialized in a fixed state are therefore needed.
Without loss of generality, we will assume that the ancillary qubits are initialized in the state \(\ket{0}^{\otimes (n-k)}\).
The action of the encoding circuit $U$ on any \(k\)-qubit state and the ancillary system is then the same as the encoding isometry, i.e., for any \(\ket{\psi} \in \calH^{2^k}\)  

\[
\calE \ket{\psi} = U \left(\ket{\psi}\otimes \ket{0}^{\otimes (n-k)}\right).
\]

The stabilizers of the code defined by the encoding isometry can be obtained from the encoding circuit by using Pauli propagation rules to commute the single-qubit \(Z\)-stabilizers of the qubits initialized in \(\ket{0}\).
Specifically, the \(i\)-th stabilizer generator \(S_i\) is given as \(U Z_{k+i} U^\dagger\).
Consequently, given an encoding circuit \(U\), the symplectic representation of the corresponding code $\calC$ can be obtained by propagating Pauli operators:
\[
C =
\begin{pmatrix}
\{U X_i U^\dagger \mid i \in [k]\} \\
\{U Z_i U^\dagger \mid i \in [k]\} \\
\{U Z_{k+i} U^\dagger \mid i \in [n-k]\}
\end{pmatrix}.
\]
Concretely, this means that, on the $k$ input qubits that are not initialized to $\ket{0}$,
the encoding circuit propagates both a single $X$ and a single $Z$ to the corresponding logical $X$ and logical $Z$ at the same time.
As an example, one may consider a magic state $\rho_H = \ket{H}\bra{H} = (I + (X+Z)/\sqrt{2})/2$, for which the coherent superposition of $X$ and $Z$ should be maintained by the encoding circuit.

\subsection{The Encoding Circuit Synthesis Problem}
\label{sec:enc-synthesis}

Implementing an encoding circuit requires synthesizing it from a set of basis gates.
For Clifford circuits, a typical choice is $\{S, H, CX\}$, as those gates are sufficient to implement any Clifford unitary.
Applying a Clifford gate to a Clifford circuit has the effect of performing some column operations on the symplectic representation of the circuit as dictated by how Pauli operators propagate through the gate: For instance, applying an \(S\) gate on qubit $i$ to a stabilizer tableau \(T = \left(T_X T_Z\right)\) is realized via the column operation

\[T_Z[i] = T_Z[i] + T_X[i]~,\]

where addition is performed over $\F_2$, i.e., addition modulo $2$.

Two-qubit gates act on the columns of multiple qubits simultaneously.
Applying a \(CX\) with control qubit \(i\) and target qubit \(j\) to \(T\) corresponds to the following column operations

\[T_X[j] = T_X[i] + T_X[j] \qquad T_Z[i] = T_Z[i] + T_Z[j]~.\]

The $n$-qubit identity Clifford operator has the $2n$-dimensional identity matrix $I_{2n}$ as its symplectic representation. Synthesizing a general Clifford circuit \(U\) from a set of basis Clifford gates \(\Gamma\), then boils down to finding a sequence of gates \(g_1, \ldots, g_m \in \Gamma\) such that

\[g_m\cdots g_1 \overline{U} = I_{2n}\]

or, equivalently,

\[\overline{U} = g_1^\dagger \cdots g_m^\dagger I_{2n}~.\]

The synthesis problem is similar for encoding circuits.
The difference is that the given Pauli operators do not need to propagate precisely to the given set of logicals and stabilizers. 
Instead, the set of propagated logicals and stabilizers of the encoding circuit only needs to be equivalent to the given operators up to multiplication by some stabilizer operator of the encoder.
Synthesizing a given encoding isometry \(\calE\) encoding $k$ qubits into $n$ qubits from a set of basis Clifford gates \(\Gamma\), then requires finding a sequence of gates \(g_1, \ldots, g_m \in \Gamma\) such that

\[g_m\cdots g_1 \overline{\calE} = A\]

where \(A\) is a \((k+n) \times 2n\) matrix that can be transformed into the form

\begin{equation}
  \label{eq:tableau-init}
\left(
\begin{array}{cc|cc}
I_k & 0 & 0 & 0\\
0 & 0 & I_k & 0\\\hline
0 & 0 & 0 & I_{n-k}
\end{array}
\right)
\end{equation}

via row additions of the lower \((n-k)\) rows representing the stabilizer generators.
The matrix in~\Cref{eq:tableau-init} represents the identity isometry which encodes $k$ qubits into $n$ qubits by simply padding the $k$ input qubits with $(n-k)$ qubits in the $\ket{0}$ state.

Note that Pauli operators usually also carry a sign~\cite{aaronson2004improvedsimulationstabilizer}.
Ignoring the sign of the Pauli operators in the stabilizer tableau in synthesis might result in circuits that implement the target encoding isometry up to correction of signs.
Since these signs can always be corrected via single-qubit Pauli operators, we ignore them for the synthesis.

\subsection{Encoding Isometries as ZX-diagrams}
\label{sec:ZX}

Another way of representing encoding isometries is through ZX-diagrams, a diagrammatic language for linear maps of qubits~\cite{vandewetering2020zxcalculusworkingquantumcomputer}.
ZX-diagrams are composed of two types of nodes, the so-called $Z$- and $X$-spiders:

\medskip

\input{zx/z_spider.tikz} $= \ket{0}^{\otimes n} \bra{0}^{\otimes m} + e^{i\alpha} \ket{1}^{\otimes n} \bra{1}^{\otimes m}$

\medskip

\input{zx/x_spider.tikz} $= \ket{+}^{\otimes n} \bra{+}^{\otimes m} + e^{i\alpha} \ket{-}^{\otimes n} \bra{-}^{\otimes m}$

\medskip

\begin{figure}[t]
\begin{center}
\resizebox{.45\columnwidth}{!}{\input{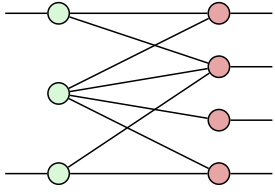}}  
\end{center}
\caption{A ZX diagram representation of the $\code{4,2,2}$ code's encoding isometry.}
\label{fig:422-zx}
\end{figure}

These building blocks are sufficient to represent any quantum linear map. 
In particular, any encoding isometry can be represented as a ZX-diagram.

Unlike the stabilizer tableau, a diagrammatic representation of an encoder can provide insight into the entanglement structure of the code~\cite{lu2024universalgraphrepresentationstabilizer}.
For example, the ZX-diagram representation of the encoding isometry for the $\code{4,2,2}$ is given in~\Cref{fig:422-zx}.
For details on the notation and terminology required for understanding the ZX diagrams in this work, please consult \Cref{sec:zx-diagrams}.

ZX-diagrams and the ZX-calculus have enjoyed increased attention over the recent years as tools in quantum circuit optimization~\cite{kissinger2020reducingnumbernonclifford,riu2025reinforcement,villorial2026optimisationsynthesis} and as a language for quantum error correction and fault-tolerant quantum computing~\cite{kissinger2022phasefreezxdiagramscss,wan2026holographiccodesseenzxcalculus,rodatz2025faulttoleranceconstruction,townsend2023floquetifying}.
While stabilizer tableaus are convenient for synthesis based on matrix elimination, the insights into the structure of a code are useful for scaling encoding circuit synthesis for structured codes, as we will see in~\Cref{sec:encod-circ-conc}.

\subsection{Related Work}
\label{sec:related_work}

The literature on Clifford circuit synthesis is extensive, but most of it focuses on \emph{unitary} Clifford operators rather than encoding isometries. Early work established asymptotically efficient normal-form constructions for Clifford circuits. In particular, Ref.~\cite{aaronson2004improvedsimulationstabilizer} gives a normal-form construction for arbitrary Clifford unitaries with $O(n^2/\log n)$ gates, matching the asymptotic complexity known for CNOT circuits~\cite{patel2008optimalsynthesis}. Beyond gate count, depth bounds are also known: Ref.~\cite{jiang2020optimalspacedepth} shows that any Clifford circuit can be implemented with depth $O(n/\log n)$ without ancillae and with depth $O\!\left(\max\!\left\{\log n,\frac{n^2}{(n+m)\log(n+m)}\right\}\right)$ using $m$ ancillary qubits.

In practice, we are interested in not only obtaining asymptotically optimal circuits, but circuits that are optimal in the absolute sense.
More recently, exact and heuristic methods have been developed for optimizing Clifford unitaries with respect to gate count and depth~\cite{bravyi2022qubitoptimalclifford,peham2023depthoptimalsynthesis,webster2025heuristicoptimalsynthesis}.
These works provide important background on Clifford synthesis, but they do not directly address the additional freedom present in encoding-circuit synthesis.

A related but narrower line of work considers special cases of encoder synthesis.
State-preparation circuits, corresponding to the preparation of particular logical basis states, have been optimized using automated heuristic methods~\cite{peham2024automatedsynthesisfaulttolerantstate} and reinforcement learning~\cite{zen2025quantumcircuitdiscovery}.
These approaches target an important subclass of the problem, but do not address the synthesis of general-state encoders for arbitrary stabilizer codes.

\begin{figure*}[t]
  \centering
  \includegraphics[width=\linewidth]{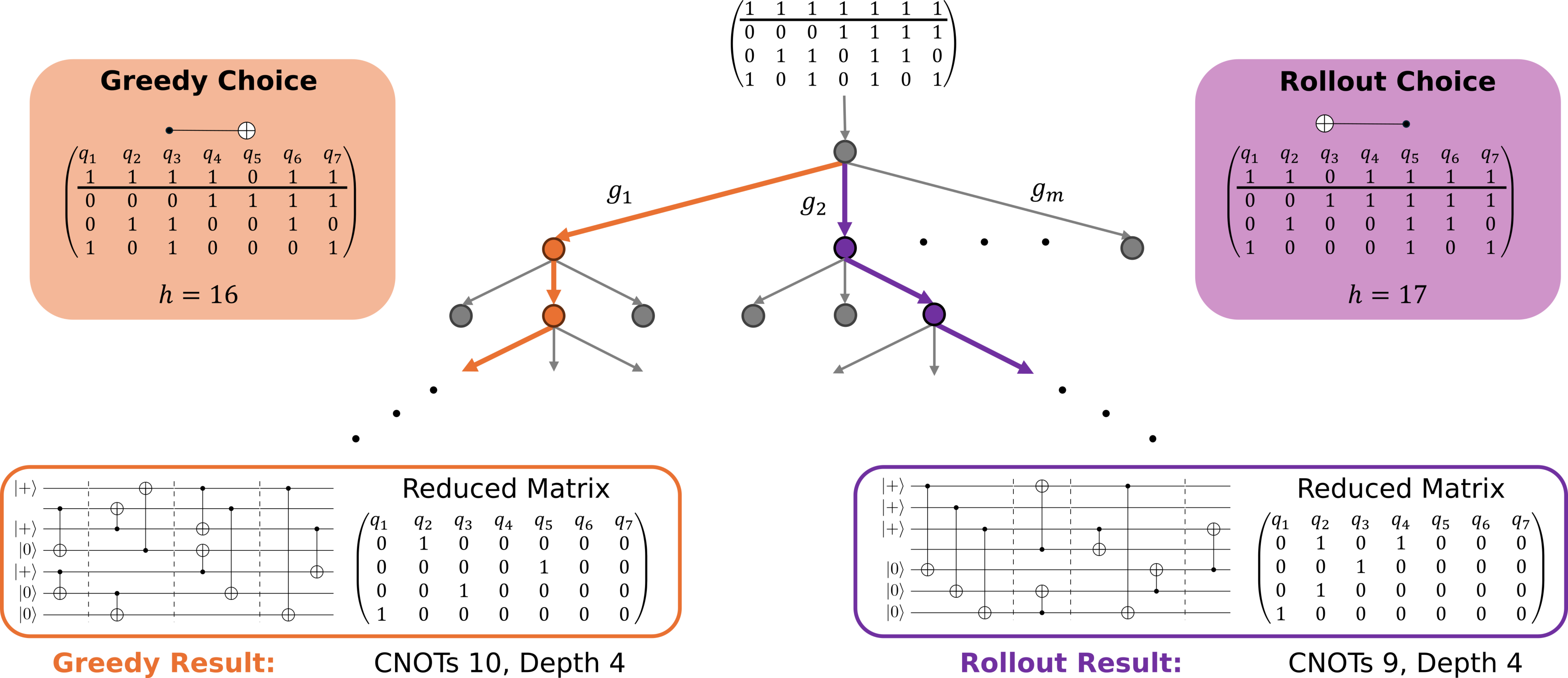}
  \caption{Encoding circuit synthesis as state space search. Every state corresponds to an intermediate circuit and a stabilizer tableau, and states are connected via gate applications.
    Goal states represent valid implementations of the target encoding isometry. A heuristic score $h$ is used to estimate the cost of synthesizing the remaining circuit for a given node.
    The orange path is obtained by greedily choosing the gate that minimizes $h$.
    The purple path is obtained by performing this greedy synthesis from the second-best node on the second level.
    The rollout approach fully expands the greedy path of both states and picks the best one.
    This results in the rollout search picking the purple path as the resulting circuit requires fewer CNOTs even though locally, the heuristic score estimates that the orange path results in a circuit requiring fewer gates.}
  \label{fig:statespace}
\end{figure*}

Closer to our setting are methods that explicitly synthesize encoding isometries. Reference~\cite{gottesman2026survivingquantumcomputer} shows that encoding circuits can be obtained by a tableau-based elimination procedure closely analogous to Gaussian elimination. The result is a simple constructive synthesis method but because columns are eliminated one at a time, the resulting circuits can incur substantial gate-count and depth overhead. Reference~\cite{lu2024universalgraphrepresentationstabilizer} studies normal forms and ZX-diagram representations for encoding isometries, giving a graphical perspective on encoder synthesis. A particularly relevant recent approach is the graph-state-based synthesis method of Ref.~\cite{goubaultdebrugiere2025graphstatebased}, which extracts an underlying graph-state structure from the stabilizer description and synthesizes the encoder by transforming this graph state. This technique provides a dedicated optimization framework for encoding circuits and serves as an important point of comparison for our methods.

Related structured constructions also appear in work on holographic codes. In particular, graph states and ZX-diagram representations have been used to derive encoding isometries for holographic-code families~\cite{angles2024engineering,wan2026holographiccodesseenzxcalculus}. In these approaches, a graph state representing a logical basis state is combined with the bulk logical qubits, and the remaining auxiliary qubits are measured to obtain an encoding map for arbitrary inputs. These works highlight the usefulness of graphical structure for deriving encoders, but do not study the optimization of general-state encoding circuits via local resynthesis of constituent encoders.

Overall, existing work either studies general Clifford unitary synthesis, gives constructive but not strongly optimized methods for encoding isometries, or focuses on special cases such as state preparation. In contrast, our work develops search-based methods for synthesizing and optimizing \emph{general-state} encoding circuits, explicitly exploits the freedom among stabilizer-equivalent realizations, and complements this with a modular synthesis framework for structured code families.

\section{Heuristic Encoding Circuit Synthesis}
\label{sec:encod-circ-synth}

Even for relatively small system sizes ($n<10$), synthesizing the gate- or depth-optimal general state encoder becomes intractable. 
Since we want to construct encoders for larger system sizes while keeping the number of required two-qubit gate operations and circuit depth low, we now describe a heuristic approach for synthesizing encoding circuits. 
Our method constructs the circuit sequentially while maintaining a stabilizer tableau representing the transformation implemented so far.
The novelty of our synthesis approach is that we explicitly exploit the degrees of freedom afforded by stabilizer row operations to construct Clifford isometries.

The heuristic synthesis process can be viewed as a search over stabilizer tableaus (see~\Cref{fig:statespace}). At every step of the algorithm, a Clifford gate is applied, transforming the current tableau into a new one. The goal of the search is therefore to apply a sequence of gates that transforms the tableau of the target encoding isometry into the terminal tableau defined in~\Cref{eq:tableau-init}.

\subsection{Encoding Circuits using Symplectic Transvections}
\label{sec:heur-encod-circ}

The search described above utilizes a set of Clifford operations that transform one tableau into another. Following previous work on Clifford circuit synthesis, we consider two-qubit symplectic transvections~\cite{pllaha2021decompositioncliffordgates}. A two-qubit transvection acting on qubits \(i\) and \(j\) is a Clifford operator of the form \[ \exp \left(\frac{i\pi}{4}(I - P_i P_j)\right), \] where \(P_i, P_j \in \{X,Y,Z\}\).

Since there are three non-identity Pauli operators on each qubit, there are nine distinct two-qubit transvections for every ordered pair of qubits. Consequently, for an \(n\)-qubit stabilizer tableau, there are \(9n(n-1)\) possible transitions at each step of the search. To determine whether a given transvection represents a promising step toward a goal state, the search requires a measure of progress, or \emph{heuristic score}.

A useful heuristic for guiding the search was introduced by Webster et al.~\cite{webster2025heuristicoptimalsynthesis}. Their method evaluates the structural properties of a stabilizer tableau to estimate how close it is to the terminal tableau. 
Concretely, the heuristic assigns to each tableau $T$ a score $h(T)$ that measures the number of remaining two-qubit interactions required to reach the terminal tableau.

For a tableau $T$, consider the $2\times2$ submatrices
\begin{equation}
  \label{eq:f}
F_{ij} =
\begin{pmatrix}
  T_{ij} & T_{i(j+n)} \\
  T_{(i+n)j} & T_{(i+n)(j+n)}
\end{pmatrix},
\end{equation}
which capture the interaction between qubits $i$ and $j$ in the
symplectic representation. Using these blocks, define the indicator matrices
\begin{align}
  \label{eq:r1-r2}
  R_1(T)_{ij} &=
  \begin{cases}
    1, & \text{if } \mathrm{rk}(F_{ij}) = 1, \\
    0, & \text{otherwise},
  \end{cases} \\
  R_2(T)_{ij} &=
  \begin{cases}
    1, & \text{if } \mathrm{rk}(F_{ij}) = 2, \\
    0, & \text{otherwise}.
  \end{cases}
\end{align}

Webster et al.~show that if \(R_1\) is all zero and \(R_2\) is a permutation matrix for a given $i$ and $j$, then the remaining stabilizer tableau is equivalent to the identity up to single-qubit Cliffords and qubit permutations.
This observation leads to a heuristic score that estimates how far a given stabilizer tableau is from the identity.
We define
\begin{align}\label{eq:cost-non-css}
v(T) &\coloneqq \mathrm{colSums}\!\left(R_2(T) + \tfrac{1}{n} R_1(T)\right), \\
h_{\mathrm{greedy}}(T) &\coloneqq \mathrm{sorted}\big(v(T), v(T^\top)\big).
\end{align}
Here, $\mathrm{colSums}(M)$ denotes the vector whose $j$-th entry is the sum of the entries in the $j$-th column of $M$, and $\mathrm{sorted}(u,v)$ denotes the pair obtained by sorting the entries of $u$ and $v$ in nonincreasing order before comparison.

This heuristic is formulated for the synthesis of Clifford unitaries and therefore assumes a complete \(2n\times 2n \) stabilizer tableau.
Encoding circuits, however, correspond to isometries and do not uniquely specify the full tableau. To apply the heuristic in this setting, we complete the tableau of the encoder to a valid Clifford tableau and perform the synthesis on this extended representation.
The symplectic representation of a code can be completed to a $2n\times 2n$ symplectic matrix in the following fashion.

Given $(n-k)$ stabilizer generators $S=\{s_1, \ldots, s_{n-k}\}$, we need to define destabilizer operators $D=\{d_1, \ldots, d_{(n-k)}\}$ such that
\begin{itemize}
\item All $d_i$ commute with each other.
\item $d_i$ anticommutes with $s_i$.
\item $d_i$ commutes with $s_j$ if $i \neq j$.
\end{itemize}

Such a set $D$ can be constructed using~\Cref{alg:destabilizers}.

\begin{algorithm}[t]
  \caption{Complete stabilizer tableau}
  \label{alg:destabilizers}
\KwIn{Independent stabilizer generators $\{s_1,\dots,s_{n-k}\}$ and their symplectic rows $\{\bar s_1,\dots,\bar s_{n-k}\}\subseteq \F_2^{2n}$.}
\KwOut{Destabilizer operators $\{d_1,\dots,d_{n-k}\}$.}

$\mathcal D \leftarrow \emptyset$\;
\For{$i=(n-k)$ \KwTo $1$}{
    \eIf{$i=n-k$}{
        Find a weight-one Pauli operator $b$ that anticommutes with $s_i$\;
    }{
        Compute a basis $B$ for the orthogonal complement of $\mathrm{span}\{\bar s_{i+1},\dots,\bar s_{n-k}\}$\;
        Pick $\bar b \in B$ such that the corresponding Pauli operator $b$ anticommutes with $s_i$\;
    }
    $A \leftarrow \{s_j  \mid d_j \in \calD, d_j \text{ anticommutes with } b\}$\;
    $d_i \leftarrow b \cdot \prod_{p\in A} p$\;
    $\mathcal D \leftarrow \mathcal D \cup \{d_i\}$\;
}
\Return $\mathcal D$\;
\end{algorithm}

This completion is not unique. Different choices of stabilizer generators yield different but equivalent tableau representations of the same encoding isometry. 
Since the heuristic depends on the tableau structure, these choices can significantly influence the behavior of the greedy search and the quality of the resulting circuit. 
In particular, a stabilizer generator \(s_i\) may be replaced by \(s_i s_j\) for \(i \neq j\) without changing the stabilizer group and therefore without modifying the encoded isometry. 
Such transformations correspond to elementary row operations on the stabilizer part of the tableau. 
To maintain the symplectic structure, the corresponding destabilizer generators must be updated accordingly; replacing \(s_i\) by \(s_i s_j\) requires replacing \(d_j\) by \(d_i d_j\).
Similarly, logical generators may be multiplied by stabilizers, so \(\calX_i\) or \(\calZ_i\) may be replaced by \(\calX_i s_j\) or \(\calZ_i s_j\) without changing the encoded isometry.

We exploit this freedom by introducing a preprocessing step that searches for a tableau representation with a lower heuristic score before starting the greedy synthesis. 
In terms of the search-space interpretation discussed above, these row transformations correspond to moving between equivalent starting nodes that represent the same encoding isometry. 
The heuristic, therefore, provides a measure of how promising a given starting point appears.
Starting from the completed tableau of the encoder, we greedily apply such transformations whenever they reduce the heuristic score. 
In this way, the preprocessing step selects a starting representation that is heuristically closer to the goal state before the synthesis begins.
The resulting preprocessing procedure is summarized in~\Cref{alg:preprocess}.
Here, $X_i$, $Z_i$, $S_i$, and $D_i$ denote the $i$-th rows of the matrices $X$, $Z$, $S$, and $D$, respectively.

\begin{algorithm}[t]
  \footnotesize
\caption{Pre-optimize stabilizer tableau}
\label{alg:preprocess}
\KwIn{Logical row matrices $X,Z \in \F_2^{k\times 2n}$, stabilizer matrix $S \in \F_2^{(n-k)\times 2n}$, heuristic $h$.}
\KwOut{Preprocessed completed tableau $T \in \F_2^{2n\times 2n}$.}

Compute a compatible destabilizer matrix $D$ using Algorithm~1\;
$T \leftarrow (X^\top\ D^\top\ Z^\top\ S^\top)^\top$\;
\Repeat{no improving update is found}{
    $T_{\mathrm{best}} \leftarrow T$\;
    $h_{\mathrm{best}} \leftarrow h(T)$\;
    Let $\mathcal U(T)$ be the set of all updates of the forms
    \[
    S_j \leftarrow S_j + S_i,\qquad D_i \leftarrow D_i + D_j \qquad (i\neq j),
    \]
    \[
    X_i \leftarrow X_i + S_j,\qquad D_j \leftarrow D_j + Z_i,
    \]
    \[
    Z_i \leftarrow Z_i + S_j,\qquad D_j \leftarrow D_j + X_i.
    \]
    \ForEach{$U \in \mathcal U(T)$}{
        $T' \leftarrow U(T)$\;
        \If{$h(T') < h_{\mathrm{best}}$}{
            $T_{\mathrm{best}} \leftarrow T'$\;
            $h_{\mathrm{best}} \leftarrow h(T')$\;
        }
    }
    $T \leftarrow T_{\mathrm{best}}$\;
}
\Return $T$\;
\end{algorithm}

After preprocessing, the synthesis proceeds by greedily exploring the search space.
Starting from the optimized tableau, the algorithm repeatedly applies the two-qubit transvection that minimizes the heuristic score of the resulting tableau.
Concretely, for the current tableau $T$, we evaluate all candidate transvections $\tau$ and compute the heuristic score $h_{\mathrm{greedy}}(\tau(T))$. The transvection that yields the smallest heuristic score is then applied, and the process repeats until the terminal tableau defined in~\Cref{eq:tableau-init} is reached.

We can also use a greedy heuristic to optimize depth,  by using an idea introduced in~\cite{peham2024automatedsynthesisfaulttolerantstate}, where circuits are constructed layerwise:
Instead of picking the two-qubit gates that optimize \(h\), we only consider two-qubit gates not involving qubits already used in this layer. 
As soon as no two-qubit gate involving qubits not already used in the current layer improves $h$, the synthesis continues with a new circuit layer. While this means the search may pick CNOTs that do not reduce $h$ by much, the layered approach steers the search towards shallower circuits.

\subsection{CNOT Encoding Circuit Synthesis for CSS Codes}
\label{sec:css}

Calderbank-Shor-Steane (CSS) codes~\cite{calderbank1996goodquantumerrorcorrecting,steane1996multipleparticleinterference,shor1996faulttolerantquantum} are stabilizer codes whose generators can be partitioned into $X$-type and $Z$-type operators.
Their stabilizer tableau can be formulated in a block structure
\[
  \begin{pmatrix}
    L_X & \mathbf{0}\\
    \mathbf{0} & L_Z\\
    H_X & \mathbf{0}\\
    \mathbf{0} & H_Z
  \end{pmatrix},
\]
i.e., the stabilizer generators and the logical operators can be partitioned into $X$-type and $Z$-type operators.
Let $m_X$ and $m_Z$ denote the numbers of $X$-type and $Z$-type stabilizer generators, respectively, i.e., the numbers of rows of $H_X$ and $H_Z$.

The $\F_2$ matrices $H_X$ and $H_Z$ are usually referred to as the \emph{check matrices} of the code, as they can be interpreted as the check matrices of two classical linear codes.
The \emph{CSS condition} states that \[H_X \cdot H_Z^T = 0~.\]
We also have \[H_X\cdot L_Z^T=0, \quad H_Z\cdot L_X^T=0, \text{ and } \quad L_X \cdot L_Z^T = I_{k}~\]
by the definition of stabilizer codes.

We now show that an encoder for any CSS code can be reduced to a matrix elimination problem of the matrix $\begin{pmatrix}  L_X^\top & H_X^\top \end{pmatrix}^\top$.
CNOT gates realize column additions over $\mathbb F_2$ on this matrix.
Since these operations generate the required invertible linear transformations, encoders for CSS codes can be implemented by CNOT circuits.

Consider a linear map $C$ implemented by a CNOT circuit such that

\begin{equation}
    \label{eq:enc_circ_matrix}
    \begin{pmatrix}  L_X \\ H_X\end{pmatrix} \cdot  C^{-1} =
\begin{pmatrix}
  B & \mathbf{0} & I_k \\
  A & \mathbf{0} & \mathbf{0}
\end{pmatrix}~,
\end{equation}
with $A \in \F_2^{m_X \times m_X}$ invertible $B \in \F_2^{k \times m_X}$, and the equality holds up to a reordering of rows.
$C$ then also satisfies
\vspace{.5\baselineskip} 
\begin{equation}
    \label{eq:enc_circ_matrix_z}
    \begin{pmatrix}  L_Z \\H_Z\end{pmatrix} \cdot  C^{-1} = 
    \begin{pmatrix}
      \mathbf{0} & E & I_k  \\
      \mathbf{0} & D & \mathbf{0}
\end{pmatrix}~,
\end{equation}
with $D \in \F_2^{m_Z \times m_Z}$ invertible $E \in \F_2^{k \times m_Z}$, and the equality holds up to a reordering of rows.
For a proof of this fact, see~\Cref{sec:proofs}.

These reduced check matrices are then the check matrices of the identity encoding isometry up to Hadamard gates. 
To see this, consider the stabilizer tableau defined by the reduced check matrices:

\[\left(\begin{array}{ccc|ccc}

    B & \mathbf{0} & I_k &  \mathbf{0} & \mathbf{0} & \mathbf{0} \\
    \mathbf{0} & \mathbf{0} & \mathbf{0} & \mathbf{0} & E & I_k \\ \hline
    A & \mathbf{0} & \mathbf{0} & \mathbf{0} & \mathbf{0} & \mathbf{0} \\
    \mathbf{0} & \mathbf{0} & \mathbf{0} & \mathbf{0} & D & \mathbf{0} 
\end{array}\right)~.\]
Since $A$ and $D$ are invertible, we can eliminate $B$ and $E$ from the tableau and reduce $A$ and $D$ to identity via stabilizer row operations, obtaining the stabilizer-equivalent tableau

\[\left(\begin{array}{ccc|ccc}

    \mathbf{0} & \mathbf{0} & I_k &  \mathbf{0} & \mathbf{0} & \mathbf{0} \\
    \mathbf{0} & \mathbf{0} & \mathbf{0} & \mathbf{0} & \mathbf{0} & I_k \\ \hline
    I_{m_X} & \mathbf{0} & \mathbf{0} & \mathbf{0} & \mathbf{0} & \mathbf{0} \\
    \mathbf{0} & \mathbf{0} & \mathbf{0} & \mathbf{0} & I_{m_Z} & \mathbf{0} 
\end{array}\right)~.\]

Applying Hadamard gates to the first $m_X$ qubits gives the tableau

\[\left(\begin{array}{ccc|ccc}

    \mathbf{0} & \mathbf{0} & I_k &  \mathbf{0} & \mathbf{0} & \mathbf{0} \\
    \mathbf{0} & \mathbf{0} & \mathbf{0} & \mathbf{0} & \mathbf{0} & I_k \\ \hline
     \mathbf{0} & \mathbf{0} & \mathbf{0} &I_{m_X} & \mathbf{0} & \mathbf{0} \\
    \mathbf{0} & \mathbf{0} & \mathbf{0} & \mathbf{0} & I_{m_Z} & \mathbf{0} 
\end{array}\right)~,\]

which is the tableau of the identity isometry from~\Cref{eq:tableau-init} up to qubit relabelling.

If we have a linear circuit $G$ that implements $C$, an encoding circuit for the input CSS code is then constructed as follows.

\begin{enumerate}
  \item Initialize the first $m_X$ qubits in $\ket{+}$.
  \item Initialize the next $m_Z$ qubits in $\ket{0}$.
  \item The last $k$ qubits of the circuit are the inputs that map physical qubit $\ket{\psi}_i$  to logical qubit $\ket{\overline{\psi}}_i$.
  \item Apply $G$ to these qubits.
  \end{enumerate}

  Encoding circuit synthesis for CSS codes is therefore equivalent to a matrix elimination problem over $\F_2$: The task is to find a sequence of CNOT gates (column additions) that transform the matrix $\begin{pmatrix}  L_X^\top & H_X^\top \end{pmatrix}^\top$  into the form of~\Cref{eq:enc_circ_matrix}. 
  This reduction is not unique; different elimination sequences yield encoding circuits with different CNOT counts or depths. 
  Our goal is therefore to synthesize circuits that minimize these resources.

\begin{example}[Steane Encoder]
  Consider the $\code{7,1,3}$ Steane code~\cite{steane1996errorcorrectingcodes} with check matrix

  \[H_X=
\begin{pmatrix}
  0&1&1&0&1&1&0\\
  1&0&1&0&1&0&1\\
  0&0&0&1&1&1&1
\end{pmatrix}~, \]
and logicals
\[L_X = L_Z = \begin{pmatrix}
  1&0&0&1&0&0&1
\end{pmatrix}~.\]

Following the reasoning in the previous section, we can construct an encoding circuit for the Steane code by reducing the matrix

  \[\begin{pmatrix}  L_X \\\hline H_X \end{pmatrix}=
    \begin{pmatrix}
      1&0&0&1&0&0&1 \\\hline
      0&1&1&0&1&1&0\\
      1&0&1&0&1&0&1\\
      0&0&0&1&1&1&1 \\ 

\end{pmatrix} \]

using the following sequence of column additions:

\begin{align*}
  &\begin{pmatrix}
1&0&0&1&0&0&1\\ \hline
0&1&1&0&1&1&0\\
1&0&1&0&1&0&1\\
0&0&0&1&1&1&1\\
\end{pmatrix}
  \xrightarrow{\;\mathrm{CX}_{5,6}\;}
    \begin{pmatrix}
      1&0&0&1&0&0&1\\\hline
      0&1&1&0&1&0&0\\
      1&0&1&0&1&1&1\\
      0&0&0&1&1&0&1\\
\end{pmatrix}
  \\
  \xrightarrow{\;\mathrm{CX}_{7,4}\;}
&\begin{pmatrix}
1&0&0&0&0&0&1\\\hline
0&1&1&0&1&0&0\\
1&0&1&1&1&1&1\\
0&0&0&0&1&0&1\\ 
\end{pmatrix}
  \xrightarrow{\;\mathrm{CX}_{6,4}\;}
  \begin{pmatrix}
    1&0&0&0&0&0&1\\\hline
    0&1&1&0&1&0&0\\
    1&0&1&0&1&1&1\\
    0&0&0&0&1&0&1\\ 
\end{pmatrix}
  \\
  \xrightarrow{\;\mathrm{CX}_{3,2}\;}
  &\begin{pmatrix}
    1&0&0&0&0&0&1\\\hline
    0&0&1&0&1&0&0\\
    1&1&1&0&1&1&1\\
    0&0&0&0&1&0&1\\ 
\end{pmatrix}
  \xrightarrow{\;\mathrm{CX}_{7,5}\;}
    \begin{pmatrix}
      1&0&0&0&1&0&1\\ \hline
      0&0&1&0&1&0&0\\
      1&1&1&0&0&1&1\\
      0&0&0&0&0&0&1\\ 
\end{pmatrix}
 \\
  \xrightarrow{\;\mathrm{CX}_{1,3}\;}
  &\begin{pmatrix}
    1&0&1&0&1&0&1 \\ \hline
    0&0&1&0&1&0&0\\
    1&1&0&0&0&1&1\\
    0&0&0&0&0&0&1\\
\end{pmatrix}
  \xrightarrow{\;\mathrm{CX}_{5,3}\;}
    \begin{pmatrix}
      1&0&0&0&1&0&1\\\hline
      0&0&0&0&1&0&0\\
      1&1&0&0&0&1&1\\
      0&0&0&0&0&0&1\\
\end{pmatrix}
  \\
  \xrightarrow{\;\mathrm{CX}_{2,1}\;}
  &\begin{pmatrix}
    1&0&0&0&1&0&1\\ \hline
    0&0&0&0&1&0&0\\
    0&1&0&0&0&1&1\\
    0&0&0&0&0&0&1\\ 
\end{pmatrix}
  \xrightarrow{\;\mathrm{CX}_{6,2}\;}
    \begin{pmatrix}
      1&0&0&0&1&0&1\\\hline
      0&0&0&0&1&0&0\\
      0&0&0&0&0&1&1\\
      0&0&0&0&0&0&1\\ 
\end{pmatrix}
\end{align*}
The synthesis is done at this point because the check matrix is of the form~\Cref{eq:enc_circ_matrix} up to reordering of qubits.
From this form, we can read off that qubits $5$, $6$, and $7$ are initialized in $\ket{+}$, qubits $2$, $3$ and $4$ are initialized in $\ket{0}$, and qubit $1$ serves as the input qubit for the state to be encoded.
\end{example}

Greedy synthesis of CNOT circuits---just like greedy synthesis of general Clifford circuits---makes use of a greedy heuristic guiding the search by evaluating which CNOT (i.e., column addition) to apply in each step during the elimination.
Since encoding circuits for CSS codes can be synthesized as CNOT circuits based on one of the check matrices of the code, we can employ techniques from CNOT circuit synthesis in our setting.
CNOT circuit synthesis has also been discussed in Ref.~\cite{webster2025heuristicoptimalsynthesis}, where the authors propose both a greedy best-first approach to CNOT circuit synthesis as well as an improved method using A* search~\cite{hart1968formalbasisheuristic}. 
But they do not make use of the degrees of freedom of row operations, i.e., stabilizer group operations, during the search.
Reference~\cite{peham2024automatedsynthesisfaulttolerantstate} specifically applies a greedy best-first approach to state preparation synthesis where arbitrary row operations are allowed.

The greedy synthesis from both Ref.~\cite{webster2025heuristicoptimalsynthesis} and Ref.~\cite{peham2024automatedsynthesisfaulttolerantstate} is based on the heuristic

\begin{equation}
  \label{eq:heuristic-css}
  h(H) =\sum_{i,j}H_{ij}~,
\end{equation}
which counts the number of non-zero entries in the intermediate matrix.
At every step of the reduction, the CNOT is chosen that locally minimizes $h$, i.e., removes the most non-zero entries.

We can reuse this greedy approach for encoding circuit synthesis.
Starting from the stacked matrix $
\begin{pmatrix}
  L_X^\top & H_X^\top
\end{pmatrix}^\top
$ we also always pick the CNOT that locally minimizes $h$; this is the same heuristic as in Refs.~\cite{peham2024automatedsynthesisfaulttolerantstate,webster2025heuristicoptimalsynthesis}.
What changes is how a goal state is identified.
Instead of requiring a full reduction of the matrix, we explicitly check whether the intermediate check matrix is of the form of~\Cref{eq:enc_circ_matrix}.
Incorporating this check directly into the search avoids unnecessary CNOT gates that reduce entries in the check matrix but do not affect the encoding isometry.
Thus, the independence of the encoding circuit from the stabilizer generators is directly incorporated into the terminating condition of the search.

The greedy search may reach a local minimum, in the sense that no single CNOT gate can reduce the heuristic score of the current check matrix. 
To escape such local minima, we use a sequence of fallback steps ordered by computational cost. 
We first apply stabilizer row operations that reduce the number of non-zero entries in the current matrix.
Since row operations preserve the represented encoding isometry and only change the basis of the row space, this might suffice to escape a local minimum while simultaneously reducing the number of non-zero entries in the check matrix.
If this does not suffice, we transform the check matrix into row-reduced echelon form and continue the synthesis from there.
If the search still fails to make progress, we finally consider short multi-gate sequences. 
This is the most expensive option, but it is necessary when no single CNOT is locally improving, even though a short sequence of CNOTs reduces the heuristic overall. 

The greedy encoding circuit synthesis algorithms are clearly suboptimal, since a locally suboptimal choice of CNOT during elimination might yield a globally smaller circuit.
In the following, we will describe how to guide the search towards more favorable circuit implementations using more sophisticated search strategies.

\subsection{Rollout Synthesis}
\label{sec:rollout-synthesis}

Returning to the state-space interpretation of encoding circuit synthesis, the greedy approaches discussed earlier efficiently traverse the search space by evaluating all possible gates at each step and selecting the one that minimizes a greedy heuristic $h$.
The path through the search space thus obtained is certainly not optimal, but it can be computed efficiently.
Reference~\cite{webster2025heuristicoptimalsynthesis} also proposes an A* search of this state space.
The idea is to use the heuristics of~\Cref{eq:cost-non-css,eq:heuristic-css} to estimate the objective value of a state and expand the state in the search queue that minimizes the sum of the state's cost and the heuristic.
The authors of Ref.~\cite{webster2025heuristicoptimalsynthesis} restrict the number of states stored in the search queue to avoid the potentially exponential space requirement of A* search.
This strategy enables a wider exploration of the search space but remains effectively informed by a locally computed heuristic score.

Here, we propose an alternative method for traversing the search space.
A basic issue with greedy heuristics is that the heuristic score can only serve as a proxy for the optimization objective we are ultimately interested in, i.e., the number of two-qubit gates or circuit depth.
We can further improve the synthesis by using a score that better reflects the actual circuit cost we seek to minimize.
Such a score can be constructed using a rollout scheme~\cite{bertsekas1997rolloutalgorithms}.
Instead of picking the gate that optimizes \(h\), we can roll out the entire path through the search space that we \emph{would} obtain if we performed greedy search.
We can then directly compute the score of this path from the synthesized circuit and choose the state for which the greedy approach would give the best circuit. 
Concretely, we define
\begin{equation}
  \label{eq:cost-rollout-non-css}
  h_\mathrm{rollout}(T) \coloneqq \mathrm{TQG}(\textsc{GreedySynth}(T))~,
\end{equation}
where $\textsc{GreedySynth}(T)$ denotes the circuit synthesized greedily using the heuristics of~\Cref{eq:cost-non-css} or~\Cref{eq:heuristic-css} for a given (intermediate) matrix $T$, and $\mathrm{TQG}$ denotes the number of two-qubit gates in the circuit.

Considering only two-qubit gates, computing this rollout score for all $O(n^2)$ two-qubit gates at each synthesis step is computationally expensive.
Instead of computing the score for all candidates, we can therefore perform rollout only on subsets of possible two-qubit gates at each step.
Since \(h_\mathrm{greedy}\) already gives an estimate of which two-qubit gates are more promising, we can pick the \(t\) candidates that locally minimize \(h_\mathrm{greedy}\), where \(t\) is an input to the synthesis.

Depending on the size of the synthesized encoding circuit, the rollout approach can be recursively applied.
This procedure can only be realized to a limited extent as the number of circuits required to be synthesized using \(\textsc{GreedySynth}\) grows exponentially with the number of rollout levels:
If we consider \(t\) candidates at every level, and assuming the synthesis requires $g$ gates, then evaluating the \(\ell\)-level rollout score requires performing greedy synthesis for \((g\cdot t)^\ell\) circuits.
The number of expanded candidates can also be chosen differently for each level, giving control over the potential runtime of this approach.
This overhead of greedily synthesizing many circuits quickly becomes intractable, but carefully choosing \(t\) and \(\ell\) can potentially lead to considerable improvements to the encoding circuit synthesis by encouraging deeper exploration of the search space.

The rollout approach can also be used to optimize for other metrics, such as depth, by changing what is evaluated on the circuit synthesized by \(\textsc{GreedySynth}\).
Since we have access to the entire synthesized circuit of a path, any optimization objective can be used to guide the search.
To optimize both the two-qubit gate count and depth, one can also minimize over tuples
\[\left( \mathrm{TQG}(G_\mathrm{greedy}), \mathrm{depth}(G_\mathrm{greedy}) \right),\]
where $G_\mathrm{greedy}$ is the circuit obtained from the greedy synthesis.
Therefore, this rollout score uses depth to break ties for paths yielding circuits with equal two-qubit gate count.

A practical approach to limit the runtime of the rollout synthesis is to keep track of the current best circuit and prune candidates that do not promise to improve it.
If at any point during the search, no rollout offers an improvement, the search can be \emph{terminated early}.
In practice, gates chosen early in the search have a much greater impact on the resulting circuit size than gates chosen later.
Therefore, pruning encourages aggressive exploration of the search space early in the search and speeds up the synthesis as it progresses.

The proposed search methods apply to arbitrary non-CSS and CSS codes.
While the greedy construction is fast and scalable, more extensive rollout can become computationally expensive on larger instances.
When the target isometry admits a modular decomposition, however, the synthesis task can be split into smaller subproblems, allowing more aggressive optimization on the resulting local instances.
We turn to this setting next.

\section{Encoding Circuits for Generalized Concatenated \& Holographic Quantum Codes}
\label{sec:encod-circ-conc}

The methods of \Cref{sec:encod-circ-synth} treat encoder synthesis as a search over the full stabilizer tableau of the target isometry. For structured code families whose encoders admit a modular decomposition, however, it is more effective to exploit this structure directly by decomposing the encoder into local constituent isometries, optimizing these local building blocks, and then composing the resulting circuits into a global encoder.

This viewpoint is particularly natural for generalized concatenated~\cite{grassl2009generalized} and holographic codes~\cite{pastawski2015holographic,fan2024lego_hqec,jahn2021holographic,harris2018calderbank,harris2020decoding,steinberg2025far,steinberg2025universal,fan2024biased}, whose encoding maps are built recursively from smaller isometries.
To describe and manipulate such decompositions, we use ZX-diagrams, which provide a convenient representation of encoding isometries and their composition~\cite{wu2024zxcalculusapproachconstructiongraph}.
We begin with a simple concatenated construction that illustrates the modular viewpoint, then describe how the resulting local instances can be optimized exactly using SMT-based synthesis, and finally discuss the partitioning problem and the trade-offs induced by the choice of decomposition for codes with more intricate composition structures.

\subsection{Concatenated Codes from Local Isometries}
\label{sec:conc-codes-from}

\begin{figure}[t]
\begin{center}
\resizebox{.45\columnwidth}{!}{\input{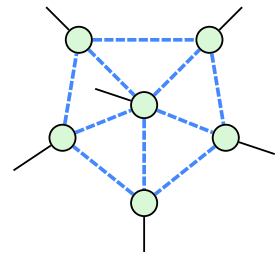}}  
\end{center}
\caption{A ZX diagram representation of the $\code{5,1,3}$ code's encoding isometry. The logical input wire is connected to the $Z$-spider at the center of the diagram.
The blue edges correspond to wires with a Hadamard box~\cite{vandewetering2020zxcalculusworkingquantumcomputer}.}
\label{fig:513_fig}
\end{figure}

\begin{figure*}[t]
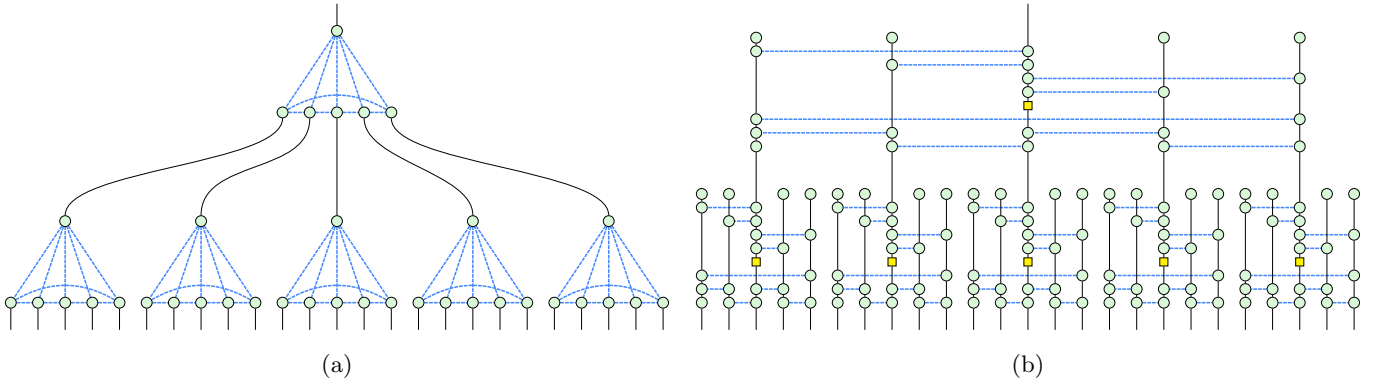

  \centering
   \begin{subfigure}[b]{.49\linewidth}
     \resizebox{\linewidth}{!}{\rotatebox{-90}{\input{zx/tree_concatenated_5_qubit.tikz}}}
      \caption{}
      \label{fig:tree-style-concat-diagram}
    \end{subfigure}
    \hfill
    \begin{subfigure}[b]{.49\linewidth}
      \resizebox{\linewidth}{!}{\rotatebox{-90}{\input{zx/tree_concatenated_5_qubit_extracted.tikz}}}
      \caption{}
      \label{fig:tree-style-concat-circuit}
    \end{subfigure}

    \caption{Encoding circuit synthesis of the $\code{25,1,9}$ concatenated $\code{5,1,3}$ code. \textbf{(a)} The ZX-diagram for the $25$-qubit encoder is constructed by connecting the outputs of one copy of the $5$-qubit encoding isometry with the inputs of five more copies of the $5$-qubit encoder.
      \textbf{(b)} An encoding circuit is constructed by replacing each $5$-qubit encoding diagram with a circuit extracted from the isometry.
    }
  \label{fig:tree-style-concat}
\end{figure*}

Any Clifford isometry can be represented by a Clifford ZX-diagram.
In this representation, the composition of encoding maps is expressed directly by connecting diagrams along shared wires.
This makes ZX-diagrams a natural language for describing recursively constructed encoders, since both the local constituent isometries and their composition into a global encoder remain explicit.

As a simple example, \Cref{fig:513_fig} shows a ZX-diagram representation of the $\code{5,1,3}$ encoding isometry.
In this representation, the inner open wire is taken to be the logical input, while the outer open wires correspond to the physical outputs.
More generally, however, a given ZX-diagram need not define a valid encoding isometry for an arbitrary choice of inputs and outputs.
The admissible choices depend specifically on the properties of the underlying quantum many-body state.\footnote{For example, if the underlying state is represented by a perfect tensor~\cite{pastawski2015holographic}, then any input can be taken, but if the codestate is not absolutely maximally entangled (AME), then this is not possible in total generality~\cite{steinberg2023holographic,harris2018calderbank,steinberg2025far}.}
Different valid choices of inputs and outputs can therefore give rise to distinct encoding isometries, corresponding to related codes obtained, for example, by code shortening~\cite{gottesman1997stabilizer,pastawski2015holographic,steinberg2025far}.

By wiring together copies of the $\code{5,1,3}$ encoding isometry, one obtains the encoder of the $\code{25,1,9}$ concatenated-five-qubit code shown in~\Cref{fig:tree-style-concat-diagram}.
Replacing each isometry block by a circuit implementation of the $\code{5,1,3}$ encoder then yields an encoding circuit for the full $25$-qubit code.
In this case, we can extract an encoding circuit from the diagram simply by unfusing some spiders:
\begin{equation}
\label{eq:513_zx_encoder_manipulation}
\begin{aligned}
  \resizebox{!}{25pt}{\input{zx/holography_toy_zx_encoder_5_1.tikz}}
  &=
  \resizebox{!}{25pt}{\input{zx/holography_toy_zx_encoder_5_2.tikz}}
  \\[4pt]
  =\resizebox{!}{25pt}{\input{zx/holography_toy_zx_encoder_5_3.tikz}}
  &=
  \resizebox{!}{25pt}{\input{zx/holography_toy_zx_encoder_5_4.tikz}}
\end{aligned}
\end{equation}
Replacing each constituent isometry by a circuit implementation, therefore, yields a global encoding circuit constructed from local building blocks as shown in~\Cref{fig:tree-style-concat-circuit}.

This example illustrates the basic idea of modular synthesis: For recursively constructed code families, the global encoder can be assembled from a small number of reusable constituent encoders.
Optimizing these local blocks is therefore a natural way to improve the resulting global circuit.

\subsection{Optimal Synthesis of Local Isometries}
\label{sec:optim-synth-local}

Once the synthesis problem has been reduced to local constituent encoders, these smaller instances can often be treated using exact methods.
To this end, we use a symbolic synthesis approach.
Since unitary Clifford synthesis is a special case of encoding-circuit synthesis, the problem is at least as hard as optimal Clifford synthesis, which is known to be NP-hard~\cite{cabello2011optimalpreparationgraph,peham2023depthoptimalsynthesis}.
Our encoding combines ideas from SAT-based optimal Clifford synthesis~\cite{peham2023depthoptimalsynthesis} and SMT-based optimal state-preparation synthesis for CSS codes~\cite{peham2024automatedsynthesisfaulttolerantstate}, and generalizes them to local encoding isometries.
We only sketch the basic idea here and defer the full encoding to~\Cref{sec:app-sat-encoding}.

The symbolic encoding follows a bounded-model-checking view of the synthesis problem.
For each circuit depth \(d=0,\dots,d_{\max}\), we introduce a symbolic stabilizer tableau (or, for CSS codes, a symbolic check matrix) \(T^{(d)}\).
Since these are \(\mathbb{F}_2\)-matrices, every matrix entry is represented by a Boolean variable.
For each layer \(d<d_{\max}\), we additionally introduce symbolic gate variables \(g^{(d)}\) describing the parallel gate applications performed in that layer.

Abstractly speaking, we need to encode the following constraints:

\begin{itemize}
\item[\(\Phi_{\mathrm{init}}\):] The initial symbolic tableau \(T^{(0)}\) must be equal to the tableau of the target encoding isometry.
\item[\(\Phi_{\mathrm{cons}}\):] The assignments to the gate variables are \emph{consistent}, i.e., gate applications at each symbolic layer of the circuit represent a valid set of parallel gate applications (e.g.~no two gates act on the same qubits).
\item[\(\Phi_{\mathrm{trans}}\):] Consecutive symbolic tableaus are related by the corresponding gate layer.
\item[\(\Phi_{\mathrm{goal}}\):] The final tableau represents the identity tableau according to~\Cref{eq:tableau-init} (or~\Cref{eq:enc_circ_matrix} in the CSS case).
\end{itemize}

The resulting synthesis problem is then encoded as the Boolean formula
\[
\Phi
=
\Phi_{\mathrm{init}}
\;\wedge\;
\Phi_{\mathrm{cons}}
\;\wedge\;
\Phi_{\mathrm{trans}}
\;\wedge\;
\Phi_{\mathrm{goal}}.
\]
Thus, \(\Phi\) is satisfiable if and only if there exists an encoding circuit of depth at most \(d_{\max}\) for the chosen gate set.

In practice, we instantiate \(\Phi\) and solve the resulting formulas using an off-the-shelf SMT solver such as Z3~\cite{demoura2008z3efficientsmt}.
A satisfying assignment fixes the symbolic gate variables \(g^{(d)}\) and therefore determines an encoding circuit layer by layer.
Solving the encoding for increasing values of \(d_{\max}\) until the first satisfiable instance is found yields a depth-optimal circuit. 

Gate-optimal synthesis can be handled in two ways.
First, one may keep the layer-based encoding and add cardinality constraints on the selected gate variables, thereby minimizing the number of gates within a fixed depth bound.
This procedure yields a natural lexicographic optimization, where depth is minimized first and gate count second.
Second, one may replace the layer-based encoding with a gate-based encoding, in which each symbolic transition corresponds to a single gate application rather than an entire layer.
In that case, the time horizon directly equals the gate budget.
Both variants are described in~\Cref{sec:app-sat-encoding}.

\subsection{Modular Synthesis}
\label{sec:modular-synthesis}

While the tree-style concatenation above provides a simple example of modular encoder synthesis, the same idea applies more broadly to structured encoding isometries.
Given the ZX-diagram of an encoding isometry $D$, the goal is to partition it into sub-isometries $D_1,\dots,D_r$ such that $D$ is obtained by composing these local blocks.
One can then optimize the constituent encoders individually and compose the resulting circuits into a global encoder.

\begin{figure}[t]
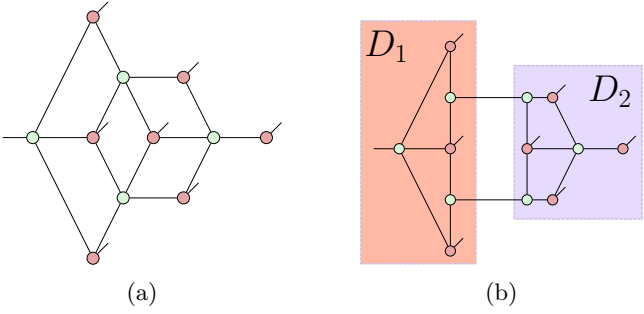

  \centering
  \begin{subfigure}[b]{0.45\linewidth}{
      \resizebox{\linewidth}{!}{\input{zx/steane_encoder_cube.tikz}}
      \caption{}
      \label{fig:steane-encoder-a}
    }
  \end{subfigure}
  \hfill
  \begin{subfigure}[b]{.45\linewidth}
    \resizebox{\linewidth}{!}{\input{zx/steane_encoder_unfused.tikz}}
    \caption{}
    \label{fig:steane-encoder-b}
  \end{subfigure}
  \caption{\textbf{(a)} ZX-diagram of the encoding isometry of the $\code{7,1,3}$ Steane code. \textbf{(b)} Partitioning of the Steane encoder into two smaller encoding isometries.}
  \label{fig:steane-encoder}
\end{figure}

For an arbitrary ZX-diagram, a difficult part of this procedure is to partition the diagram into smaller subdiagrams that still represent valid encoding isometries.
Consider, for example, the encoding isometry of the $\code{7,1,3}$ Steane code shown in~\Cref{fig:steane-encoder-a}~\cite{duncan2014verifyingsteanecode,kissinger2022phasefreezxdiagramscss}.
Not every cut through the diagram partitions it into two valid encoders from which we can extract an encoding circuit. 
In particular, for the fused diagram in~\Cref{fig:steane-encoder-a}, no direct cut yields two valid encoding isometries.
We can, however, unfuse two of the inner $Z$-spiders and cut between them as shown in~\Cref{fig:steane-encoder-b}.
The resulting cut partitions the encoder into a \(1\)-input-\(5\)-output encoding isometry and a \(2\)-input-\(4\)-output encoding isometry\footnote{$D_1$ and $D_2$ are actually both the same diagram as in~\Cref{fig:422-zx}. We have therefore re-derived the fact that the Steane code can be obtained by concatenating two copies of the $\code{4,2,2}$ code~\cite{cao2022quantumlego}.}.

For generalized concatenated and holographic codes, this partitioning problem is considerably more structured, since their recursive construction already specifies a natural decomposition into constituent encoders.
At a higher level, holographic codes can be viewed as code constructions inspired by the AdS/CFT correspondence, in which logical degrees of freedom reside in the bulk and physical qubits on the boundary~\cite{pastawski2015holographic,pastawski2017code,almheiri2015bulk}.
Logical qubits closer to the boundary are typically less protected, but remain more directly accessible for logical measurements and gates.
Accordingly, one may associate different effective distances with different logical degrees of freedom.
More generally, decoding and recovery are governed by boundary regions whose size is determined by geodesics in the underlying hyperbolic geometry (see, for instance, Ref.~\cite{angles2024engineering} for conceptual visualizations). For our purposes, this bulk--boundary structure is precisely what makes the global encoding map naturally amenable to a modular description in terms of constituent local isometries.

From the perspective of encoder synthesis, this structure manifests itself as a composition of local tensors or isometries whose inputs and outputs correspond naturally to bulk logical and boundary physical degrees of freedom.\footnote{One may envision using the \emph{quantum LEGO formalism}~\cite{cao2022quantumlego,cao2024quantum}, as the tensor-network formulation of the codespace allows for easily repartitioning the state space into suitable isometries; however, such exploration is outside of our scope, and we reserve it for future work.}
Cuts placed along the inputs and outputs of these constituent encoders, therefore, again yield valid local encoding isometries.
This modular viewpoint is particularly effective because the global encoder is often assembled from only a small number of distinct constituent encoders that are reused many times.
Hence, only a small number of local isometries must be synthesized, while improvements in these local building blocks propagate throughout the full construction.
This makes exact or near-exact synthesis of the constituent encoders worthwhile even when direct optimization of the full encoding isometry is infeasible.

\begin{figure}[t]
  \centering
  \resizebox{\linewidth}{!}{\input{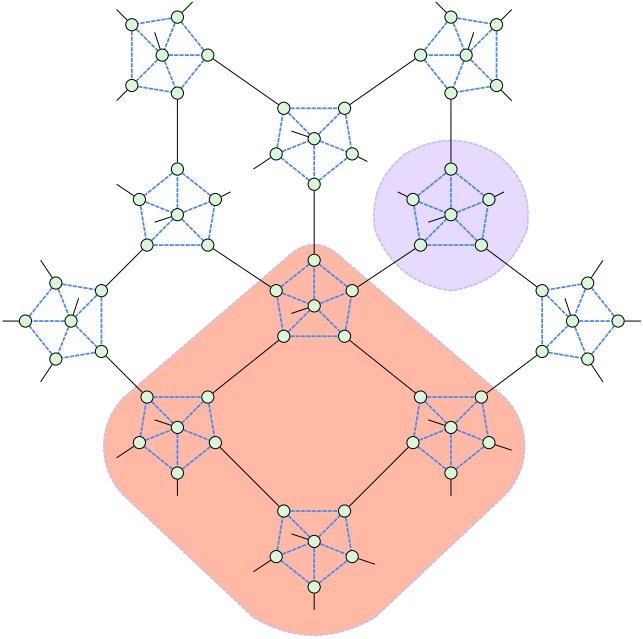}}
  \caption{ZX-diagram of a holographic HaPPY~\cite{pastawski2015holographic} code. Logical (bulk) input legs are connected to the center spiders of the constituent five-qubit encoding isometries, while physical output (boundary) legs are connected to the outer spiders.
    The highlighted subdiagrams illustrate two constituent local encoding isometries: a smaller $2$-input-$4$-output block (blue) and a larger $4$-input-$12$-output block (orange).}
  \label{fig:happy}
\end{figure}

Consider, for example, the HaPPY code~\cite{pastawski2015holographic} shown in~\Cref{fig:happy}.
As in the tree-style construction of~\Cref{fig:tree-style-concat-diagram}, the global encoder is assembled from multiple copies of the encoding diagram of the $\code{5,1,3}$ code.
The composition pattern is more intricate, however, since these constituent encoders are not connected only from logical to physical legs.
Nevertheless, valid local encoding isometries can still be identified systematically.
The subdiagram marked in blue, for example, can be rewritten to define the following encoding isometry from two input qubits to four output qubits:

\begin{equation}
\label{holography_toy_encoder_subdiagram}
\begin{aligned}
\resizebox{!}{25pt}{\input{zx/holography_toy_zx_encoder_subdiagram.tikz}}
&=
\resizebox{!}{25pt}{\input{zx/holography_toy_zx_encoder_4_1.tikz}}
\end{aligned}.
\end{equation}

The choice of partition also induces a trade-off.
Finer partitions produce smaller subproblems and therefore allow stronger local optimization, but they may also introduce additional composition layers, thereby incurring undesired depth overhead.
Coarser partitions reduce this composition overhead at the cost of more difficult local synthesis problems.
For example, the subdiagram marked in orange in~\Cref{fig:happy} defines an encoding isometry from four logical qubits to $12$ physical qubits.
Resynthesizing such a larger block can reduce composition overhead and therefore improve circuit depth, but it also yields a substantially harder local synthesis problem.
This trade-off provides a practical design space for modular encoder synthesis, which we evaluate in~\Cref{sec:encod-circ-synth-1}.

\section{Evaluation}
\label{sec:circuits}

\begin{table*}[t]
\centering
\caption{Best depth-optimized circuits for each synthesis task. For each code, we report the gate count $G$ and depth $D$ obtained by our method and by \textsc{Rustiq}~\cite{goubaultdebrugiere2025graphstatebased} for the best encoder, zero-state preparation circuit, and plus-state preparation circuit.}
\label{tab:best-depth-summary}
\providecommand{\RustiqCell}[1]{\hspace{9pt}#1\hspace{9pt}}
\providecommand{\RustiqCodeCell}[1]{\hspace{12pt}#1\hspace{12pt}}
\providecommand{\RustiqHdr}[1]{\multicolumn{1}{c}{#1}}
\setlength{\tabcolsep}{0pt}
\renewcommand{\arraystretch}{1.12}
\begin{tabular}{lrrrrrrrrrrrr}
\toprule
& \multicolumn{4}{c}{\textbf{Encoder}} & \multicolumn{4}{c}{\textbf{Zero}} & \multicolumn{4}{c}{\textbf{Plus}} \\
\cmidrule(r{0.4em}){2-5}
\cmidrule(lr{0.4em}){6-9}
\cmidrule(l{0.4em}){10-13}
\multicolumn{1}{c}{Code} & \multicolumn{2}{c}{Ours} & \multicolumn{2}{c}{\textsc{RustiQ}} & \multicolumn{2}{c}{Ours} & \multicolumn{2}{c}{\textsc{RustiQ}} & \multicolumn{2}{c}{Ours} & \multicolumn{2}{c}{\textsc{RustiQ}} \\
& \RustiqHdr{$G$} & \RustiqHdr{$D$} & \RustiqHdr{$G$} & \RustiqHdr{$D$} & \RustiqHdr{$G$} & \RustiqHdr{$D$} & \RustiqHdr{$G$} & \RustiqHdr{$D$} & \RustiqHdr{$G$} & \RustiqHdr{$D$} & \RustiqHdr{$G$} & \RustiqHdr{$D$} \\
\midrule
\rowcolor{black!4}
\RustiqCodeCell{$[[8,3,3]]$} & \RustiqCell{\textbf{14}} & \RustiqCell{\textbf{5}} & \RustiqCell{19} & \RustiqCell{7} & \RustiqCell{13} & \RustiqCell{5} & \RustiqCell{\textbf{11}} & \RustiqCell{\textbf{4}} & \RustiqCell{\textbf{9}} & \RustiqCell{\textbf{5}} & \RustiqCell{12} & \RustiqCell{\textbf{5}} \\
\RustiqCodeCell{$[[15,1,3]]$} & \RustiqCell{\textbf{24}} & \RustiqCell{\textbf{4}} & \RustiqCell{25} & \RustiqCell{6} & \RustiqCell{\textbf{22}} & \RustiqCell{\textbf{4}} & \RustiqCell{24} & \RustiqCell{5} & \RustiqCell{\textbf{23}} & \RustiqCell{\textbf{4}} & \RustiqCell{25} & \RustiqCell{\textbf{4}} \\
\rowcolor{black!4}
\RustiqCodeCell{$[[15,3,5]]$} & \RustiqCell{\textbf{43}} & \RustiqCell{10} & \RustiqCell{51} & \RustiqCell{\textbf{9}} & \RustiqCell{45} & \RustiqCell{10} & \RustiqCell{\textbf{36}} & \RustiqCell{\textbf{7}} & \RustiqCell{41} & \RustiqCell{9} & \RustiqCell{\textbf{35}} & \RustiqCell{\textbf{7}} \\
\RustiqCodeCell{$[[15,7,3]]$} & \RustiqCell{\textbf{31}} & \RustiqCell{\textbf{6}} & \RustiqCell{49} & \RustiqCell{15} & \RustiqCell{\textbf{22}} & \RustiqCell{\textbf{4}} & \RustiqCell{24} & \RustiqCell{6} & \RustiqCell{\textbf{22}} & \RustiqCell{\textbf{4}} & \RustiqCell{23} & \RustiqCell{5} \\
\rowcolor{black!4}
\RustiqCodeCell{$[[17,1,5]]$} & \RustiqCell{\textbf{25}} & \RustiqCell{\textbf{5}} & \RustiqCell{\textbf{25}} & \RustiqCell{\textbf{5}} & \RustiqCell{\textbf{23}} & \RustiqCell{\textbf{4}} & \RustiqCell{24} & \RustiqCell{\textbf{4}} & \RustiqCell{\textbf{23}} & \RustiqCell{\textbf{4}} & \RustiqCell{24} & \RustiqCell{5} \\
\RustiqCodeCell{$[[19,1,5]]$} & \RustiqCell{\textbf{29}} & \RustiqCell{\textbf{5}} & \RustiqCell{34} & \RustiqCell{6} & \RustiqCell{\textbf{27}} & \RustiqCell{\textbf{4}} & \RustiqCell{33} & \RustiqCell{6} & \RustiqCell{\textbf{27}} & \RustiqCell{\textbf{4}} & \RustiqCell{30} & \RustiqCell{5} \\
\rowcolor{black!4}
\RustiqCodeCell{$[[23,1,7]]$} & \RustiqCell{\textbf{53}} & \RustiqCell{\textbf{8}} & \RustiqCell{55} & \RustiqCell{\textbf{8}} & \RustiqCell{\textbf{57}} & \RustiqCell{\textbf{7}} & \RustiqCell{58} & \RustiqCell{\textbf{7}} & \RustiqCell{\textbf{57}} & \RustiqCell{\textbf{7}} & \RustiqCell{62} & \RustiqCell{\textbf{7}} \\
\RustiqCodeCell{$[[30,6,5]]$} & \RustiqCell{\textbf{95}} & \RustiqCell{\textbf{9}} & \RustiqCell{109} & \RustiqCell{15} & \RustiqCell{\textbf{68}} & \RustiqCell{\textbf{7}} & \RustiqCell{80} & \RustiqCell{8} & \RustiqCell{\textbf{68}} & \RustiqCell{\textbf{7}} & \RustiqCell{81} & \RustiqCell{\textbf{7}} \\
\rowcolor{black!4}
\RustiqCodeCell{$[[31,1,7]]$} & \RustiqCell{\textbf{48}} & \RustiqCell{\textbf{5}} & \RustiqCell{50} & \RustiqCell{7} & \RustiqCell{\textbf{46}} & \RustiqCell{\textbf{4}} & \RustiqCell{52} & \RustiqCell{6} & \RustiqCell{\textbf{46}} & \RustiqCell{\textbf{4}} & \RustiqCell{48} & \RustiqCell{6} \\
\RustiqCodeCell{$[[31,21,3]]$} & \RustiqCell{\textbf{109}} & \RustiqCell{\textbf{11}} & \RustiqCell{194} & \RustiqCell{37} & \RustiqCell{\textbf{52}} & \RustiqCell{\textbf{5}} & \RustiqCell{56} & \RustiqCell{7} & \RustiqCell{\textbf{52}} & \RustiqCell{\textbf{5}} & \RustiqCell{58} & \RustiqCell{8} \\
\rowcolor{black!4}
\RustiqCodeCell{$[[37,1,7]]$} & \RustiqCell{\textbf{63}} & \RustiqCell{\textbf{6}} & \RustiqCell{68} & \RustiqCell{\textbf{6}} & \RustiqCell{\textbf{59}} & \RustiqCell{\textbf{5}} & \RustiqCell{70} & \RustiqCell{6} & \RustiqCell{\textbf{59}} & \RustiqCell{\textbf{5}} & \RustiqCell{74} & \RustiqCell{7} \\
\RustiqCodeCell{$[[72,12,6]]$} & \RustiqCell{\textbf{261}} & \RustiqCell{\textbf{19}} & \RustiqCell{383} & \RustiqCell{24} & \RustiqCell{\textbf{223}} & \RustiqCell{\textbf{9}} & \RustiqCell{273} & \RustiqCell{12} & \RustiqCell{\textbf{214}} & \RustiqCell{\textbf{9}} & \RustiqCell{251} & \RustiqCell{10} \\
\rowcolor{black!4}
\RustiqCodeCell{$[[90,8,10]]$} & \RustiqCell{\textbf{183}} & \RustiqCell{\textbf{12}} & \RustiqCell{242} & \RustiqCell{14} & \RustiqCell{\textbf{147}} & \RustiqCell{\textbf{8}} & \RustiqCell{178} & \RustiqCell{\textbf{8}} & \RustiqCell{\textbf{147}} & \RustiqCell{\textbf{8}} & \RustiqCell{180} & \RustiqCell{\textbf{8}} \\
\RustiqCodeCell{$[[108,8,10]]$} & \RustiqCell{667} & \RustiqCell{24} & \RustiqCell{\textbf{634}} & \RustiqCell{\textbf{23}} & \RustiqCell{\textbf{548}} & \RustiqCell{\textbf{15}} & \RustiqCell{597} & \RustiqCell{16} & \RustiqCell{\textbf{460}} & \RustiqCell{\textbf{14}} & \RustiqCell{592} & \RustiqCell{16} \\
\rowcolor{black!4}
\RustiqCodeCell{$[[144,12,12]]$} & \RustiqCell{1081} & \RustiqCell{29} & \RustiqCell{\textbf{856}} & \RustiqCell{\textbf{26}} & \RustiqCell{\textbf{687}} & \RustiqCell{15} & \RustiqCell{749} & \RustiqCell{\textbf{14}} & \RustiqCell{\textbf{330}} & \RustiqCell{16} & \RustiqCell{740} & \RustiqCell{\textbf{15}} \\
\bottomrule
\end{tabular}

\end{table*}%

The proposed encoding circuit synthesis methods have been implemented in the open-source library MQT QECC, which is part of the Munich Quantum Toolkit~\cite{wille2024mqthandbook}.

Using this implementation, we synthesized general-state encoding circuits for a range of stabilizer codes. Since state-preparation circuits arise as a special case of general encoding circuits and are independently relevant for resource-state preparation in fault-tolerant protocols, we also synthesized circuits for preparing logical all-zero and all-plus states; these additional results are reported in~\Cref{sec:results-state-prep}.

We evaluate the proposed methods along three axes: Overall circuit quality on a benchmark set of stabilizer codes, the effect of rollout on the search, and a case study of modular synthesis for a 12-qubit holographic code. Additional information on the impact of early termination is reported in~\Cref{sec:impact-early-term}.

\subsection{Setup}
\label{sec:setup}

\paragraph{Synthesis of arbitrary encoding circuits.}

We evaluate the rollout synthesis from~\Cref{sec:rollout-synthesis} on a variety of stabilizer codes. For CSS codes, we consider 2D~\cite{bombin2006topologicalquantumdistillation} and 3D color codes~\cite{bombin2007exacttopologicalquantum}, quantum Hamming codes~\cite{steane1996simplequantum}, the $\code{23,1,7}$ Golay code~\cite{steane1996simplequantum}, the $\code{30,6,5}$ symplectic double code from Ref.~\cite{kanomata2025fault} and bivariate bicycle codes~\cite{bravyi2024highthreshold}.
For non-CSS codes, we consider the $\code{8,3,3}$ Gottesman code~\cite{gottesman1996classquantumerror} and the $\code{15,3,5}$ code from Ref.~\cite{kanomata2025fault}.

The heuristic encoding circuit synthesis methodology proposed in~\Cref{sec:encod-circ-synth} has two settings that can be adjusted to influence how thoroughly the search space for the circuit constructions is explored during the synthesis:

\begin{itemize}
\item[$\ell$, $t$:] The number of decision layers $\ell$ for which rollout is performed and the number of candidates $t_1, \ldots, t_\ell$ that are evaluated using the rollout synthesis for each placed gate in the search. 
\item[ET:] Whether early termination is enabled, meaning that the search is stopped once rollout no longer produces an improvement during synthesis.
\end{itemize}

Since $t_1, \cdots, t_\ell$ candidates are evaluated at each stage of the search, raising $\ell$ has an exponential impact on the runtime.
However, a higher degree of rollout naturally explores the search space more thoroughly, potentially yielding more compact circuits.
For the evaluations, we set $\ell \in \{0,1,2\}$, $t_1 \in \{10,20,30,40,50, 100, 200\}$, and $t_2 \in \{t_1/2, t_1/5\}$.
The parameter $\ell$ denotes the rollout depth: Instead of choosing the next gate directly from the greedy heuristic, the search evaluates candidate gates by greedily completing the remaining synthesis task recursively for $\ell$ decision layers. 
If $\ell = 0$, no rollout is performed and gates are chosen directly using the greedy heuristic from~\Cref{sec:encod-circ-synth}.

The same trade-off is made when deciding whether to terminate the search early.
If early termination during the search is \emph{disabled}, the rollout is performed for each gate until the entire circuit has been constructed, even if many decisions do not yield an immediate benefit to the search objective. 
For each synthesis instance, we construct two circuits, one optimized for two-qubit gate count and one optimized for depth.
If $\ell \geq 1$, the secondary optimization objective is used to break ties.

We benchmark our rollout synthesis with the state-of-the-art graph-state-based Clifford isometry synthesis tool Rustiq, which implements the methods described in Ref.~\cite{goubaultdebrugiere2025graphstatebased}.
The syndrome-decoding-based algorithm implemented by Rustiq internally solves a syndrome-decoding problem during synthesis, repeatedly trying to solve a system of equations.
The number of iterations for this is given as an input to the synthesis.
For our comparison, we set the number of iterations to $10^6$.

\paragraph{Modular synthesis of generalized concatenated codes.}

To demonstrate the modular synthesis of concatenated codes proposed in~\Cref{sec:modular-synthesis}, we work out how to construct an encoding circuit for the $\code{12,4,2}$ holographic HaPPY code in detail and compare with existing constructions from the literature.

\subsection{Best Found Encoding Circuits}
\label{sec:comp-with-state}

\begin{table*}[t]
\centering
\caption{Best gate-optimized circuits for each synthesis task. For each code, we report the gate count $G$ and depth $D$ obtained by our method and by \textsc{Rustiq}~\cite{goubaultdebrugiere2025graphstatebased} for the best encoder, zero-state preparation circuit, and plus-state preparation circuit. }
\label{tab:best-gates-summary}
\providecommand{\RustiqCell}[1]{\hspace{9pt}#1\hspace{9pt}}
\providecommand{\RustiqCodeCell}[1]{\hspace{12pt}#1\hspace{12pt}}
\providecommand{\RustiqHdr}[1]{\multicolumn{1}{c}{#1}}
\setlength{\tabcolsep}{0pt}
\renewcommand{\arraystretch}{1.12}
\begin{tabular}{lrrrrrrrrrrrr}
\toprule
& \multicolumn{4}{c}{\textbf{Encoder}} & \multicolumn{4}{c}{\textbf{Zero}} & \multicolumn{4}{c}{\textbf{Plus}} \\
\cmidrule(r{0.4em}){2-5}
\cmidrule(lr{0.4em}){6-9}
\cmidrule(l{0.4em}){10-13}
\multicolumn{1}{c}{Code} & \multicolumn{2}{c}{Ours} & \multicolumn{2}{c}{\textsc{RustiQ}} & \multicolumn{2}{c}{Ours} & \multicolumn{2}{c}{\textsc{RustiQ}} & \multicolumn{2}{c}{Ours} & \multicolumn{2}{c}{\textsc{RustiQ}} \\
& \RustiqHdr{$G$} & \RustiqHdr{$D$} & \RustiqHdr{$G$} & \RustiqHdr{$D$} & \RustiqHdr{$G$} & \RustiqHdr{$D$} & \RustiqHdr{$G$} & \RustiqHdr{$D$} & \RustiqHdr{$G$} & \RustiqHdr{$D$} & \RustiqHdr{$G$} & \RustiqHdr{$D$} \\
\midrule
\rowcolor{black!4}
\RustiqCodeCell{$[[8,3,3]]$} & \RustiqCell{\textbf{12}} & \RustiqCell{\textbf{8}} & \RustiqCell{15} & \RustiqCell{10} & \RustiqCell{11} & \RustiqCell{8} & \RustiqCell{\textbf{10}} & \RustiqCell{\textbf{7}} & \RustiqCell{\textbf{9}} & \RustiqCell{\textbf{5}} & \RustiqCell{\textbf{9}} & \RustiqCell{\textbf{5}} \\
\RustiqCodeCell{$[[15,1,3]]$} & \RustiqCell{\textbf{24}} & \RustiqCell{\textbf{5}} & \RustiqCell{\textbf{24}} & \RustiqCell{9} & \RustiqCell{\textbf{22}} & \RustiqCell{\textbf{5}} & \RustiqCell{\textbf{22}} & \RustiqCell{12} & \RustiqCell{\textbf{23}} & \RustiqCell{\textbf{5}} & \RustiqCell{\textbf{23}} & \RustiqCell{9} \\
\rowcolor{black!4}
\RustiqCodeCell{$[[15,3,5]]$} & \RustiqCell{36} & \RustiqCell{\textbf{11}} & \RustiqCell{\textbf{35}} & \RustiqCell{16} & \RustiqCell{36} & \RustiqCell{13} & \RustiqCell{\textbf{28}} & \RustiqCell{\textbf{8}} & \RustiqCell{31} & \RustiqCell{\textbf{12}} & \RustiqCell{\textbf{27}} & \RustiqCell{13} \\
\RustiqCodeCell{$[[15,7,3]]$} & \RustiqCell{\textbf{28}} & \RustiqCell{\textbf{7}} & \RustiqCell{46} & \RustiqCell{23} & \RustiqCell{\textbf{22}} & \RustiqCell{\textbf{5}} & \RustiqCell{\textbf{22}} & \RustiqCell{7} & \RustiqCell{\textbf{22}} & \RustiqCell{\textbf{5}} & \RustiqCell{23} & \RustiqCell{13} \\
\rowcolor{black!4}
\RustiqCodeCell{$[[17,1,5]]$} & \RustiqCell{25} & \RustiqCell{\textbf{5}} & \RustiqCell{\textbf{24}} & \RustiqCell{7} & \RustiqCell{\textbf{23}} & \RustiqCell{\textbf{4}} & \RustiqCell{25} & \RustiqCell{7} & \RustiqCell{\textbf{23}} & \RustiqCell{\textbf{4}} & \RustiqCell{24} & \RustiqCell{9} \\
\RustiqCodeCell{$[[19,1,5]]$} & \RustiqCell{\textbf{29}} & \RustiqCell{\textbf{6}} & \RustiqCell{32} & \RustiqCell{10} & \RustiqCell{\textbf{27}} & \RustiqCell{\textbf{4}} & \RustiqCell{30} & \RustiqCell{15} & \RustiqCell{\textbf{27}} & \RustiqCell{\textbf{4}} & \RustiqCell{30} & \RustiqCell{11} \\
\rowcolor{black!4}
\RustiqCodeCell{$[[23,1,7]]$} & \RustiqCell{\textbf{49}} & \RustiqCell{\textbf{10}} & \RustiqCell{54} & \RustiqCell{32} & \RustiqCell{\textbf{45}} & \RustiqCell{\textbf{10}} & \RustiqCell{52} & \RustiqCell{24} & \RustiqCell{\textbf{45}} & \RustiqCell{\textbf{10}} & \RustiqCell{50} & \RustiqCell{24} \\
\RustiqCodeCell{$[[30,6,5]]$} & \RustiqCell{\textbf{82}} & \RustiqCell{\textbf{13}} & \RustiqCell{84} & \RustiqCell{29} & \RustiqCell{\textbf{61}} & \RustiqCell{\textbf{11}} & \RustiqCell{71} & \RustiqCell{28} & \RustiqCell{\textbf{62}} & \RustiqCell{\textbf{11}} & \RustiqCell{71} & \RustiqCell{27} \\
\rowcolor{black!4}
\RustiqCodeCell{$[[31,1,7]]$} & \RustiqCell{\textbf{47}} & \RustiqCell{\textbf{6}} & \RustiqCell{53} & \RustiqCell{16} & \RustiqCell{\textbf{45}} & \RustiqCell{\textbf{5}} & \RustiqCell{51} & \RustiqCell{20} & \RustiqCell{\textbf{45}} & \RustiqCell{\textbf{5}} & \RustiqCell{52} & \RustiqCell{15} \\
\RustiqCodeCell{$[[31,21,3]]$} & \RustiqCell{\textbf{94}} & \RustiqCell{\textbf{18}} & \RustiqCell{153} & \RustiqCell{45} & \RustiqCell{\textbf{52}} & \RustiqCell{\textbf{8}} & \RustiqCell{\textbf{52}} & \RustiqCell{13} & \RustiqCell{\textbf{52}} & \RustiqCell{\textbf{8}} & \RustiqCell{55} & \RustiqCell{35} \\
\rowcolor{black!4}
\RustiqCodeCell{$[[37,1,7]]$} & \RustiqCell{\textbf{60}} & \RustiqCell{\textbf{8}} & \RustiqCell{69} & \RustiqCell{21} & \RustiqCell{\textbf{57}} & \RustiqCell{\textbf{6}} & \RustiqCell{68} & \RustiqCell{27} & \RustiqCell{\textbf{57}} & \RustiqCell{\textbf{6}} & \RustiqCell{66} & \RustiqCell{17} \\
\RustiqCodeCell{$[[72,12,6]]$} & \RustiqCell{\textbf{239}} & \RustiqCell{\textbf{40}} & \RustiqCell{273} & \RustiqCell{48} & \RustiqCell{\textbf{150}} & \RustiqCell{\textbf{14}} & \RustiqCell{202} & \RustiqCell{61} & \RustiqCell{\textbf{150}} & \RustiqCell{\textbf{15}} & \RustiqCell{214} & \RustiqCell{73} \\
\rowcolor{black!4}
\RustiqCodeCell{$[[90,8,10]]$} & \RustiqCell{\textbf{178}} & \RustiqCell{\textbf{23}} & \RustiqCell{193} & \RustiqCell{46} & \RustiqCell{\textbf{134}} & \RustiqCell{\textbf{21}} & \RustiqCell{144} & \RustiqCell{26} & \RustiqCell{\textbf{135}} & \RustiqCell{\textbf{21}} & \RustiqCell{185} & \RustiqCell{54} \\
\RustiqCodeCell{$[[108,8,10]]$} & \RustiqCell{\textbf{367}} & \RustiqCell{\textbf{62}} & \RustiqCell{448} & \RustiqCell{132} & \RustiqCell{\textbf{402}} & \RustiqCell{\textbf{58}} & \RustiqCell{446} & \RustiqCell{133} & \RustiqCell{\textbf{378}} & \RustiqCell{\textbf{49}} & \RustiqCell{457} & \RustiqCell{136} \\
\rowcolor{black!4}
\RustiqCodeCell{$[[144,12,12]]$} & \RustiqCell{\textbf{544}} & \RustiqCell{\textbf{110}} & \RustiqCell{775} & \RustiqCell{224} & \RustiqCell{\textbf{330}} & \RustiqCell{\textbf{39}} & \RustiqCell{581} & \RustiqCell{134} & \RustiqCell{\textbf{330}} & \RustiqCell{\textbf{32}} & \RustiqCell{578} & \RustiqCell{166} \\
\bottomrule
\end{tabular}

\end{table*}%

  \Cref{tab:best-depth-summary,tab:best-gates-summary} report the best circuits\footnote{Note that the tables only report the best circuit found with respect to the primary optimization objective, even though other synthesized circuits might arguably strike a better balance between two-qubit gate count and depth.
  For the $\code{144,12,12}$ gross code, for example, we found a logical $\ket{0}^{\otimes 12}$ circuit with $330$ gates and a depth of $17$ with the greedy heuristic when optimizing for depth, but~\Cref{tab:best-depth-summary} lists the $687$ gate, depth $15$ circuit. See~\Cref{sec:results-state-prep} for more details.} we obtained for both general-state encoders and the corresponding state-preparation tasks using the proposed rollout search and the graph-state-based synthesis tool Rustiq~\cite{goubaultdebrugiere2025graphstatebased}.

Across the benchmark set, the proposed synthesis framework produces competitive and often improved circuits compared to Rustiq, with particularly strong gains for high-rate and larger instances.
The main exceptions are the depth-optimized general-state encoders for the $\code{108,8,10}$ and $\code{144,12,12}$ bivariate bicycle codes, for which Rustiq achieves both lower gate count and lower depth.
The comparison also reveals a qualitative difference between non-CSS and CSS codes.
For non-CSS codes, Rustiq tends to perform better on state-preparation synthesis, whereas our methods are slightly more favorable for general-state encoder synthesis.
A plausible explanation is that, in the non-CSS setting, our method first completes the target isometry to a full stabilizer tableau by introducing destabilizers, making state preparation and general-state encoder synthesis structurally very similar.
In this regime, the preprocessing step in~\Cref {alg:preprocess} appears to yield less favorable tableau completions for state preparation than for encoder synthesis.
By contrast, for CSS codes, our methods perform particularly well, suggesting that restricting the synthesis to a single check matrix substantially simplifies the elimination problem.

Reference~\cite{webster2025heuristicoptimalsynthesis} compares the two-qubit gate counts of their proposed A* search and greedy synthesis for logical $\ket{0}$ states against various other CNOT circuit synthesis tools, determining that the A* approach outperforms all other methods on all instances (Ref.~\cite{webster2025heuristicoptimalsynthesis}, Table 5).
Our gate-counts for the same circuit instances shown in~\Cref{tab:best-gates-summary} either match those reported by Ref.~\cite{webster2025heuristicoptimalsynthesis} or are lower.
For the $\code{23,1,7}$ Golay code, we find an encoder using only $45$ CNOT gates, a reduction of $19\%$ compared to Ref.~\cite{webster2025heuristicoptimalsynthesis} and $21\%$ compared to Ref.~\cite{paetznick2013faulttolerantancillapreparationnoise}.
The circuit from Ref.~\cite{paetznick2013faulttolerantancillapreparationnoise}, however, achieves $57$ CNOT gates at depth $7$, whereas our $45$-CNOT circuit has depth $10$.
When targeting depth as the primary optimization objective, our rollout synthesis also yields a circuit of depth $7$ using $57$ two-qubit gates, matching the circuit metrics of Ref.~\cite{paetznick2013faulttolerantancillapreparationnoise} exactly.

For the $\code{30,6,5}$ high-rate symplectic double code, Ref.~\cite{kanomata2025fault} proposes an optimized state preparation circuit for the logical $\ket{+}$ state using $67$ CNOT gates at a depth of $11$.
Our automated synthesis approach produces a circuit using $62$ gates at a depth of $11$ when optimizing for gates, and a circuit using $68$ two-qubit gates at a depth of $7$ when optimizing for depth.
Ref.~\cite{kanomata2025fault} constructs an encoding circuit for the $\code{30,6,5}$ code by reusing the state preparation circuit for the $\ket{+}$ state and prepending a CNOT circuit to ensure the correct encoding of an arbitrary $6$-qubit state into the code.
This method gives a circuit using $101$ two-qubit gates at a depth of $17$\footnote{The circuit depicted in Ref.~\cite{kanomata2025fault} uses an unoptimized $\ket{+}$ state preparation circuit. The listed circuit metrics are for the optimized encoding circuit.}.
In comparison, our synthesis approach directly constructs a general state encoding circuit using $82$ two-qubit gates at a depth of $13$ when optimizing for gates and a circuit using $95$ two-qubit gates at a depth of $9$ when optimizing for depth.

For the $\code{17,1,5}$ color code, the best circuit we obtain uses $25$ CNOT gates at depth $5$, one CNOT more than the encoder from Ref.~\cite{rodriguez2024experimentaldemonstrationlogicalmagic} and the one found by Rustiq.
Even with extensive rollout, the search does not recover the $24$-CNOT circuit when initialized with our default minimal-weight logical representative. Switching to a different representative, however, guides the synthesis toward a circuit matching the gate count and depth reported in Ref.~\cite{rodriguez2024experimentaldemonstrationlogicalmagic}.

\begin{table*}[t]
\centering
\caption{Depth-optimized encoding circuits for different rollout levels. $G$ denotes two-qubit gate count, $D$ circuit depth, and $\Delta G$ / $\Delta D$ denote the relative improvement (\%) with respect to the greedy baseline.}
\label{tab:depth-rollout-summary}
\providecommand{\RolloutCell}[1]{\hspace{7pt}#1\hspace{7pt}}
\providecommand{\RolloutHdr}[1]{\multicolumn{1}{c}{#1}}
\setlength{\tabcolsep}{0pt}
\renewcommand{\arraystretch}{1.1}
\begin{tabular}{lrrrcrrrrrcrrrr}
\toprule
& \multicolumn{2}{c}{\textbf{Greedy}} & \multicolumn{6}{c}{\textbf{Rollout 1}} & \multicolumn{6}{c}{\textbf{Rollout 2}} \\
\cmidrule(r{0.4em}){2-3}
\cmidrule(lr{0.4em}){4-9}
\cmidrule(l{0.4em}){10-15}
Code & \RolloutHdr{$G$} & \RolloutHdr{$D$} & \RolloutHdr{\hspace{2pt}$t$\hspace{6pt}} & \RolloutHdr{\hspace{6pt}ET\hspace{2pt}} & \RolloutHdr{$G$} & \RolloutHdr{$\Delta G$} & \RolloutHdr{$D$} & \RolloutHdr{$\Delta D$} & \RolloutHdr{\hspace{2pt}$(t_1,t_2)$\hspace{6pt}} & \RolloutHdr{\hspace{6pt}ET\hspace{2pt}} & \RolloutHdr{$G$} & \RolloutHdr{$\Delta G$} & \RolloutHdr{$D$} & \RolloutHdr{$\Delta D$} \\
\midrule
\rowcolor{black!6}
\RolloutCell{$[[8,3,3]]$} & \RolloutCell{15} & \RolloutCell{6} & \RolloutCell{10} & \RolloutCell{n} & \RolloutCell{\textbf{14}} & \RolloutCell{6.7} & \RolloutCell{\textbf{5}} & \RolloutCell{16.7} & \RolloutCell{10,2} & \RolloutCell{y} & \RolloutCell{\textbf{14}} & \RolloutCell{6.7} & \RolloutCell{\textbf{5}} & \RolloutCell{16.7} \\
\RolloutCell{$[[15,1,3]]$} & \RolloutCell{\textbf{24}} & \RolloutCell{\textbf{4}} & \RolloutCell{10} & \RolloutCell{y} & \RolloutCell{\textbf{24}} & \RolloutCell{0.0} & \RolloutCell{\textbf{4}} & \RolloutCell{0.0} & \RolloutCell{10,2} & \RolloutCell{y} & \RolloutCell{\textbf{24}} & \RolloutCell{0.0} & \RolloutCell{\textbf{4}} & \RolloutCell{0.0} \\
\rowcolor{black!6}
\RolloutCell{$[[15,3,5]]$} & \RolloutCell{53} & \RolloutCell{12} & \RolloutCell{10} & \RolloutCell{n} & \RolloutCell{45} & \RolloutCell{15.1} & \RolloutCell{\textbf{10}} & \RolloutCell{16.7} & \RolloutCell{40,20} & \RolloutCell{y} & \RolloutCell{\textbf{43}} & \RolloutCell{18.9} & \RolloutCell{\textbf{10}} & \RolloutCell{16.7} \\
\RolloutCell{$[[15,7,3]]$} & \RolloutCell{36} & \RolloutCell{9} & \RolloutCell{30} & \RolloutCell{n} & \RolloutCell{33} & \RolloutCell{8.3} & \RolloutCell{7} & \RolloutCell{22.2} & \RolloutCell{30,15} & \RolloutCell{y} & \RolloutCell{\textbf{31}} & \RolloutCell{13.9} & \RolloutCell{\textbf{6}} & \RolloutCell{33.3} \\
\rowcolor{black!6}
\RolloutCell{$[[17,1,5]]$} & \RolloutCell{\textbf{25}} & \RolloutCell{\textbf{5}} & \RolloutCell{10} & \RolloutCell{y} & \RolloutCell{\textbf{25}} & \RolloutCell{0.0} & \RolloutCell{\textbf{5}} & \RolloutCell{0.0} & \RolloutCell{10,2} & \RolloutCell{y} & \RolloutCell{\textbf{25}} & \RolloutCell{0.0} & \RolloutCell{\textbf{5}} & \RolloutCell{0.0} \\
\RolloutCell{$[[19,1,5]]$} & \RolloutCell{\textbf{29}} & \RolloutCell{6} & \RolloutCell{10} & \RolloutCell{n} & \RolloutCell{\textbf{29}} & \RolloutCell{0.0} & \RolloutCell{\textbf{5}} & \RolloutCell{16.7} & \RolloutCell{10,2} & \RolloutCell{n} & \RolloutCell{\textbf{29}} & \RolloutCell{0.0} & \RolloutCell{\textbf{5}} & \RolloutCell{16.7} \\
\rowcolor{black!6}
\RolloutCell{$[[23,1,7]]$} & \RolloutCell{63} & \RolloutCell{9} & \RolloutCell{100} & \RolloutCell{n} & \RolloutCell{\textbf{53}} & \RolloutCell{15.9} & \RolloutCell{\textbf{8}} & \RolloutCell{11.1} & \RolloutCell{50,25} & \RolloutCell{y} & \RolloutCell{55} & \RolloutCell{12.7} & \RolloutCell{\textbf{8}} & \RolloutCell{11.1} \\
\RolloutCell{$[[30,6,5]]$} & \RolloutCell{99} & \RolloutCell{13} & \RolloutCell{100} & \RolloutCell{n} & \RolloutCell{\textbf{92}} & \RolloutCell{7.1} & \RolloutCell{10} & \RolloutCell{23.1} & \RolloutCell{30,6} & \RolloutCell{n} & \RolloutCell{95} & \RolloutCell{4.0} & \RolloutCell{\textbf{9}} & \RolloutCell{30.8} \\
\rowcolor{black!6}
\RolloutCell{$[[31,1,7]]$} & \RolloutCell{\textbf{48}} & \RolloutCell{6} & \RolloutCell{10} & \RolloutCell{y} & \RolloutCell{\textbf{48}} & \RolloutCell{0.0} & \RolloutCell{6} & \RolloutCell{0.0} & \RolloutCell{50,25} & \RolloutCell{y} & \RolloutCell{\textbf{48}} & \RolloutCell{0.0} & \RolloutCell{\textbf{5}} & \RolloutCell{16.7} \\
\RolloutCell{$[[31,21,3]]$} & \RolloutCell{127} & \RolloutCell{23} & \RolloutCell{10} & \RolloutCell{n} & \RolloutCell{\textbf{109}} & \RolloutCell{14.2} & \RolloutCell{\textbf{11}} & \RolloutCell{52.2} & \RolloutCell{20,4} & \RolloutCell{n} & \RolloutCell{112} & \RolloutCell{11.8} & \RolloutCell{12} & \RolloutCell{47.8} \\
\rowcolor{black!6}
\RolloutCell{$[[37,1,7]]$} & \RolloutCell{65} & \RolloutCell{8} & \RolloutCell{40} & \RolloutCell{y} & \RolloutCell{\textbf{63}} & \RolloutCell{3.1} & \RolloutCell{7} & \RolloutCell{12.5} & \RolloutCell{30,6} & \RolloutCell{n} & \RolloutCell{\textbf{63}} & \RolloutCell{3.1} & \RolloutCell{\textbf{6}} & \RolloutCell{25.0} \\
\RolloutCell{$[[72,12,6]]$} & \RolloutCell{274} & \RolloutCell{33} & \RolloutCell{50} & \RolloutCell{n} & \RolloutCell{\textbf{261}} & \RolloutCell{4.7} & \RolloutCell{\textbf{19}} & \RolloutCell{42.4} & \RolloutCell{50,10} & \RolloutCell{y} & \RolloutCell{268} & \RolloutCell{2.2} & \RolloutCell{20} & \RolloutCell{39.4} \\
\rowcolor{black!6}
\RolloutCell{$[[90,8,10]]$} & \RolloutCell{187} & \RolloutCell{20} & \RolloutCell{50} & \RolloutCell{n} & \RolloutCell{\textbf{183}} & \RolloutCell{2.1} & \RolloutCell{\textbf{12}} & \RolloutCell{40.0} & \RolloutCell{10,2} & \RolloutCell{n} & \RolloutCell{187} & \RolloutCell{0.0} & \RolloutCell{\textbf{12}} & \RolloutCell{40.0} \\
\RolloutCell{$[[108,8,10]]$} & \RolloutCell{750} & \RolloutCell{30} & \RolloutCell{100} & \RolloutCell{y} & \RolloutCell{724} & \RolloutCell{3.5} & \RolloutCell{\textbf{24}} & \RolloutCell{20.0} & \RolloutCell{40,8} & \RolloutCell{y} & \RolloutCell{\textbf{667}} & \RolloutCell{11.1} & \RolloutCell{\textbf{24}} & \RolloutCell{20.0} \\
\rowcolor{black!6}
\RolloutCell{$[[144,12,12]]$} & \RolloutCell{798} & \RolloutCell{83} & \RolloutCell{200} & \RolloutCell{y} & \RolloutCell{1081} & \RolloutCell{-35.5} & \RolloutCell{\textbf{29}} & \RolloutCell{65.1} & \RolloutCell{20,10} & \RolloutCell{y} & \RolloutCell{\textbf{700}} & \RolloutCell{12.3} & \RolloutCell{70} & \RolloutCell{15.7} \\
\bottomrule
\end{tabular}

\end{table*}%

Interestingly, our methods find zero-state preparation circuits for the $\code{108,8,10}$ and $\code{144,12,12}$ bivariate bicycle codes requiring $430$ and $330$ two-qubit gates, respectively, even though the latter code has more physical qubits and stabilizers. 
A related anomaly appears for the $\code{108,8,10}$ code: The gate-optimized general-state encoder requires only $367$ two-qubit gates, whereas the synthesized logical $\ket{0}^{\otimes 8}$ state-preparation circuit requires $402$, despite the fact that the latter could in principle be obtained from the encoder simply by fixing the inputs to $\ket{0}$.
A closer inspection suggests that this behavior is caused by repeated local minima in the greedy search.
As described in~\Cref{sec:encod-circ-synth}, escaping such minima may require row operations, row reduction, or short multi-gate sequences.
While these steps allow the synthesis to proceed, they can also leave the intermediate check matrix in a form that is less favorable for subsequent greedy optimization.
For the $\code{108,8,10}$ code, this appears to happen repeatedly during state-preparation synthesis, leading to the observed overhead.
Local minima are encountered in other instances as well --- for example, in the synthesis of the $\code{31,21,3}$ quantum Hamming code --- but there such minima do not lead to a similarly large overhead.
This finding suggests that more sophisticated strategies for escaping local minima could further improve synthesis quality for such instances.

\subsection{Impact of Rollout on the Search}
\label{sec:impact-look-search}

To illustrate the impact of the rollout depth $\ell$ on the search,~\Cref{tab:depth-rollout-summary} (\Cref{tab:gates-rollout-summary}) lists the \emph{best} circuit obtained for each choice of $\ell$ when optimizing for depth (two-qubit gates).
Greedy corresponds to the case of $\ell=0$.
If $\ell > 0$ the $t$ columns correspond to the \emph{lowest} setting of $t$ for which the corresponding circuit was obtained, meaning no further improvements could be obtained by increasing the number of candidate gates explored during the search.
The ET column denotes whether the circuit was obtained with early search termination enabled.
A \enquote{n} indicates that a better circuit was found when not enabling early termination, and a \enquote{y} indicates that no improvements could be made by disabling early termination.

From these results, we can see that the rollout is particularly helpful in optimizing the secondary objective.
Even when additional rollout yields no (or little) benefit for the primary objective, significant improvements can be achieved for the secondary objective.
For synthesis with gate optimization, this approach reduces depth by up to $63\%$ in some instances.

\begin{table*}[t]
\centering
\caption{Gate-optimized encoding circuits for different rollout levels. $G$ denotes two-qubit gate count, $D$ circuit depth, and $\Delta G$ / $\Delta D$ denote the relative improvement (\%) with respect to the greedy baseline.}
\label{tab:gates-rollout-summary}
\providecommand{\RolloutCell}[1]{\hspace{7pt}#1\hspace{7pt}}
\providecommand{\RolloutHdr}[1]{\multicolumn{1}{c}{#1}}
\setlength{\tabcolsep}{0pt}
\renewcommand{\arraystretch}{1.1}
\begin{tabular}{lrrrcrrrrrcrrrr}
\toprule
& \multicolumn{2}{c}{\textbf{Greedy}} & \multicolumn{6}{c}{\textbf{Rollout 1}} & \multicolumn{6}{c}{\textbf{Rollout 2}} \\
\cmidrule(r{0.4em}){2-3}
\cmidrule(lr{0.4em}){4-9}
\cmidrule(l{0.4em}){10-15}
Code & \RolloutHdr{$G$} & \RolloutHdr{$D$} & \RolloutHdr{\hspace{2pt}$t$\hspace{6pt}} & \RolloutHdr{\hspace{6pt}ET\hspace{2pt}} & \RolloutHdr{$G$} & \RolloutHdr{$\Delta G$} & \RolloutHdr{$D$} & \RolloutHdr{$\Delta D$} & \RolloutHdr{\hspace{2pt}$(t_1,t_2)$\hspace{6pt}} & \RolloutHdr{\hspace{6pt}ET\hspace{2pt}} & \RolloutHdr{$G$} & \RolloutHdr{$\Delta G$} & \RolloutHdr{$D$} & \RolloutHdr{$\Delta D$} \\
\midrule
\rowcolor{black!6}
\RolloutCell{$[[8,3,3]]$} & \RolloutCell{13} & \RolloutCell{7} & \RolloutCell{10} & \RolloutCell{n} & \RolloutCell{13} & \RolloutCell{0.0} & \RolloutCell{\textbf{6}} & \RolloutCell{14.3} & \RolloutCell{40,8} & \RolloutCell{y} & \RolloutCell{\textbf{12}} & \RolloutCell{7.7} & \RolloutCell{8} & \RolloutCell{-14.3} \\
\RolloutCell{$[[15,1,3]]$} & \RolloutCell{\textbf{24}} & \RolloutCell{10} & \RolloutCell{20} & \RolloutCell{n} & \RolloutCell{\textbf{24}} & \RolloutCell{0.0} & \RolloutCell{6} & \RolloutCell{40.0} & \RolloutCell{10,2} & \RolloutCell{n} & \RolloutCell{\textbf{24}} & \RolloutCell{0.0} & \RolloutCell{\textbf{5}} & \RolloutCell{50.0} \\
\rowcolor{black!6}
\RolloutCell{$[[15,3,5]]$} & \RolloutCell{48} & \RolloutCell{25} & \RolloutCell{20} & \RolloutCell{n} & \RolloutCell{39} & \RolloutCell{18.8} & \RolloutCell{12} & \RolloutCell{52.0} & \RolloutCell{30,6} & \RolloutCell{n} & \RolloutCell{\textbf{36}} & \RolloutCell{25.0} & \RolloutCell{\textbf{11}} & \RolloutCell{56.0} \\
\RolloutCell{$[[15,7,3]]$} & \RolloutCell{35} & \RolloutCell{18} & \RolloutCell{20} & \RolloutCell{n} & \RolloutCell{31} & \RolloutCell{11.4} & \RolloutCell{8} & \RolloutCell{55.6} & \RolloutCell{50,10} & \RolloutCell{n} & \RolloutCell{\textbf{28}} & \RolloutCell{20.0} & \RolloutCell{\textbf{7}} & \RolloutCell{61.1} \\
\rowcolor{black!6}
\RolloutCell{$[[17,1,5]]$} & \RolloutCell{\textbf{25}} & \RolloutCell{7} & \RolloutCell{10} & \RolloutCell{y} & \RolloutCell{\textbf{25}} & \RolloutCell{0.0} & \RolloutCell{7} & \RolloutCell{0.0} & \RolloutCell{10,5} & \RolloutCell{n} & \RolloutCell{\textbf{25}} & \RolloutCell{0.0} & \RolloutCell{\textbf{5}} & \RolloutCell{28.6} \\
\RolloutCell{$[[19,1,5]]$} & \RolloutCell{30} & \RolloutCell{13} & \RolloutCell{30} & \RolloutCell{y} & \RolloutCell{\textbf{29}} & \RolloutCell{3.3} & \RolloutCell{7} & \RolloutCell{46.1} & \RolloutCell{20,10} & \RolloutCell{n} & \RolloutCell{\textbf{29}} & \RolloutCell{3.3} & \RolloutCell{\textbf{6}} & \RolloutCell{53.9} \\
\rowcolor{black!6}
\RolloutCell{$[[23,1,7]]$} & \RolloutCell{57} & \RolloutCell{27} & \RolloutCell{50} & \RolloutCell{n} & \RolloutCell{50} & \RolloutCell{12.3} & \RolloutCell{13} & \RolloutCell{51.9} & \RolloutCell{40,8} & \RolloutCell{n} & \RolloutCell{\textbf{49}} & \RolloutCell{14.0} & \RolloutCell{\textbf{10}} & \RolloutCell{63.0} \\
\RolloutCell{$[[30,6,5]]$} & \RolloutCell{101} & \RolloutCell{29} & \RolloutCell{200} & \RolloutCell{n} & \RolloutCell{85} & \RolloutCell{15.8} & \RolloutCell{18} & \RolloutCell{37.9} & \RolloutCell{30,15} & \RolloutCell{n} & \RolloutCell{\textbf{82}} & \RolloutCell{18.8} & \RolloutCell{\textbf{13}} & \RolloutCell{55.2} \\
\rowcolor{black!6}
\RolloutCell{$[[31,1,7]]$} & \RolloutCell{49} & \RolloutCell{13} & \RolloutCell{30} & \RolloutCell{n} & \RolloutCell{\textbf{47}} & \RolloutCell{4.1} & \RolloutCell{8} & \RolloutCell{38.5} & \RolloutCell{20,10} & \RolloutCell{n} & \RolloutCell{\textbf{47}} & \RolloutCell{4.1} & \RolloutCell{\textbf{6}} & \RolloutCell{53.9} \\
\RolloutCell{$[[31,21,3]]$} & \RolloutCell{125} & \RolloutCell{47} & \RolloutCell{200} & \RolloutCell{n} & \RolloutCell{100} & \RolloutCell{20.0} & \RolloutCell{20} & \RolloutCell{57.5} & \RolloutCell{50,25} & \RolloutCell{n} & \RolloutCell{\textbf{94}} & \RolloutCell{24.8} & \RolloutCell{\textbf{18}} & \RolloutCell{61.7} \\
\rowcolor{black!6}
\RolloutCell{$[[37,1,7]]$} & \RolloutCell{61} & \RolloutCell{14} & \RolloutCell{30} & \RolloutCell{n} & \RolloutCell{61} & \RolloutCell{0.0} & \RolloutCell{10} & \RolloutCell{28.6} & \RolloutCell{20,4} & \RolloutCell{n} & \RolloutCell{\textbf{60}} & \RolloutCell{1.6} & \RolloutCell{\textbf{8}} & \RolloutCell{42.9} \\
\RolloutCell{$[[72,12,6]]$} & \RolloutCell{262} & \RolloutCell{88} & \RolloutCell{40} & \RolloutCell{n} & \RolloutCell{241} & \RolloutCell{8.0} & \RolloutCell{\textbf{40}} & \RolloutCell{54.5} & \RolloutCell{10,5} & \RolloutCell{n} & \RolloutCell{\textbf{239}} & \RolloutCell{8.8} & \RolloutCell{\textbf{40}} & \RolloutCell{54.5} \\
\rowcolor{black!6}
\RolloutCell{$[[90,8,10]]$} & \RolloutCell{179} & \RolloutCell{36} & \RolloutCell{40} & \RolloutCell{n} & \RolloutCell{179} & \RolloutCell{0.0} & \RolloutCell{24} & \RolloutCell{33.3} & \RolloutCell{20,4} & \RolloutCell{n} & \RolloutCell{\textbf{178}} & \RolloutCell{0.6} & \RolloutCell{\textbf{23}} & \RolloutCell{36.1} \\
\RolloutCell{$[[108,8,10]]$} & \RolloutCell{625} & \RolloutCell{107} & \RolloutCell{50} & \RolloutCell{n} & \RolloutCell{\textbf{367}} & \RolloutCell{41.3} & \RolloutCell{\textbf{62}} & \RolloutCell{42.1} & \RolloutCell{40,20} & \RolloutCell{y} & \RolloutCell{371} & \RolloutCell{40.6} & \RolloutCell{97} & \RolloutCell{9.3} \\
\rowcolor{black!6}
\RolloutCell{$[[144,12,12]]$} & \RolloutCell{682} & \RolloutCell{152} & \RolloutCell{40} & \RolloutCell{n} & \RolloutCell{\textbf{544}} & \RolloutCell{20.2} & \RolloutCell{\textbf{110}} & \RolloutCell{27.6} & \RolloutCell{20,4} & \RolloutCell{y} & \RolloutCell{556} & \RolloutCell{18.5} & \RolloutCell{157} & \RolloutCell{-3.3} \\
\bottomrule
\end{tabular}

\end{table*}%

We also see that, for $\ell=2$ and larger codes, we obtain worse results than for $\ell=1$.
The reason for this is that the extensive rollout exceeds the $24\si{h}$ timeout.
The best circuits found are either those where the search terminated early, missing optimization potential during synthesis, or those where the number of considered candidates was too small to find an improvement over the $\ell=1$ rollout.

\subsection{Encoding Circuit Synthesis for the $12$-qubit HaPPY Code}
\label{sec:encod-circ-synth-1}

We now evaluate the modular synthesis workflow on the $12$-qubit HaPPY sub-isometry highlighted in~\Cref{fig:happy}. This code is obtained by composing four copies of the five-qubit encoding isometry in the non-tree-like pattern described in~\Cref{sec:modular-synthesis}, yielding an encoding isometry from four bulk logical qubits to twelve boundary qubits~\cite{angles2024engineering}.
Our goal is to compare direct extraction of the local constituent encoders, exact resynthesis of these local encoders, and rollout-based resynthesis of the full $12$-qubit block.

The $12$-qubit code has previously been analyzed using ZX-diagram rewrites in Ref.~\cite{wu2024zxcalculusapproachconstructiongraph}, where the encoder is optimized at the diagrammatic level and a circuit is extracted afterwards~\cite{backens2021therebackagain,duncan2020graphtheoretic}. Here, by contrast, we use the partition from~\Cref{sec:modular-synthesis} to construct the global encoder from optimized local circuits.

To make the modular structure explicit, we first reorder the ZX-diagram so that the flow of intermediate wires between constituent encoders becomes visible:

\begin{equation}
    \resizebox{\linewidth}{!}{
  \label{eq:holographic_toy_zx}
\resizebox{!}{50pt}{\input{zx/holography_toy_zx.tikz}}  = \resizebox{!}{50pt}{\input{zx/holography_toy_zx_ordered.tikz}}  }
\end{equation}

Reading the reordered diagram from left to right, we identify three distinct local isometries (up to spider rearrangements).
We can identify the stabilizer tableau of these local encoders and extract circuits from them by repeated application of the spider fusion rule.

{
  \centering
  \setlength{\tabcolsep}{6pt}
  \renewcommand{\arraystretch}{1.30}
  \begin{tabular}{@{}ccc
                  @{}}
    \toprule
    Sub-encoder & Tableau & Extracted circuit \\
    \midrule

    \SubEncoderFig{24pt}{zx/holography_toy_zx_encoder_5_1}
    &
    \Tableau[24pt]{ccccc}{
      Z&Z&Z&Z&Z\\
      X&X&X&X&X\\\midrule
      X&Z&Z&X&I\\
      I&X&Z&Z&X\\
      X&I&X&Z&Z\\
      Z&X&I&X&Z
    }
    &
    \ExtractedFig{20pt}{zx/holography_toy_zx_encoder_5_4}
    \\[2.5em]

    \SubEncoderFig{20pt}{zx/holography_toy_zx_encoder_4_1}
    &
    \Tableau[24pt]{cccc}{
      Z&I&X&X\\
      Z&Z&X&I\\
      I&Z&X&Z\\
      Z&X&I&X\\\midrule
      X&Z&Z&X\\
      Z&Y&Y&Z
    }
    &
    \ExtractedFig{15pt}{zx/holography_toy_zx_encoder_4_4}
    \\[2.5em]

    \SubEncoderFig{22pt}{zx/holography_toy_zx_encoder_3_1}
    &
    \Tableau[24pt]{ccc}{
      Z&Z&X\\
      X&Z&Z\\
      Y&Z&Y\\
      Y&Y&Z\\
      Z&Y&Y\\
      Z&X&Z
    }
    &
    \ExtractedFig{18pt}{zx/holography_toy_zx_encoder_3_3}
    \\
    \bottomrule
  \end{tabular}
}

\begin{figure*}[t]
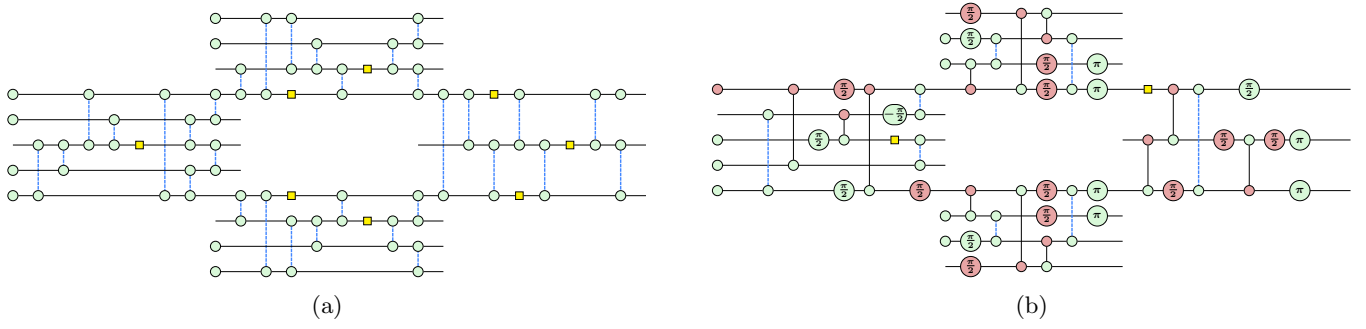

  \centering
    \begin{subfigure}[b]{.48\linewidth}
      \resizebox{\linewidth}{!}{\input{zx/holography_toy_zx_extracted.tikz}}
      \caption{}
      \label{fig:holography-toy-a}
    \end{subfigure}
    \hfill
    \begin{subfigure}[b]{.48\linewidth}
      \resizebox{\linewidth}{!}{\input{zx/holography_toy_zx_extracted_optimized.tikz}}
      \caption{}
      \label{fig:holography-toy-b}
    \end{subfigure}

    \caption{Encoding circuits for the $12$-qubit HaPPY code obtained from two modular synthesis strategies. \textbf{(a)} Circuit obtained by directly extracting local encoding circuits from the constituent sub-isometries, requiring $32$ two-qubit gates. \textbf{(b)} Circuit obtained by optimally resynthesizing the local constituent encoders, reducing the gate count to $20$ two-qubit gates at depth $10$. For comparison, direct rollout-based resynthesis of the full $12$-qubit encoder yields a circuit with $22$ two-qubit gates at depth $9$ when optimizing for gate count, and a circuit with $24$ two-qubit gates at depth $8$ when optimizing for depth. This illustrates the trade-off between modular exact local synthesis and full heuristic resynthesis of the larger encoding block.} \label{fig:holography-toy}
  \end{figure*}
  
As before, Pauli operators above the horizontal line are logical operators.
Note that the last isometry is a Clifford unitary as it \enquote{encodes} three logical qubits into three physical qubits. 
In particular, this means that the isometry has no stabilizers.

Composing the directly extracted local circuits yields the encoder shown in~\Cref{fig:holography-toy-a}.
This circuit directly implements the $12$-qubit holographic encoder using $32$ two-qubit gates, compared to $48$ two-qubit gates in Ref.~\cite{wu2024zxcalculusapproachconstructiongraph}, and does not require additional qubits.

Applying the exact local resynthesis method from the previous section further improves this. Resynthesizing the local constituent encoders yields the circuit shown in~\Cref{fig:holography-toy-b}, which requires only $20$ two-qubit gates at depth $10$. For comparison, the encoding map in Refs.~\cite{wu2024zxcalculusapproachconstructiongraph,angles2024engineering} already requires $20$ CZ gates merely to prepare the logical $\ket{0000}_L$ graph state that is subsequently entangled with the logical inputs.

For the same $12$-qubit instance, direct rollout-based resynthesis of the full encoder yields a circuit with $22$ two-qubit gates at depth $9$ when optimizing for gate count, and a circuit with $24$ two-qubit gates at depth $8$ when optimizing for depth.
Thus, modular exact synthesis of the local constituent encoders gives the smallest gate count, whereas direct resynthesis of the full encoder yields shallower circuits.
This tradeoff directly illustrates the trade-off discussed in~\Cref{sec:modular-synthesis}: Optimizing smaller local pieces is more effective for reducing gate count, while optimizing larger connected blocks can reduce composition overhead and thereby improve depth.

\section{Conclusions and outlook}
\label{sec:conclusion}

We developed methods for synthesizing and optimizing encoding circuits for arbitrary stabilizer codes.
For codes lacking modular structure, we formulated encoder synthesis as a search over stabilizer tableaus and introduced greedy and rollout-based methods that optimize two-qubit gate count and circuit depth. For modular code families, such as generalized concatenated and holographic codes, we exploited decompositions into local constituent encoders and used SMT-based exact synthesis to optimize these constituent encoders.

Across a broad benchmark set, these methods yield competitive and often improved circuits compared to recent encoder-synthesis methods and existing constructions from the literature, with improvements of up to \(43\%\) in two-qubit gate count and up to \(70\%\) in depth.
The results indicate that exploiting the freedom among stabilizer-equivalent realizations can substantially improve general-state encoder synthesis, and that modular exact synthesis provides an effective complementary strategy for structured code families.

Several directions for future work remain.
On the algorithmic side, it would be interesting to optimize for objectives beyond two-qubit gate count and depth, such as logical error rate, especially in settings such as magic state preparation~\cite{goto2016minimizingresourceoverheads, chamberland2019faulttolerantmagic, postler2022demonstration, dasu2025breakingmagicdemonstrationhighfidelity}.
Since arbitrary-input encoding circuits are not circuit-level fault-tolerant in general, optimizing them in the context of a specific FTQC protocol and under circuit-level noise is a particularly promising direction.
Incorporating noise directly into the rollout search would likely be computationally demanding, but the independent synthesis tasks appearing in such an approach may offer opportunities for parallelization.

More broadly, we envision our circuits also finding application in quantum communication settings where the goal is to maximize transmission rates by employing $\code{n,k,d}$ codes with high encoding rates $k/n$ for large $n$~\cite{bonilla2025constant, belzig2026constantspaceoverheadfaulttolerantquantuminputoutput}. 
Regarding concatenated codes, we have so far only partially exploited the internal mathematical structure of generalized concatenated and holographic codes.
It would therefore be interesting to investigate whether additional structure in these code families beyond modularity can be leveraged to achieve more efficient synthesis.
One may imagine applying our methods to other types of quantum circuits, for instance, realizing logical gates \cite{chen2025tailoring, baspin2025fast, popov2026optimizedcompilationlogicalclifford}. 
Further related directions include the synthesis of morphing circuits~\cite{vasmer2022morphing, wolanski2026automatedcompilationincludingdropouts, shaw2026optimisingquantumerrorcorrection} or the extension of rollout-based search to other compilation tasks like the orchestration of physical control sequences that realize FTQC on the hardware level \cite{wang2024atomique, conta2026toolchainshuttlingtrappedionqubits}.

Overall, our results show that automated synthesis can substantially improve the design of encoding circuits and provide a practical route to lower-overhead encoded-state preparation across a broad range of stabilizer-code constructions.

\section*{Author contributions}

TP, MS, and SH conceptualized and developed the project. TP implemented the rollout scheme and wrote the manuscript with input, contributions, and feedback from MS and SH. RW provided formal supervision. 

\section*{Acknowledgments}
We would like to thank Michael A. Perlin, Joseph Sullivan, and Ludwig Schmid for proofreading and their valuable feedback on the manuscript.

While writing this manuscript, Doherty et al.~\cite{doherty2026faststabilizerstatepreparation} published work on synthesizing state preparation using reinforcement learning (among other methods).
While the work focuses on state preparation, the methods could be extended to the synthesis of Clifford isometries as the authors acknowledge.
With their proposed methods, Doherty et al. manage to obtain a logical $\ket{0}$ state preparation circuit for the Golay code using $44$ two-qubit gates at a depth of $8$---an improvement to the $45$ CNOT, depth $10$ circuit obtained with our methods.
For the Gross code, they find a logical $\ket{0}^{\otimes 12}$ preparation circuit using $357$ two-qubit gates at a depth of $18$---using more CNOTs and depth than the $330$ CNOT, depth $17$ circuit we found.
These results provide encouraging evidence that future work might further improve the synthesis of encoding and state preparation circuits. 

TP and RW acknowledge funding from the European Research Council (ERC) under the European Union’s Horizon 2020 research and innovation program (grant agreement No.\ 101001318) and Millenion (grant agreement No.\ 101114305). Their work was carried out as part of Munich Quantum Valley, which is supported by the Bavarian state government with funds from the Hightech Agenda Bayern Plus. TP and RW also acknowledge funding from the Deutsche Forschungsgemeinschaft (DFG, German Research Foundation, No.\ 563402549).

SH gratefully acknowledges funding from the German Federal Ministry of Research, Technology and Space (BMFTR) as part of the Research Program Quantum Systems, research projects 13N17317 (”SQale”) and 13N17421 ("snaQCs2025").

\section*{Disclaimer}
This paper was prepared for informational purposes with contributions from the Global Technology Applied Research center of JPMorgan Chase \& Co. This paper is not a product of the Research Department of JPMorgan Chase \& Co. or its affiliates. Neither JPMorgan Chase \& Co. nor any of its affiliates makes any explicit or implied representation or warranty and none of them accept any liability in connection with this paper, including, without limitation, with respect to the completeness, accuracy, or reliability of the information contained herein and the potential legal, compliance, tax, or accounting effects thereof. This document is not intended as investment research or investment advice, or as a recommendation, offer, or solicitation for the purchase or sale of any security, financial instrument, financial product or service, or to be used in any way for evaluating the merits of participating in any transaction.

\clearpage
\bibliographystyle{bibstyle}
\bibliography{references}

\clearpage
\appendix

\section{ZX-calculus}
\label{sec:zx-diagrams}

A ZX-diagram is constructed from two types of nodes or \enquote{spiders}, the $Z$- and the $X$-spider:

\medskip

\input{zx/z_spider.tikz} $= \ket{0}^{\otimes n} \bra{0}^{\otimes m} + e^{i\alpha} \ket{1}^{\otimes n} \bra{1}^{\otimes m}$

\medskip

\input{zx/x_spider.tikz} $= \ket{+}^{\otimes n} \bra{+}^{\otimes m} + e^{i\alpha} \ket{-}^{\otimes n} \bra{-}^{\otimes m}$

\medskip

Clifford ZX-diagrams are comprised of spiders which have phases of $k\frac{\pi}{2}$ for $k\in \{0,1,2,3\}$.
The ZX-diagram for a CNOT gate is composed of a phase-free $Z$- and a phase-free $X$-spider:

\begin{center}
  \input{zx/cnot_gate.tikz}
\end{center}

A Hadamard gate is often denoted by a yellow box:

\begin{center}
  \begin{tikzpicture}
	\begin{pgfonlayer}{nodelayer}
		\node [style=none] (2) at (-1, 0) {};
		\node [style=none] (3) at (1, 0) {};
		\node [style=hadamard] (6) at (0, 0) {};
	\end{pgfonlayer}
	\begin{pgfonlayer}{edgelayer}
		\draw (2.center) to (6);
		\draw (3.center) to (6);
	\end{pgfonlayer}
\end{tikzpicture}

\end{center}

Wires carrying a Hadamard box are often abbreviated using a blue-dashed line. Consequently, the CZ gate has multiple representations as a ZX-diagram.

\begin{center}
  \input{zx/cz_double_h_box.tikz} $=$ \input{zx/cz_single_h_box.tikz} $=$ \input{zx/cz_dashed.tikz}
\end{center}

The ZX-calculus is a graphical calculus for rewriting ZX-diagrams.
Only seven axioms are required to derive any equality between ZX-diagrams~\cite{vandewetering2020zxcalculusworkingquantumcomputer}, but for our purposes, we only require a subset of these:

\begin{itemize}
  \item The identity rule:

\begin{center}
  \input{zx/identity_rule.tikz}
\end{center}

  \item The spider-fusion rule:

\begin{center}
  \input{zx/fusion.tikz}
\end{center}

\item The Hadamard (or color change) rule:

\begin{center}
  \input{zx/hadamard_rule.tikz}
\end{center}

\item The $\pi$-copy rule:

\begin{center}
  \input{zx/pi_copy.tikz}
\end{center}
\end{itemize}

The rules also work with all the colors reversed.

A Clifford ZX-diagram with $m$ inputs and $n>m$ outputs can be used to define an encoding isometry for a $\code{n,m,d}$ code.
An example for a $1$-input, $5$-output encoder is the isometry of the $5$-qubit code shown in~\Cref{fig:513_fig}.
The encoded logical operator can be obtained from this diagram via \enquote{Pauli pushing} by applying the $\pi$-copy rule repeatedly to move a $\pi$ spider (the ZX version of Pauli $X$ and $Z$) from the input to the output:

\[
\resizebox{\linewidth}{!}{$
\begin{aligned}
&\input{zx/five_qubit_encoder_pushing.tikz}
= \input{zx/five_qubit_encoder_pushing_x_2.tikz}
 = \input{zx/five_qubit_encoder_pushing_x_3.tikz}
 = {} \\
&\input{zx/five_qubit_encoder_pushing_x_4.tikz}
 = \input{zx/five_qubit_encoder_pushing_x_5.tikz}
\end{aligned}
$}
\]

We can check the correctness of the stabilizers via Pauli pushing:

\[
\resizebox{\linewidth}{!}{$
\begin{aligned}
&\input{zx/five_qubit_encoder_pushing_stab_1.tikz}
= \input{zx/five_qubit_encoder_pushing_stab_2.tikz}
 = \input{zx/five_qubit_encoder_pushing_stab_3.tikz}
 = {} \\
&\input{zx/five_qubit_encoder_pushing_stab_4.tikz}
 = \input{zx/five_qubit_encoder_pushing_stab_5.tikz}
\end{aligned}
$}
\]

We can also diagrammatically shorten the code by pushing \enquote{logical} Paulis on an input leg.

\[
\resizebox{\linewidth}{!}{$
\begin{aligned}
&\input{zx/five_qubit_encoder_shorten_x_1.tikz}
= \input{zx/five_qubit_encoder_shorten_x_2.tikz}
 = \input{zx/five_qubit_encoder_shorten_x_3.tikz}
 = {} \\
&\input{zx/five_qubit_encoder_shorten_x_4.tikz}
  = \input{zx/five_qubit_encoder_shorten_x_5.tikz}
   = \input{zx/five_qubit_encoder_shorten_x_6.tikz}
  = {} \\
  &\input{zx/five_qubit_encoder_shorten_x_7.tikz}
  = \input{zx/five_qubit_encoder_shorten_x_8.tikz}
   = \input{zx/five_qubit_encoder_shorten_x_9.tikz}
 = {} \\
\end{aligned}
$}
\]

Hence, we can also view this diagram as encoding two logical qubits.

\section{SMT Encoding for Optimal Encoding Circuit Synthesis}
\label{sec:app-sat-encoding}

As discussed in~\Cref{sec:optim-synth-local}, optimal synthesis of encoding circuits can be formulated as a symbolic reachability problem over stabilizer tableaus. The synthesis task is to find a sequence of Clifford gates that transforms the tableau of a target encoding isometry into the canonical tableau from~\Cref{sec:enc-synthesis}. Since stabilizer tableaus are binary matrices, this reachability problem can be encoded using Boolean variables for tableau entries and Boolean gate-selection variables for the transitions between successive tableaus.

The symbolic encoding follows the bounded model-checking viewpoint presented in the main text. For a fixed depth bound \(d_{\max}\), we introduce a symbolic tableau \(T^{(d)}\) for every layer \(d=0,\dots,d_{\max}\). A satisfying assignment to all tableau and gate variables then describes a depth-\(d_{\max}\) circuit candidate together with the symbolic evolution of the tableau under that circuit. In this appendix, we spell out the constraints
\[
\Phi_{\mathrm{init}},\qquad
\Phi_{\mathrm{cons}},\qquad
\Phi_{\mathrm{trans}},\qquad
\Phi_{\mathrm{goal}}
\]
for arbitrary stabilizer codes and for the CSS specialization.

\subsection{Arbitrary Stabilizer Codes}

Let
\[
T_{\mathrm{tar}}=
\begin{pmatrix}
X_{\mathrm{tar}}\\
Z_{\mathrm{tar}}\\
S_{\mathrm{tar}}
\end{pmatrix}
\in \F_2^{(n+k)\times 2n}
\]
be the tableau of the target encoding isometry, where \(X_{\mathrm{tar}}\in\F_2^{k\times 2n}\), \(Z_{\mathrm{tar}}\in\F_2^{k\times 2n}\), and \(S_{\mathrm{tar}}\in\F_2^{(n-k)\times 2n}\) denote the logical-\(X\), logical-\(Z\), and stabilizer row blocks, respectively. Let \(T_{\mathrm{can}}\) denote the canonical tableau from the synthesis condition in~\Cref{sec:enc-synthesis}.

For every depth \(0\le d\le d_{\max}\), we introduce a symbolic tableau
\[
T^{(d)}=
\begin{pmatrix}
X^{(d)}\\
Z^{(d)}\\
S^{(d)}
\end{pmatrix}
=
\bigl(T_X^{(d)} \mid T_Z^{(d)}\bigr)
\in \F_2^{(n+k)\times 2n}.
\]
Thus, \(X^{(d)},Z^{(d)},S^{(d)}\) refer to the row blocks of the tableau at depth \(d\), while \(T_X^{(d)}\) and \(T_Z^{(d)}\) denote its \(X\)- and \(Z\)-column blocks.

\paragraph*{\(\Phi_{\mathrm{init}}\): initial tableau.}
For every row \(r\in\{1,\dots,n+k\}\), qubit \(q\in Q:=\{1,\dots,n\}\), and depth \(0\le d\le d_{\max}\), we introduce Boolean variables
\[
x_{r,q}^d,\qquad z_{r,q}^d,
\]
which encode the \(X\)- and \(Z\)-parts of tableau entry \((r,q)\) at depth \(d\). The initial symbolic tableau is fixed to the target encoding isometry:
\[
\Phi_{\mathrm{init}}
\;:=\;
\bigl(T^{(0)} = T_{\mathrm{tar}}\bigr).
\]

\medskip

\paragraph*{Gate-selection variables.}
We explicitly encode the gate set
\[
\{I,H,S,\mathrm{CNOT}\}.
\]
For every qubit \(i\in Q\) and layer \(0\le d<d_{\max}\), we introduce Boolean variables
\[
\mathit{id}_i^d,\qquad h_i^d,\qquad s_i^d,
\]
indicating that qubit \(i\) is acted on by the identity, a Hadamard gate, or an \(S\) gate in layer \(d\), respectively. For every ordered pair of distinct qubits \(i,j\in Q\) and layer \(0\le d<d_{\max}\), we introduce a Boolean variable
\[
c_{i,j}^d,
\]
indicating that a CNOT with control \(i\) and target \(j\) is applied in layer \(d\).

This explicit encoding is sufficient to synthesize any Clifford isometry.
Other single- or two-qubit Clifford gates can be incorporated analogously by introducing the corresponding gate-selection variables and the associated tableau-update constraints.

\medskip

\paragraph*{\(\Phi_{\mathrm{cons}}\): consistency constraints.}
For every depth \(d\) and qubit \(i\), exactly one operation must be selected on that qubit. Here the identity counts as an operation, so qubits on which no nontrivial gate acts are represented explicitly. Thus, for every \(i\in Q\) and \(0\le d<d_{\max}\), we impose
\[
\mathit{id}_i^d + h_i^d + s_i^d
+ \sum_{\substack{j\in Q\\ j\neq i}} c_{i,j}^d
+ \sum_{\substack{j\in Q\\ j\neq i}} c_{j,i}^d
= 1.
\]
Since all variables are Boolean, this is an exactly-one constraint.
It ensures that each layer represents a valid set of parallel gate applications and, in particular, that no two gates act on the same qubit in the same layer.
We write the conjunction of all such constraints as \(\Phi_{\mathrm{cons}}\).

\medskip

\paragraph*{\(\Phi_{\mathrm{trans}}\): transition relation.}
The transition relation \(\Phi_{\mathrm{trans}}\) connects \(T^{(d)}\) and \(T^{(d+1)}\) by the tableau action of the selected gates.

If qubit \(i\) is assigned the identity in layer \(d\), then the corresponding tableau column remains unchanged:
\[
\mathit{id}_i^d \Longrightarrow
\bigwedge_{r=1}^{n+k}
\left(
x_{r,i}^{d+1} \Leftrightarrow x_{r,i}^{d}
\right)
\land
\left(
z_{r,i}^{d+1} \Leftrightarrow z_{r,i}^{d}
\right).
\]

A Hadamard gate on qubit \(i\) swaps the \(X\)- and \(Z\)-parts of that column:
\[
h_i^d \Longrightarrow
\bigwedge_{r=1}^{n+k}
\left(
x_{r,i}^{d+1} \Leftrightarrow z_{r,i}^{d}
\right)
\land
\left(
z_{r,i}^{d+1} \Leftrightarrow x_{r,i}^{d}
\right).
\]

An \(S\) gate on qubit \(i\) performs the update \(Z_i \leftarrow Z_i + X_i\):
\[
s_i^d \Longrightarrow
\bigwedge_{r=1}^{n+k}
\left(
x_{r,i}^{d+1} \Leftrightarrow x_{r,i}^{d}
\right)
\land
\left(
z_{r,i}^{d+1} \Leftrightarrow z_{r,i}^{d} \oplus x_{r,i}^{d}
\right).
\]

A CNOT with control \(i\) and target \(j\) performs the updates \(X_j \leftarrow X_j + X_i\) and \(Z_i \leftarrow Z_i + Z_j\):
\[
\begin{aligned}
c_{i,j}^d \Longrightarrow
\bigwedge_{r=1}^{n+k}\Bigl(
&\left(
x_{r,i}^{d+1} \Leftrightarrow x_{r,i}^{d}
\right)
\land
\left(
z_{r,i}^{d+1} \Leftrightarrow z_{r,i}^{d} \oplus z_{r,j}^{d}
\right)
\\
&\land
\left(
x_{r,j}^{d+1} \Leftrightarrow x_{r,j}^{d} \oplus x_{r,i}^{d}
\right)
\land
\left(
z_{r,j}^{d+1} \Leftrightarrow z_{r,j}^{d}
\right)
\Bigr).
\end{aligned}
\]

Together with the consistency constraints, the identity variables make all unchanged columns explicit: no tableau entry is left unchanged only implicitly. The full transition relation is the conjunction of all identity, Hadamard, \(S\), and CNOT update constraints over all layers.

\medskip

\paragraph*{\(\Phi_{\mathrm{goal}}\): reducibility to canonical form.}
The final tableau need not equal \(T_{\mathrm{can}}\) entrywise. Instead, \(\Phi_{\mathrm{goal}}\) characterizes all final tableaus that can be transformed into \(T_{\mathrm{can}}\) by stabilizer row operations together with a permutation of qubits.

Let
\[
T^{(d_{\max})}=
\begin{pmatrix}
X^{(d_{\max})}\\
Z^{(d_{\max})}\\
S^{(d_{\max})}
\end{pmatrix}
\]
be the symbolic tableau at the final depth, and write
\[
S^{(d_{\max})}
=
\bigl(S_X^{(d_{\max})} \mid S_Z^{(d_{\max})}\bigr)
\]
for the \(X\)- and \(Z\)-column blocks of its stabilizer row block.

We first require that the stabilizer rows are \(Z\)-type only:
\[
x_{2k+r,q}^{d_{\max}} = 0
\qquad
\text{for all } r\in\{1,\dots,n-k\},\ q\in Q.
\]

Let
\[
r_{\mathrm{stab}} := \mathrm{rank}(S_{\mathrm{tar}})
\]
denote the rank of the stabilizer block of the target tableau. Since Clifford gates induce invertible transformations on the tableau, this rank is preserved throughout the synthesis.

We now introduce Boolean pivot indicators
\[
p_q,\qquad q\in Q,
\]
where \(p_q=1\) indicates that column \(q\) of \(S_Z^{(d_{\max})}\) is nonzero. These variables are constrained by
\[
p_q \Longleftrightarrow \bigvee_{r=1}^{n-k} z_{2k+r,q}^{d_{\max}}
\qquad
\text{for all } q\in Q,
\]
together with
\[
\sum_{q\in Q} p_q = r_{\mathrm{stab}}.
\]
Hence, the final stabilizer block has exactly \(r_{\mathrm{stab}}\) nonzero columns. Since its rank is also \(r_{\mathrm{stab}}\), these nonzero columns are linearly independent and can therefore be reduced by stabilizer row operations to identity on the pivot columns.

Next, for every logical qubit \(i\in\{1,\dots,k\}\) and column \(q\in Q\), we introduce Boolean selector variables
\[
\lambda_{i,q},
\]
where \(\lambda_{i,q}=1\) indicates that column \(q\) is the distinguished non-pivot column carrying the canonical \(X/Z\) pair of logical qubit \(i\). These selector variables satisfy
\[
\sum_{q\in Q} \lambda_{i,q} = 1
\qquad
\text{for all } i\in\{1,\dots,k\},
\]
\[
\lambda_{i,q} \Longrightarrow \neg p_q
\qquad
\text{for all } i\in\{1,\dots,k\},\ q\in Q,
\]
and
\[
\sum_{i=1}^{k} \lambda_{i,q} \le 1
\qquad
\text{for all } q\in Q.
\]
Thus, each logical qubit selects exactly one non-pivot column, and distinct logical qubits select distinct columns.

If \(\lambda_{i,q}=1\), then column \(q\) carries the canonical logical \(X/Z\) pair for logical qubit \(i\):
\[
\lambda_{i,q} \Longrightarrow
x_{i,q}^{d_{\max}} = 1
\land
z_{i,q}^{d_{\max}} = 0
\land
x_{k+i,q}^{d_{\max}} = 0
\land
z_{k+i,q}^{d_{\max}} = 1.
\]

If \(q\) is a non-pivot column not selected for logical qubit \(i\), then both logical rows of qubit \(i\) vanish on that column:
If \(q\) is a non-pivot column not selected for logical qubit \(i\), then both logical rows of qubit \(i\) vanish on that column:
\[
\neg p_q \land \neg \lambda_{i,q}
\Longrightarrow
\begin{aligned}[t]
&x_{i,q}^{d_{\max}} = 0
\land
z_{i,q}^{d_{\max}} = 0
\\
&\land
x_{k+i,q}^{d_{\max}} = 0
\land
z_{k+i,q}^{d_{\max}} = 0.
\end{aligned}
\]

Finally, on pivot columns only \(Z\)-support is allowed in the logical rows:
\[
\begin{aligned}
p_q \Longrightarrow\;&
x_{i,q}^{d_{\max}} = 0
\land
x_{k+i,q}^{d_{\max}} = 0,
\\
&\text{for all } i\in\{1,\dots,k\},\ q\in Q.
\end{aligned}
\]
Any remaining \(Z\)-support of the logical rows on pivot columns can then be removed by adding stabilizer rows once the stabilizer block has been reduced to identity on those pivot columns.
Collecting these constraints yields \(\Phi_{\mathrm{goal}}\).

\medskip

\paragraph*{Full encoding.}
Putting everything together, we obtain the SMT formula
\[
\Phi
=
\Phi_{\mathrm{init}}
\land
\Phi_{\mathrm{cons}}
\land
\Phi_{\mathrm{trans}}
\land
\Phi_{\mathrm{goal}}.
\]
The formula \(\Phi\) is satisfiable if and only if there exists a circuit of depth at most \(d_{\max}\) that reduces \(T_{\mathrm{tar}}\) to a tableau equivalent to \(T_{\mathrm{can}}\) under the allowed row operations.
A satisfying assignment, therefore, determines a reduction circuit from \(T_{\mathrm{tar}}\) to \(T_{\mathrm{can}}\), and reversing this circuit yields an encoding circuit for the target isometry.

To obtain a depth-optimal circuit, one solves \(\Phi\) for increasing values of \(d_{\max}\) until the first satisfiable instance is found.

\subsection{CSS Codes}

For CSS codes, the encoding simplifies considerably. As shown in the main text, it is sufficient to track only one of the two check matrices together with the corresponding logical operators. We use the \(X\)-side representation
\[
M_X=
\begin{pmatrix}
L_X\\
H_X
\end{pmatrix}
\in \F_2^{n \times n},
\]
where the first \(k\) rows correspond to the logical \(X\) operators and the remaining \(n-k\) rows correspond to the rows of the \(X\)-check matrix \(H_X\).

For every depth \(0\le d\le d_{\max}\), we introduce a symbolic matrix
\[
M_X^{(d)} \in \F_2^{n\times n}.
\]
For every row \(r\in\{1,\dots,n\}\), column \(q\in Q\), and depth \(0\le d\le d_{\max}\), let \(m_{r,q}^d\) denote the Boolean variable encoding the \((r,q)\)-entry of \(M^{(d)}\).

Since only CNOT circuits are needed in the CSS case, the gate set reduces to
\[
\{I,\mathrm{CNOT}\}.
\]
Accordingly, for every qubit \(i\in Q\) and layer \mbox{\(0\le d<d_{\max}\)}, we introduce an identity gate variable
\[
\mathit{id}_i^d,
\]
and for every ordered pair of distinct qubits \(i,j\in Q\),
\[
c_{i,j}^d,
\]
where $c_{i,j}^d$ indicates whether a CNOT with control $i$ and target $j$ is applied in layer $d$. This procedure eliminates all gate variables for single-qubit Clifford gates and, since only one check matrix has to be tracked, also considerably reduces the number of symbolic matrix variables.

The consistency constraints become
\[
\mathit{id}_i^d
+
\sum_{\substack{j\in Q\\ j\neq i}} c_{i,j}^d
+
\sum_{\substack{j\in Q\\ j\neq i}} c_{j,i}^d
=1
\quad
\text{for all } i\in Q,\ 0\le d<d_{\max}.
\]
Thus, every qubit either participates in exactly one CNOT or is acted on by the identity in that layer.

The transition relation is the corresponding specialization of the general case. An identity on qubit \(i\) leaves column \(i\) unchanged, while a CNOT \(c_{i,j}^d\) adds column \(i\) into column \(j\) over \(\F_2\):
\[
c_{i,j}^d \Longrightarrow
\bigwedge_{r=1}^{n}
\left(
m_{r,i}^{d+1} \Leftrightarrow m_{r,i}^{d}
\right)
\land
\left(
m_{r,j}^{d+1} \Leftrightarrow m_{r,j}^{d} \oplus m_{r,i}^{d}
\right).
\]
As before, the identity variables ensure that qubits not acted on by a nontrivial gate are updated explicitly.

The goal constraint is the corresponding CSS specialization of the general case. Let
\[
r_X := \mathrm{rank}(H_X).
\]
Then the final symbolic matrix must have exactly \(r_X\) nonzero columns in the part corresponding to the rows of \(H_X\). These columns serve as pivot columns and can be reduced by row operations to identity. Each logical \(X\) row must then have exactly one nonzero entry on a distinct non-pivot column, while any remaining support may only occur on pivot columns. Equivalently, the final symbolic matrix is reducible to the canonical CSS form from~\Cref{eq:enc_circ_matrix}.

Overall, the CSS encoding is substantially smaller than the general encoding: it tracks only one \(n\times n\) matrix instead of a full stabilizer tableau and requires only \(\mathit{id}\)- and CNOT-selection variables.

\subsection{Gate-Optimal Synthesis}

The encoding above is formulated for depth-optimal synthesis, since each symbolic transition corresponds to one layer of parallel gate applications. Gate-optimal synthesis can be handled in two ways.

First, one may keep the layer-based encoding and add pseudo-Boolean constraints on the selected gate variables.
To constrain the number of two-qubit gates to be at most $g_{\max}$, we enforce the constraint
\[
\sum_{d=0}^{d_{\max}-1}
\sum_{\substack{i,j\in Q\\ i\neq j}}
c_{i,j}^d
\le g_{\max}.
\]
If one wishes to minimize the total number of gates rather than only the number of two-qubit gates, the single-qubit variables \(h_i^d\) and \(s_i^d\) can be counted analogously, while the identity variables are excluded. Solving the encoding for decreasing values of the corresponding bound yields a minimum-gate circuit among all circuits of depth at most \(d_{\max}\). This gives a natural lexicographic optimization: one first determines the optimal depth and then minimizes the gate count within that depth bound.

Second, one may replace the layer-based time-like encoding by a gate-based encoding in which each symbolic transition corresponds to a single gate application rather than an entire circuit layer. In that case, the time horizon directly equals the gate budget. This is the approach used for gate-optimal CSS state-preparation circuits in Ref.~\cite{peham2024automatedsynthesisfaulttolerantstate}: consecutive symbolic matrices are related by a single column addition, so that every step contains at most one CNOT.
The same idea carries over directly to general encoding isometries by replacing symbolic check matrices with symbolic stabilizer tableaus and by using single-gate tableau updates instead of layer updates.

\begin{table*}[t]
\centering
\caption{Depth-optimized state preparation circuits for different rollout levels. $G$ denotes two-qubit gate count, $D$ circuit depth, and $\Delta G$ / $\Delta D$ denote the relative improvement (\%) with respect to the greedy baseline.}
\label{tab:depth-stateprep-rollout-summary}
\providecommand{\RolloutCell}[1]{\hspace{7pt}#1\hspace{7pt}}
\providecommand{\RolloutHdr}[1]{\multicolumn{1}{c}{#1}}
\setlength{\tabcolsep}{0pt}
\renewcommand{\arraystretch}{1.1}
\begin{tabular}{llrrrcrrrrrcrrrr}
\toprule
& & \multicolumn{2}{c}{\textbf{Greedy}} & \multicolumn{6}{c}{\textbf{Rollout 1}} & \multicolumn{6}{c}{\textbf{Rollout 2}} \\
\cmidrule(r{0.4em}){3-4}
\cmidrule(lr{0.4em}){5-10}
\cmidrule(l{0.4em}){11-16}
Code & State & \RolloutHdr{$G$} & \RolloutHdr{$D$} & \RolloutHdr{\hspace{2pt}$t$\hspace{6pt}} & \RolloutHdr{\hspace{6pt}ET\hspace{2pt}} & \RolloutHdr{$G$} & \RolloutHdr{$\Delta G$} & \RolloutHdr{$D$} & \RolloutHdr{$\Delta D$} & \RolloutHdr{\hspace{2pt}$(t_1,t_2)$\hspace{6pt}} & \RolloutHdr{\hspace{6pt}ET\hspace{2pt}} & \RolloutHdr{$G$} & \RolloutHdr{$\Delta G$} & \RolloutHdr{$D$} & \RolloutHdr{$\Delta D$} \\
\midrule
\rowcolor{black!6}
\RolloutCell{$[[8,3,3]]$} & \RolloutCell{$\ket{+}_L^{\otimes k}$} & \RolloutCell{10} & \RolloutCell{\textbf{5}} & \RolloutCell{10} & \RolloutCell{y} & \RolloutCell{\textbf{9}} & \RolloutCell{10.0} & \RolloutCell{\textbf{5}} & \RolloutCell{0.0} & \RolloutCell{10,2} & \RolloutCell{y} & \RolloutCell{\textbf{9}} & \RolloutCell{10.0} & \RolloutCell{\textbf{5}} & \RolloutCell{0.0} \\
\rowcolor{black!6}
\RolloutCell{$[[8,3,3]]$} & \RolloutCell{$\ket{0}_L^{\otimes k}$} & \RolloutCell{15} & \RolloutCell{6} & \RolloutCell{10} & \RolloutCell{y} & \RolloutCell{\textbf{13}} & \RolloutCell{13.3} & \RolloutCell{\textbf{5}} & \RolloutCell{16.7} & \RolloutCell{10,2} & \RolloutCell{y} & \RolloutCell{\textbf{13}} & \RolloutCell{13.3} & \RolloutCell{\textbf{5}} & \RolloutCell{16.7} \\
\RolloutCell{$[[15,1,3]]$} & \RolloutCell{$\ket{+}_L^{\otimes k}$} & \RolloutCell{\textbf{23}} & \RolloutCell{5} & \RolloutCell{10} & \RolloutCell{y} & \RolloutCell{24} & \RolloutCell{-4.3} & \RolloutCell{\textbf{4}} & \RolloutCell{20.0} & \RolloutCell{10,2} & \RolloutCell{n} & \RolloutCell{\textbf{23}} & \RolloutCell{0.0} & \RolloutCell{\textbf{4}} & \RolloutCell{20.0} \\
\RolloutCell{$[[15,1,3]]$} & \RolloutCell{$\ket{0}_L^{\otimes k}$} & \RolloutCell{\textbf{22}} & \RolloutCell{\textbf{4}} & \RolloutCell{10} & \RolloutCell{y} & \RolloutCell{\textbf{22}} & \RolloutCell{0.0} & \RolloutCell{\textbf{4}} & \RolloutCell{0.0} & \RolloutCell{10,2} & \RolloutCell{y} & \RolloutCell{\textbf{22}} & \RolloutCell{0.0} & \RolloutCell{\textbf{4}} & \RolloutCell{0.0} \\
\rowcolor{black!6}
\RolloutCell{$[[15,3,5]]$} & \RolloutCell{$\ket{+}_L^{\otimes k}$} & \RolloutCell{48} & \RolloutCell{11} & \RolloutCell{20} & \RolloutCell{n} & \RolloutCell{\textbf{41}} & \RolloutCell{14.6} & \RolloutCell{\textbf{9}} & \RolloutCell{18.2} & \RolloutCell{20,4} & \RolloutCell{y} & \RolloutCell{\textbf{41}} & \RolloutCell{14.6} & \RolloutCell{\textbf{9}} & \RolloutCell{18.2} \\
\rowcolor{black!6}
\RolloutCell{$[[15,3,5]]$} & \RolloutCell{$\ket{0}_L^{\otimes k}$} & \RolloutCell{57} & \RolloutCell{16} & \RolloutCell{100} & \RolloutCell{y} & \RolloutCell{45} & \RolloutCell{21.1} & \RolloutCell{\textbf{10}} & \RolloutCell{37.5} & \RolloutCell{30,6} & \RolloutCell{y} & \RolloutCell{\textbf{43}} & \RolloutCell{24.6} & \RolloutCell{11} & \RolloutCell{31.2} \\
\RolloutCell{$[[15,7,3]]$} & \RolloutCell{$\ket{+}_L^{\otimes k}$} & \RolloutCell{\textbf{22}} & \RolloutCell{\textbf{4}} & \RolloutCell{10} & \RolloutCell{y} & \RolloutCell{\textbf{22}} & \RolloutCell{0.0} & \RolloutCell{\textbf{4}} & \RolloutCell{0.0} & \RolloutCell{10,2} & \RolloutCell{y} & \RolloutCell{\textbf{22}} & \RolloutCell{0.0} & \RolloutCell{\textbf{4}} & \RolloutCell{0.0} \\
\RolloutCell{$[[15,7,3]]$} & \RolloutCell{$\ket{0}_L^{\otimes k}$} & \RolloutCell{\textbf{22}} & \RolloutCell{\textbf{4}} & \RolloutCell{10} & \RolloutCell{y} & \RolloutCell{\textbf{22}} & \RolloutCell{0.0} & \RolloutCell{\textbf{4}} & \RolloutCell{0.0} & \RolloutCell{10,2} & \RolloutCell{y} & \RolloutCell{\textbf{22}} & \RolloutCell{0.0} & \RolloutCell{\textbf{4}} & \RolloutCell{0.0} \\
\rowcolor{black!6}
\RolloutCell{$[[17,1,5]]$} & \RolloutCell{$\ket{+}_L^{\otimes k}$} & \RolloutCell{\textbf{23}} & \RolloutCell{\textbf{4}} & \RolloutCell{10} & \RolloutCell{y} & \RolloutCell{\textbf{23}} & \RolloutCell{0.0} & \RolloutCell{\textbf{4}} & \RolloutCell{0.0} & \RolloutCell{10,2} & \RolloutCell{y} & \RolloutCell{\textbf{23}} & \RolloutCell{0.0} & \RolloutCell{\textbf{4}} & \RolloutCell{0.0} \\
\rowcolor{black!6}
\RolloutCell{$[[17,1,5]]$} & \RolloutCell{$\ket{0}_L^{\otimes k}$} & \RolloutCell{\textbf{23}} & \RolloutCell{\textbf{4}} & \RolloutCell{10} & \RolloutCell{y} & \RolloutCell{\textbf{23}} & \RolloutCell{0.0} & \RolloutCell{\textbf{4}} & \RolloutCell{0.0} & \RolloutCell{10,2} & \RolloutCell{y} & \RolloutCell{\textbf{23}} & \RolloutCell{0.0} & \RolloutCell{\textbf{4}} & \RolloutCell{0.0} \\
\RolloutCell{$[[19,1,5]]$} & \RolloutCell{$\ket{+}_L^{\otimes k}$} & \RolloutCell{\textbf{27}} & \RolloutCell{5} & \RolloutCell{10} & \RolloutCell{y} & \RolloutCell{\textbf{27}} & \RolloutCell{0.0} & \RolloutCell{\textbf{4}} & \RolloutCell{20.0} & \RolloutCell{10,2} & \RolloutCell{y} & \RolloutCell{\textbf{27}} & \RolloutCell{0.0} & \RolloutCell{\textbf{4}} & \RolloutCell{20.0} \\
\RolloutCell{$[[19,1,5]]$} & \RolloutCell{$\ket{0}_L^{\otimes k}$} & \RolloutCell{\textbf{27}} & \RolloutCell{5} & \RolloutCell{10} & \RolloutCell{y} & \RolloutCell{\textbf{27}} & \RolloutCell{0.0} & \RolloutCell{\textbf{4}} & \RolloutCell{20.0} & \RolloutCell{10,2} & \RolloutCell{y} & \RolloutCell{\textbf{27}} & \RolloutCell{0.0} & \RolloutCell{\textbf{4}} & \RolloutCell{20.0} \\
\rowcolor{black!6}
\RolloutCell{$[[23,1,7]]$} & \RolloutCell{$\ket{+}_L^{\otimes k}$} & \RolloutCell{63} & \RolloutCell{9} & \RolloutCell{20} & \RolloutCell{n} & \RolloutCell{\textbf{54}} & \RolloutCell{14.3} & \RolloutCell{8} & \RolloutCell{11.1} & \RolloutCell{50,10} & \RolloutCell{n} & \RolloutCell{57} & \RolloutCell{9.5} & \RolloutCell{\textbf{7}} & \RolloutCell{22.2} \\
\rowcolor{black!6}
\RolloutCell{$[[23,1,7]]$} & \RolloutCell{$\ket{0}_L^{\otimes k}$} & \RolloutCell{63} & \RolloutCell{9} & \RolloutCell{20} & \RolloutCell{n} & \RolloutCell{\textbf{54}} & \RolloutCell{14.3} & \RolloutCell{8} & \RolloutCell{11.1} & \RolloutCell{50,10} & \RolloutCell{n} & \RolloutCell{57} & \RolloutCell{9.5} & \RolloutCell{\textbf{7}} & \RolloutCell{22.2} \\
\RolloutCell{$[[30,6,5]]$} & \RolloutCell{$\ket{+}_L^{\otimes k}$} & \RolloutCell{78} & \RolloutCell{8} & \RolloutCell{100} & \RolloutCell{n} & \RolloutCell{\textbf{68}} & \RolloutCell{12.8} & \RolloutCell{\textbf{7}} & \RolloutCell{12.5} & \RolloutCell{10,5} & \RolloutCell{y} & \RolloutCell{69} & \RolloutCell{11.5} & \RolloutCell{\textbf{7}} & \RolloutCell{12.5} \\
\RolloutCell{$[[30,6,5]]$} & \RolloutCell{$\ket{0}_L^{\otimes k}$} & \RolloutCell{77} & \RolloutCell{8} & \RolloutCell{200} & \RolloutCell{n} & \RolloutCell{\textbf{68}} & \RolloutCell{11.7} & \RolloutCell{\textbf{7}} & \RolloutCell{12.5} & \RolloutCell{20,10} & \RolloutCell{y} & \RolloutCell{\textbf{68}} & \RolloutCell{11.7} & \RolloutCell{\textbf{7}} & \RolloutCell{12.5} \\
\rowcolor{black!6}
\RolloutCell{$[[31,1,7]]$} & \RolloutCell{$\ket{+}_L^{\otimes k}$} & \RolloutCell{\textbf{46}} & \RolloutCell{5} & \RolloutCell{10} & \RolloutCell{y} & \RolloutCell{\textbf{46}} & \RolloutCell{0.0} & \RolloutCell{5} & \RolloutCell{0.0} & \RolloutCell{50,25} & \RolloutCell{n} & \RolloutCell{\textbf{46}} & \RolloutCell{0.0} & \RolloutCell{\textbf{4}} & \RolloutCell{20.0} \\
\rowcolor{black!6}
\RolloutCell{$[[31,1,7]]$} & \RolloutCell{$\ket{0}_L^{\otimes k}$} & \RolloutCell{\textbf{46}} & \RolloutCell{5} & \RolloutCell{10} & \RolloutCell{y} & \RolloutCell{\textbf{46}} & \RolloutCell{0.0} & \RolloutCell{5} & \RolloutCell{0.0} & \RolloutCell{50,25} & \RolloutCell{n} & \RolloutCell{\textbf{46}} & \RolloutCell{0.0} & \RolloutCell{\textbf{4}} & \RolloutCell{20.0} \\
\RolloutCell{$[[31,21,3]]$} & \RolloutCell{$\ket{+}_L^{\otimes k}$} & \RolloutCell{54} & \RolloutCell{6} & \RolloutCell{10} & \RolloutCell{y} & \RolloutCell{\textbf{52}} & \RolloutCell{3.7} & \RolloutCell{\textbf{5}} & \RolloutCell{16.7} & \RolloutCell{10,2} & \RolloutCell{y} & \RolloutCell{\textbf{52}} & \RolloutCell{3.7} & \RolloutCell{\textbf{5}} & \RolloutCell{16.7} \\
\RolloutCell{$[[31,21,3]]$} & \RolloutCell{$\ket{0}_L^{\otimes k}$} & \RolloutCell{54} & \RolloutCell{6} & \RolloutCell{10} & \RolloutCell{y} & \RolloutCell{\textbf{52}} & \RolloutCell{3.7} & \RolloutCell{\textbf{5}} & \RolloutCell{16.7} & \RolloutCell{10,2} & \RolloutCell{y} & \RolloutCell{\textbf{52}} & \RolloutCell{3.7} & \RolloutCell{\textbf{5}} & \RolloutCell{16.7} \\
\rowcolor{black!6}
\RolloutCell{$[[37,1,7]]$} & \RolloutCell{$\ket{+}_L^{\otimes k}$} & \RolloutCell{60} & \RolloutCell{\textbf{5}} & \RolloutCell{10} & \RolloutCell{y} & \RolloutCell{60} & \RolloutCell{0.0} & \RolloutCell{\textbf{5}} & \RolloutCell{0.0} & \RolloutCell{10,2} & \RolloutCell{n} & \RolloutCell{\textbf{59}} & \RolloutCell{1.7} & \RolloutCell{\textbf{5}} & \RolloutCell{0.0} \\
\rowcolor{black!6}
\RolloutCell{$[[37,1,7]]$} & \RolloutCell{$\ket{0}_L^{\otimes k}$} & \RolloutCell{60} & \RolloutCell{\textbf{5}} & \RolloutCell{10} & \RolloutCell{y} & \RolloutCell{60} & \RolloutCell{0.0} & \RolloutCell{\textbf{5}} & \RolloutCell{0.0} & \RolloutCell{10,2} & \RolloutCell{n} & \RolloutCell{\textbf{59}} & \RolloutCell{1.7} & \RolloutCell{\textbf{5}} & \RolloutCell{0.0} \\
\RolloutCell{$[[72,12,6]]$} & \RolloutCell{$\ket{+}_L^{\otimes k}$} & \RolloutCell{\textbf{150}} & \RolloutCell{11} & \RolloutCell{100} & \RolloutCell{n} & \RolloutCell{214} & \RolloutCell{-42.7} & \RolloutCell{\textbf{9}} & \RolloutCell{18.2} & \RolloutCell{10,2} & \RolloutCell{y} & \RolloutCell{\textbf{150}} & \RolloutCell{0.0} & \RolloutCell{10} & \RolloutCell{9.1} \\
\RolloutCell{$[[72,12,6]]$} & \RolloutCell{$\ket{0}_L^{\otimes k}$} & \RolloutCell{\textbf{150}} & \RolloutCell{11} & \RolloutCell{100} & \RolloutCell{n} & \RolloutCell{223} & \RolloutCell{-48.7} & \RolloutCell{\textbf{9}} & \RolloutCell{18.2} & \RolloutCell{10,2} & \RolloutCell{y} & \RolloutCell{\textbf{150}} & \RolloutCell{0.0} & \RolloutCell{10} & \RolloutCell{9.1} \\
\rowcolor{black!6}
\RolloutCell{$[[90,8,10]]$} & \RolloutCell{$\ket{+}_L^{\otimes k}$} & \RolloutCell{150} & \RolloutCell{10} & \RolloutCell{20} & \RolloutCell{n} & \RolloutCell{\textbf{147}} & \RolloutCell{2.0} & \RolloutCell{\textbf{8}} & \RolloutCell{20.0} & \RolloutCell{10,5} & \RolloutCell{n} & \RolloutCell{148} & \RolloutCell{1.3} & \RolloutCell{\textbf{8}} & \RolloutCell{20.0} \\
\rowcolor{black!6}
\RolloutCell{$[[90,8,10]]$} & \RolloutCell{$\ket{0}_L^{\otimes k}$} & \RolloutCell{151} & \RolloutCell{10} & \RolloutCell{40} & \RolloutCell{n} & \RolloutCell{\textbf{147}} & \RolloutCell{2.6} & \RolloutCell{\textbf{8}} & \RolloutCell{20.0} & \RolloutCell{10,2} & \RolloutCell{n} & \RolloutCell{148} & \RolloutCell{2.0} & \RolloutCell{\textbf{8}} & \RolloutCell{20.0} \\
\RolloutCell{$[[108,8,10]]$} & \RolloutCell{$\ket{+}_L^{\otimes k}$} & \RolloutCell{508} & \RolloutCell{21} & \RolloutCell{200} & \RolloutCell{n} & \RolloutCell{\textbf{460}} & \RolloutCell{9.4} & \RolloutCell{\textbf{14}} & \RolloutCell{33.3} & \RolloutCell{10,2} & \RolloutCell{y} & \RolloutCell{508} & \RolloutCell{0.0} & \RolloutCell{20} & \RolloutCell{4.8} \\
\RolloutCell{$[[108,8,10]]$} & \RolloutCell{$\ket{0}_L^{\otimes k}$} & \RolloutCell{584} & \RolloutCell{24} & \RolloutCell{100} & \RolloutCell{n} & \RolloutCell{\textbf{548}} & \RolloutCell{6.2} & \RolloutCell{\textbf{15}} & \RolloutCell{37.5} & \RolloutCell{20,4} & \RolloutCell{y} & \RolloutCell{584} & \RolloutCell{0.0} & \RolloutCell{23} & \RolloutCell{4.2} \\
\rowcolor{black!6}
\RolloutCell{$[[144,12,12]]$} & \RolloutCell{$\ket{+}_L^{\otimes k}$} & \RolloutCell{\textbf{330}} & \RolloutCell{17} & \RolloutCell{30} & \RolloutCell{y} & \RolloutCell{\textbf{330}} & \RolloutCell{0.0} & \RolloutCell{\textbf{16}} & \RolloutCell{5.9} & \RolloutCell{30,6} & \RolloutCell{y} & \RolloutCell{\textbf{330}} & \RolloutCell{0.0} & \RolloutCell{\textbf{16}} & \RolloutCell{5.9} \\
\rowcolor{black!6}
\RolloutCell{$[[144,12,12]]$} & \RolloutCell{$\ket{0}_L^{\otimes k}$} & \RolloutCell{\textbf{330}} & \RolloutCell{17} & \RolloutCell{100} & \RolloutCell{n} & \RolloutCell{687} & \RolloutCell{-108.2} & \RolloutCell{\textbf{15}} & \RolloutCell{11.8} & \RolloutCell{10,2} & \RolloutCell{y} & \RolloutCell{\textbf{330}} & \RolloutCell{0.0} & \RolloutCell{17} & \RolloutCell{0.0} \\
\bottomrule
\end{tabular}

\end{table*}%

\section{Proof that~\Cref{eq:enc_circ_matrix} implies~\Cref{eq:enc_circ_matrix_z}}
\label{sec:proofs}

\begin{proof}
Let
\[
M_X :=
\begin{pmatrix}
L_X\\ H_X
\end{pmatrix},
\qquad
M_Z :=
\begin{pmatrix}
L_Z\\ H_Z
\end{pmatrix}.
\]
By assumption,
\[
\widetilde M_X
:=
M_X C^{-1}
=
\begin{pmatrix}
B & \mathbf{0} & I_k\\
A & \mathbf{0} & \mathbf{0}
\end{pmatrix},
\]
where \(A\) has full rank. Now write
\[
\widetilde M_Z
:=
M_Z C^{-1}
=
\begin{pmatrix}
F & G & H\\
J & K & L
\end{pmatrix}.
\]

Since the CSS commutation relations are preserved under applications of CNOT circuits, we have
\[
\widetilde M_X \widetilde M_Z^\top
=
\begin{pmatrix}
I_k & \mathbf{0}\\
\mathbf{0} & \mathbf{0}
\end{pmatrix}.
\]
Expanding this product gives
\[
\begin{aligned}
&\begin{pmatrix}
B & \mathbf{0} & I_k\\
A & \mathbf{0} & \mathbf{0}
\end{pmatrix}
\begin{pmatrix}
F^\top & J^\top\\
G^\top & K^\top\\
H^\top & L^\top
\end{pmatrix}
\\
&=
\begin{pmatrix}
BF^\top + H^\top & BJ^\top + L^\top\\
AF^\top & AJ^\top
\end{pmatrix}
=
\begin{pmatrix}
I_k & \mathbf{0}\\
\mathbf{0} & \mathbf{0}
\end{pmatrix}.
\end{aligned}
\]
Hence
\[
AF^\top = \mathbf{0},
\qquad
AJ^\top = \mathbf{0}.
\]
Since \(A\) is invertible, it follows that
\[
F=\mathbf{0},
\qquad
J=\mathbf{0}.
\]
Substituting this back into the upper block equations yields
\[
H=I_k,
\qquad
L=\mathbf{0}.
\]
Therefore
\[
M_Z C^{-1}
=
\begin{pmatrix}
\mathbf{0} & G & I_k\\
\mathbf{0} & K & \mathbf{0}
\end{pmatrix}.
\]

It remains to show that \(K\) has full rank. Since \(C\) is invertible, \(M_Z C^{-1}\) has the same row rank as \(M_Z\), namely \(k+m_Z\). In the displayed block form above, the top \(k\) rows already contribute rank \(k\) because of the identity block in the last \(k\) columns. Therefore the lower block \(K\) must contribute rank \(m_Z\), i.e. \(K\) has full rank.
\end{proof}

\section{Further Results on the Impact of Rollout on State Preparation Circuit Synthesis}
\label{sec:results-state-prep}

This section gives more details on the circuit synthesis for Pauli eigenstate preparation circuits using the rollout synthesis proposed in~\Cref{sec:rollout-synthesis}.
As for general state encoding circuits, we investigated the impact of the rollout on the search.
The resulting circuit metrics and improvements can be seen in~\Cref{tab:depth-stateprep-rollout-summary} for depth-optimized state preparation circuits and \Cref{tab:gates-stateprep-rollout-summary} for gate-optimized circuits.

Unlike the results for encoding circuit synthesis presented in~\Cref{tab:gates-rollout-summary,tab:depth-rollout-summary}, the simple greedy synthesis often manages to find a circuit requiring the same number of two-qubit gates or depth as the ones synthesized using rollout.
However, in these cases, the rollout synthesis is often able to substantially improve the secondary objective as well.
Generally, it seems that rollout has an even more pronounced impact on the secondary optimization objective for state preparation circuit synthesis.

\begin{table*}[t]
\centering
\caption{Gate-optimized state preparation circuits for different rollout levels. $G$ denotes two-qubit gate count, $D$ circuit depth, and $\Delta G$ / $\Delta D$ denote the relative improvement (\%) with respect to the greedy baseline.}
\label{tab:gates-stateprep-rollout-summary}
\providecommand{\RolloutCell}[1]{\hspace{7pt}#1\hspace{7pt}}
\providecommand{\RolloutHdr}[1]{\multicolumn{1}{c}{#1}}
\setlength{\tabcolsep}{0pt}
\renewcommand{\arraystretch}{1.1}
\begin{tabular}{llrrrcrrrrrcrrrr}
\toprule
& & \multicolumn{2}{c}{\textbf{Greedy}} & \multicolumn{6}{c}{\textbf{Rollout 1}} & \multicolumn{6}{c}{\textbf{Rollout 2}} \\
\cmidrule(r{0.4em}){3-4}
\cmidrule(lr{0.4em}){5-10}
\cmidrule(l{0.4em}){11-16}
Code & State & \RolloutHdr{$G$} & \RolloutHdr{$D$} & \RolloutHdr{\hspace{2pt}$t$\hspace{6pt}} & \RolloutHdr{\hspace{6pt}ET\hspace{2pt}} & \RolloutHdr{$G$} & \RolloutHdr{$\Delta G$} & \RolloutHdr{$D$} & \RolloutHdr{$\Delta D$} & \RolloutHdr{\hspace{2pt}$(t_1,t_2)$\hspace{6pt}} & \RolloutHdr{\hspace{6pt}ET\hspace{2pt}} & \RolloutHdr{$G$} & \RolloutHdr{$\Delta G$} & \RolloutHdr{$D$} & \RolloutHdr{$\Delta D$} \\
\midrule
\rowcolor{black!6}
\RolloutCell{$[[8,3,3]]$} & \RolloutCell{$\ket{+}_L^{\otimes k}$} & \RolloutCell{\textbf{9}} & \RolloutCell{6} & \RolloutCell{10} & \RolloutCell{y} & \RolloutCell{\textbf{9}} & \RolloutCell{0.0} & \RolloutCell{\textbf{5}} & \RolloutCell{16.7} & \RolloutCell{10,2} & \RolloutCell{y} & \RolloutCell{\textbf{9}} & \RolloutCell{0.0} & \RolloutCell{\textbf{5}} & \RolloutCell{16.7} \\
\rowcolor{black!6}
\RolloutCell{$[[8,3,3]]$} & \RolloutCell{$\ket{0}_L^{\otimes k}$} & \RolloutCell{13} & \RolloutCell{11} & \RolloutCell{10} & \RolloutCell{n} & \RolloutCell{13} & \RolloutCell{0.0} & \RolloutCell{\textbf{6}} & \RolloutCell{45.5} & \RolloutCell{20,4} & \RolloutCell{n} & \RolloutCell{\textbf{11}} & \RolloutCell{15.4} & \RolloutCell{8} & \RolloutCell{27.3} \\
\RolloutCell{$[[15,1,3]]$} & \RolloutCell{$\ket{+}_L^{\otimes k}$} & \RolloutCell{\textbf{23}} & \RolloutCell{8} & \RolloutCell{10} & \RolloutCell{n} & \RolloutCell{\textbf{23}} & \RolloutCell{0.0} & \RolloutCell{7} & \RolloutCell{12.5} & \RolloutCell{30,6} & \RolloutCell{n} & \RolloutCell{\textbf{23}} & \RolloutCell{0.0} & \RolloutCell{\textbf{5}} & \RolloutCell{37.5} \\
\RolloutCell{$[[15,1,3]]$} & \RolloutCell{$\ket{0}_L^{\otimes k}$} & \RolloutCell{\textbf{22}} & \RolloutCell{11} & \RolloutCell{20} & \RolloutCell{n} & \RolloutCell{\textbf{22}} & \RolloutCell{0.0} & \RolloutCell{6} & \RolloutCell{45.5} & \RolloutCell{10,2} & \RolloutCell{n} & \RolloutCell{\textbf{22}} & \RolloutCell{0.0} & \RolloutCell{\textbf{5}} & \RolloutCell{54.5} \\
\rowcolor{black!6}
\RolloutCell{$[[15,3,5]]$} & \RolloutCell{$\ket{+}_L^{\otimes k}$} & \RolloutCell{47} & \RolloutCell{21} & \RolloutCell{50} & \RolloutCell{n} & \RolloutCell{34} & \RolloutCell{27.7} & \RolloutCell{13} & \RolloutCell{38.1} & \RolloutCell{30,15} & \RolloutCell{n} & \RolloutCell{\textbf{31}} & \RolloutCell{34.0} & \RolloutCell{\textbf{12}} & \RolloutCell{42.9} \\
\rowcolor{black!6}
\RolloutCell{$[[15,3,5]]$} & \RolloutCell{$\ket{0}_L^{\otimes k}$} & \RolloutCell{54} & \RolloutCell{31} & \RolloutCell{100} & \RolloutCell{n} & \RolloutCell{39} & \RolloutCell{27.8} & \RolloutCell{\textbf{12}} & \RolloutCell{61.3} & \RolloutCell{20,10} & \RolloutCell{n} & \RolloutCell{\textbf{36}} & \RolloutCell{33.3} & \RolloutCell{13} & \RolloutCell{58.1} \\
\RolloutCell{$[[15,7,3]]$} & \RolloutCell{$\ket{+}_L^{\otimes k}$} & \RolloutCell{\textbf{22}} & \RolloutCell{8} & \RolloutCell{10} & \RolloutCell{n} & \RolloutCell{\textbf{22}} & \RolloutCell{0.0} & \RolloutCell{6} & \RolloutCell{25.0} & \RolloutCell{10,2} & \RolloutCell{n} & \RolloutCell{\textbf{22}} & \RolloutCell{0.0} & \RolloutCell{\textbf{5}} & \RolloutCell{37.5} \\
\RolloutCell{$[[15,7,3]]$} & \RolloutCell{$\ket{0}_L^{\otimes k}$} & \RolloutCell{\textbf{22}} & \RolloutCell{8} & \RolloutCell{10} & \RolloutCell{n} & \RolloutCell{\textbf{22}} & \RolloutCell{0.0} & \RolloutCell{6} & \RolloutCell{25.0} & \RolloutCell{10,2} & \RolloutCell{n} & \RolloutCell{\textbf{22}} & \RolloutCell{0.0} & \RolloutCell{\textbf{5}} & \RolloutCell{37.5} \\
\rowcolor{black!6}
\RolloutCell{$[[17,1,5]]$} & \RolloutCell{$\ket{+}_L^{\otimes k}$} & \RolloutCell{\textbf{23}} & \RolloutCell{7} & \RolloutCell{20} & \RolloutCell{n} & \RolloutCell{\textbf{23}} & \RolloutCell{0.0} & \RolloutCell{5} & \RolloutCell{28.6} & \RolloutCell{20,10} & \RolloutCell{n} & \RolloutCell{\textbf{23}} & \RolloutCell{0.0} & \RolloutCell{\textbf{4}} & \RolloutCell{42.9} \\
\rowcolor{black!6}
\RolloutCell{$[[17,1,5]]$} & \RolloutCell{$\ket{0}_L^{\otimes k}$} & \RolloutCell{\textbf{23}} & \RolloutCell{7} & \RolloutCell{20} & \RolloutCell{n} & \RolloutCell{\textbf{23}} & \RolloutCell{0.0} & \RolloutCell{5} & \RolloutCell{28.6} & \RolloutCell{20,10} & \RolloutCell{n} & \RolloutCell{\textbf{23}} & \RolloutCell{0.0} & \RolloutCell{\textbf{4}} & \RolloutCell{42.9} \\
\RolloutCell{$[[19,1,5]]$} & \RolloutCell{$\ket{+}_L^{\otimes k}$} & \RolloutCell{\textbf{27}} & \RolloutCell{10} & \RolloutCell{10} & \RolloutCell{n} & \RolloutCell{\textbf{27}} & \RolloutCell{0.0} & \RolloutCell{5} & \RolloutCell{50.0} & \RolloutCell{20,10} & \RolloutCell{n} & \RolloutCell{\textbf{27}} & \RolloutCell{0.0} & \RolloutCell{\textbf{4}} & \RolloutCell{60.0} \\
\RolloutCell{$[[19,1,5]]$} & \RolloutCell{$\ket{0}_L^{\otimes k}$} & \RolloutCell{\textbf{27}} & \RolloutCell{10} & \RolloutCell{10} & \RolloutCell{n} & \RolloutCell{\textbf{27}} & \RolloutCell{0.0} & \RolloutCell{5} & \RolloutCell{50.0} & \RolloutCell{20,10} & \RolloutCell{n} & \RolloutCell{\textbf{27}} & \RolloutCell{0.0} & \RolloutCell{\textbf{4}} & \RolloutCell{60.0} \\
\rowcolor{black!6}
\RolloutCell{$[[23,1,7]]$} & \RolloutCell{$\ket{+}_L^{\otimes k}$} & \RolloutCell{51} & \RolloutCell{22} & \RolloutCell{20} & \RolloutCell{n} & \RolloutCell{46} & \RolloutCell{9.8} & \RolloutCell{13} & \RolloutCell{40.9} & \RolloutCell{50,25} & \RolloutCell{n} & \RolloutCell{\textbf{45}} & \RolloutCell{11.8} & \RolloutCell{\textbf{10}} & \RolloutCell{54.5} \\
\rowcolor{black!6}
\RolloutCell{$[[23,1,7]]$} & \RolloutCell{$\ket{0}_L^{\otimes k}$} & \RolloutCell{51} & \RolloutCell{22} & \RolloutCell{20} & \RolloutCell{n} & \RolloutCell{46} & \RolloutCell{9.8} & \RolloutCell{13} & \RolloutCell{40.9} & \RolloutCell{50,25} & \RolloutCell{n} & \RolloutCell{\textbf{45}} & \RolloutCell{11.8} & \RolloutCell{\textbf{10}} & \RolloutCell{54.5} \\
\RolloutCell{$[[30,6,5]]$} & \RolloutCell{$\ket{+}_L^{\otimes k}$} & \RolloutCell{70} & \RolloutCell{28} & \RolloutCell{200} & \RolloutCell{n} & \RolloutCell{63} & \RolloutCell{10.0} & \RolloutCell{12} & \RolloutCell{57.1} & \RolloutCell{20,10} & \RolloutCell{n} & \RolloutCell{\textbf{62}} & \RolloutCell{11.4} & \RolloutCell{\textbf{11}} & \RolloutCell{60.7} \\
\RolloutCell{$[[30,6,5]]$} & \RolloutCell{$\ket{0}_L^{\otimes k}$} & \RolloutCell{70} & \RolloutCell{17} & \RolloutCell{100} & \RolloutCell{n} & \RolloutCell{64} & \RolloutCell{8.6} & \RolloutCell{\textbf{11}} & \RolloutCell{35.3} & \RolloutCell{40,20} & \RolloutCell{n} & \RolloutCell{\textbf{61}} & \RolloutCell{12.9} & \RolloutCell{\textbf{11}} & \RolloutCell{35.3} \\
\rowcolor{black!6}
\RolloutCell{$[[31,1,7]]$} & \RolloutCell{$\ket{+}_L^{\otimes k}$} & \RolloutCell{46} & \RolloutCell{13} & \RolloutCell{30} & \RolloutCell{n} & \RolloutCell{\textbf{45}} & \RolloutCell{2.2} & \RolloutCell{6} & \RolloutCell{53.9} & \RolloutCell{50,25} & \RolloutCell{n} & \RolloutCell{\textbf{45}} & \RolloutCell{2.2} & \RolloutCell{\textbf{5}} & \RolloutCell{61.5} \\
\rowcolor{black!6}
\RolloutCell{$[[31,1,7]]$} & \RolloutCell{$\ket{0}_L^{\otimes k}$} & \RolloutCell{46} & \RolloutCell{13} & \RolloutCell{30} & \RolloutCell{n} & \RolloutCell{\textbf{45}} & \RolloutCell{2.2} & \RolloutCell{6} & \RolloutCell{53.9} & \RolloutCell{50,25} & \RolloutCell{n} & \RolloutCell{\textbf{45}} & \RolloutCell{2.2} & \RolloutCell{\textbf{5}} & \RolloutCell{61.5} \\
\RolloutCell{$[[31,21,3]]$} & \RolloutCell{$\ket{+}_L^{\otimes k}$} & \RolloutCell{\textbf{52}} & \RolloutCell{15} & \RolloutCell{10} & \RolloutCell{n} & \RolloutCell{\textbf{52}} & \RolloutCell{0.0} & \RolloutCell{11} & \RolloutCell{26.7} & \RolloutCell{20,4} & \RolloutCell{n} & \RolloutCell{\textbf{52}} & \RolloutCell{0.0} & \RolloutCell{\textbf{8}} & \RolloutCell{46.7} \\
\RolloutCell{$[[31,21,3]]$} & \RolloutCell{$\ket{0}_L^{\otimes k}$} & \RolloutCell{\textbf{52}} & \RolloutCell{15} & \RolloutCell{10} & \RolloutCell{n} & \RolloutCell{\textbf{52}} & \RolloutCell{0.0} & \RolloutCell{11} & \RolloutCell{26.7} & \RolloutCell{20,4} & \RolloutCell{n} & \RolloutCell{\textbf{52}} & \RolloutCell{0.0} & \RolloutCell{\textbf{8}} & \RolloutCell{46.7} \\
\rowcolor{black!6}
\RolloutCell{$[[37,1,7]]$} & \RolloutCell{$\ket{+}_L^{\otimes k}$} & \RolloutCell{58} & \RolloutCell{13} & \RolloutCell{100} & \RolloutCell{n} & \RolloutCell{\textbf{57}} & \RolloutCell{1.7} & \RolloutCell{8} & \RolloutCell{38.5} & \RolloutCell{50,25} & \RolloutCell{n} & \RolloutCell{\textbf{57}} & \RolloutCell{1.7} & \RolloutCell{\textbf{6}} & \RolloutCell{53.9} \\
\rowcolor{black!6}
\RolloutCell{$[[37,1,7]]$} & \RolloutCell{$\ket{0}_L^{\otimes k}$} & \RolloutCell{58} & \RolloutCell{13} & \RolloutCell{100} & \RolloutCell{n} & \RolloutCell{\textbf{57}} & \RolloutCell{1.7} & \RolloutCell{8} & \RolloutCell{38.5} & \RolloutCell{50,25} & \RolloutCell{n} & \RolloutCell{\textbf{57}} & \RolloutCell{1.7} & \RolloutCell{\textbf{6}} & \RolloutCell{53.9} \\
\RolloutCell{$[[72,12,6]]$} & \RolloutCell{$\ket{+}_L^{\otimes k}$} & \RolloutCell{\textbf{150}} & \RolloutCell{41} & \RolloutCell{40} & \RolloutCell{n} & \RolloutCell{\textbf{150}} & \RolloutCell{0.0} & \RolloutCell{19} & \RolloutCell{53.7} & \RolloutCell{30,6} & \RolloutCell{n} & \RolloutCell{\textbf{150}} & \RolloutCell{0.0} & \RolloutCell{\textbf{15}} & \RolloutCell{63.4} \\
\RolloutCell{$[[72,12,6]]$} & \RolloutCell{$\ket{0}_L^{\otimes k}$} & \RolloutCell{\textbf{150}} & \RolloutCell{41} & \RolloutCell{20} & \RolloutCell{n} & \RolloutCell{\textbf{150}} & \RolloutCell{0.0} & \RolloutCell{21} & \RolloutCell{48.8} & \RolloutCell{20,10} & \RolloutCell{n} & \RolloutCell{\textbf{150}} & \RolloutCell{0.0} & \RolloutCell{\textbf{14}} & \RolloutCell{65.8} \\
\rowcolor{black!6}
\RolloutCell{$[[90,8,10]]$} & \RolloutCell{$\ket{+}_L^{\otimes k}$} & \RolloutCell{141} & \RolloutCell{52} & \RolloutCell{30} & \RolloutCell{n} & \RolloutCell{136} & \RolloutCell{3.5} & \RolloutCell{46} & \RolloutCell{11.5} & \RolloutCell{30,6} & \RolloutCell{n} & \RolloutCell{\textbf{135}} & \RolloutCell{4.3} & \RolloutCell{\textbf{21}} & \RolloutCell{59.6} \\
\rowcolor{black!6}
\RolloutCell{$[[90,8,10]]$} & \RolloutCell{$\ket{0}_L^{\otimes k}$} & \RolloutCell{141} & \RolloutCell{23} & \RolloutCell{30} & \RolloutCell{n} & \RolloutCell{\textbf{134}} & \RolloutCell{5.0} & \RolloutCell{22} & \RolloutCell{4.3} & \RolloutCell{30,15} & \RolloutCell{n} & \RolloutCell{\textbf{134}} & \RolloutCell{5.0} & \RolloutCell{\textbf{21}} & \RolloutCell{8.7} \\
\RolloutCell{$[[108,8,10]]$} & \RolloutCell{$\ket{+}_L^{\otimes k}$} & \RolloutCell{425} & \RolloutCell{71} & \RolloutCell{200} & \RolloutCell{n} & \RolloutCell{380} & \RolloutCell{10.6} & \RolloutCell{\textbf{39}} & \RolloutCell{45.1} & \RolloutCell{10,2} & \RolloutCell{n} & \RolloutCell{\textbf{378}} & \RolloutCell{11.1} & \RolloutCell{49} & \RolloutCell{31.0} \\
\RolloutCell{$[[108,8,10]]$} & \RolloutCell{$\ket{0}_L^{\otimes k}$} & \RolloutCell{472} & \RolloutCell{78} & \RolloutCell{50} & \RolloutCell{n} & \RolloutCell{\textbf{402}} & \RolloutCell{14.8} & \RolloutCell{\textbf{58}} & \RolloutCell{25.6} & \RolloutCell{40,20} & \RolloutCell{y} & \RolloutCell{472} & \RolloutCell{0.0} & \RolloutCell{72} & \RolloutCell{7.7} \\
\rowcolor{black!6}
\RolloutCell{$[[144,12,12]]$} & \RolloutCell{$\ket{+}_L^{\otimes k}$} & \RolloutCell{\textbf{330}} & \RolloutCell{66} & \RolloutCell{40} & \RolloutCell{n} & \RolloutCell{\textbf{330}} & \RolloutCell{0.0} & \RolloutCell{\textbf{32}} & \RolloutCell{51.5} & \RolloutCell{40,20} & \RolloutCell{y} & \RolloutCell{\textbf{330}} & \RolloutCell{0.0} & \RolloutCell{38} & \RolloutCell{42.4} \\
\rowcolor{black!6}
\RolloutCell{$[[144,12,12]]$} & \RolloutCell{$\ket{0}_L^{\otimes k}$} & \RolloutCell{\textbf{330}} & \RolloutCell{63} & \RolloutCell{40} & \RolloutCell{n} & \RolloutCell{\textbf{330}} & \RolloutCell{0.0} & \RolloutCell{\textbf{39}} & \RolloutCell{38.1} & \RolloutCell{30,15} & \RolloutCell{y} & \RolloutCell{\textbf{330}} & \RolloutCell{0.0} & \RolloutCell{48} & \RolloutCell{23.8} \\
\bottomrule
\end{tabular}

\end{table*}%

\section{Impact of Early Termination on the Search}
\label{sec:impact-early-term}

\begin{figure*}[t]
\centering

\begin{subfigure}{\textwidth}
\includegraphics[width=.8\linewidth]{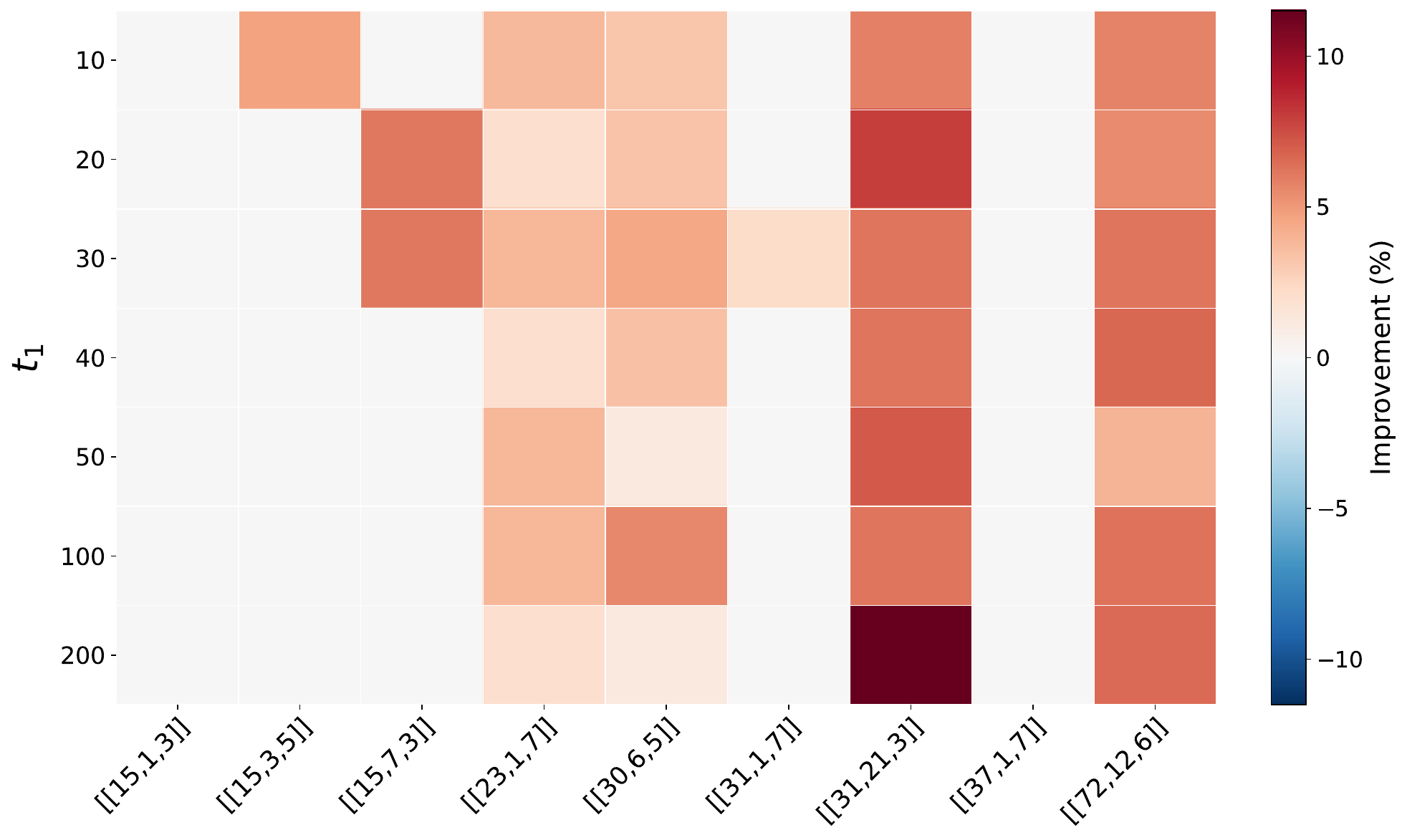}
    \caption{Relative improvement in two-qubit gate count.}
\end{subfigure}

\begin{subfigure}{\textwidth}
\includegraphics[width=.8\linewidth]{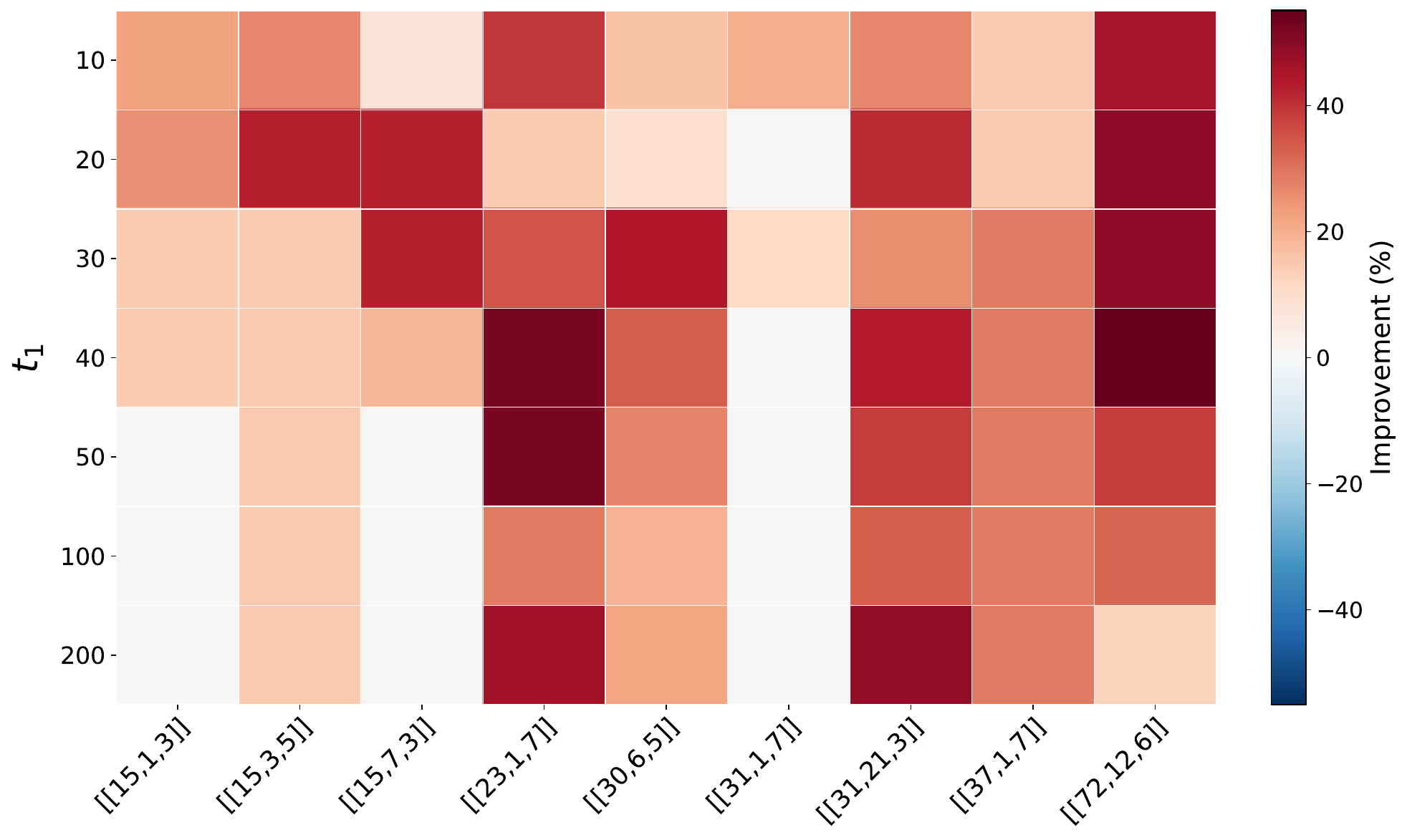}
    \caption{Relative improvement in depth.}
\end{subfigure}

\caption{Impact of early termination on encoding circuits with rollout level $\ell=1$ when two-qubit gate count is used as the primary optimization objective. $t_1$ corresponds to the number of candidates for which rollout is performed at each stage of the search. Colors indicate the relative improvement obtained by running without early termination; positive values denote an improvement. Subfigure~(a) reports the effect on two-qubit gate count and subfigure~(b) reports the effect on circuit depth.}
\label{fig:et-gates-encoding-l1}
\end{figure*}

\begin{figure*}[t]
\centering

\begin{subfigure}{\textwidth}
\includegraphics[width=.8\linewidth]{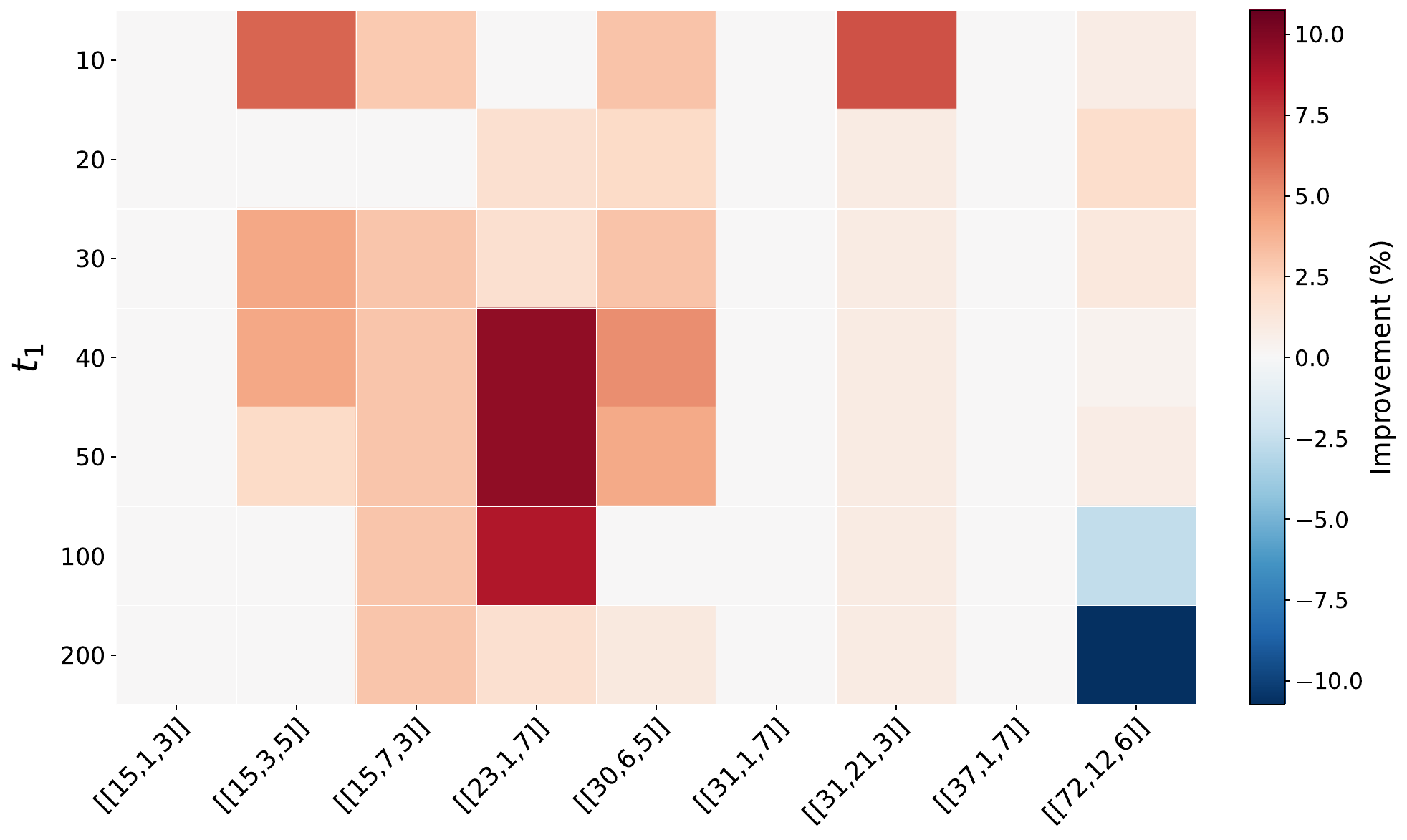}
    \caption{Relative improvement in two-qubit gate count.}
\end{subfigure}

\begin{subfigure}{\textwidth}
\includegraphics[width=.8\linewidth]{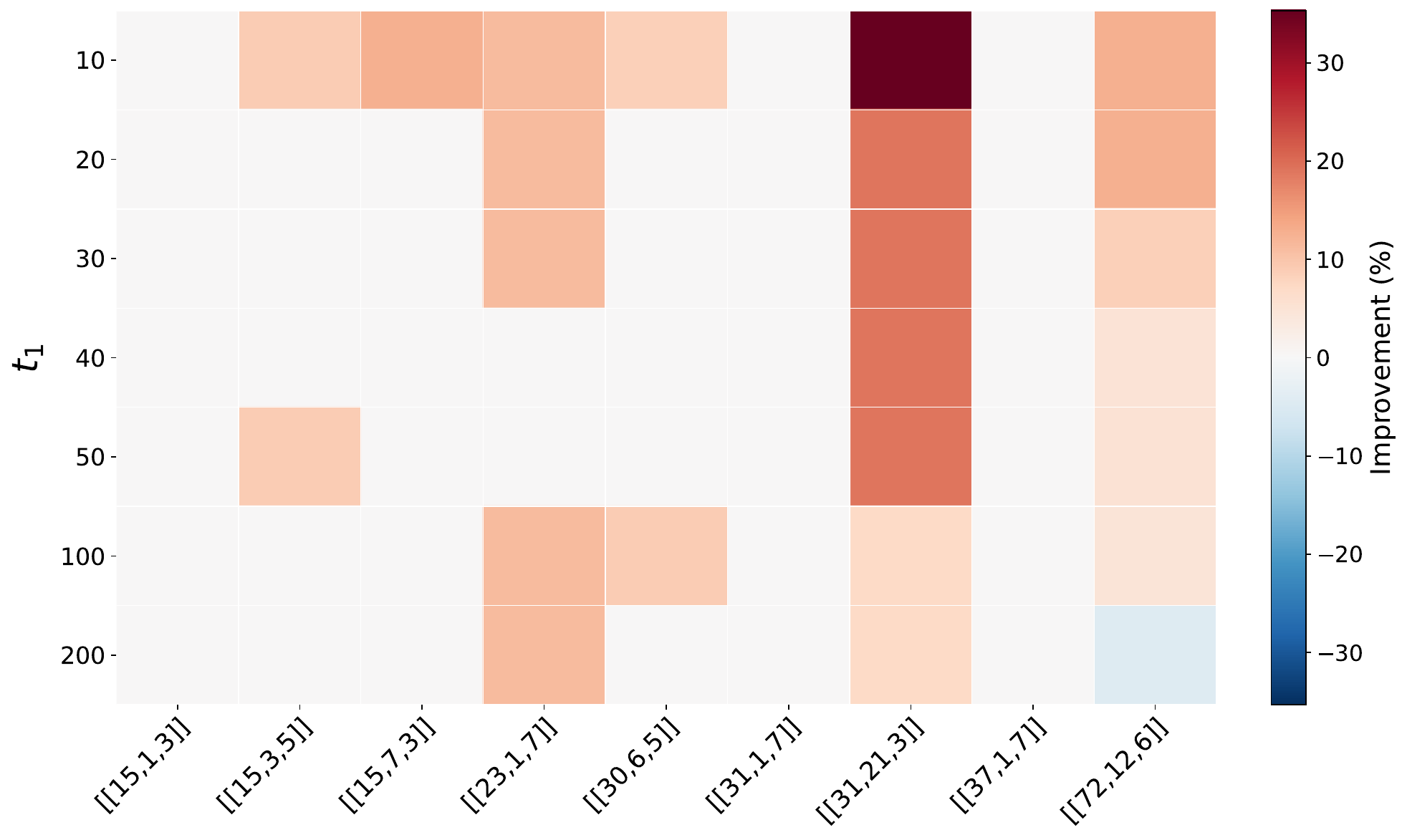}
    \caption{Relative improvement in depth.}
\end{subfigure}

\caption{Impact of early termination on encoding circuits with rollout level $\ell=1$ when depth is used as the primary optimization objective. $t_1$ corresponds to the number of candidates for which rollout is performed at each stage of the search. Colors indicate the relative improvement obtained by running without early termination; positive values denote an improvement. Subfigure~(a) reports the effect on two-qubit gate count, and subfigure~(b) reports the effect on circuit depth.}
\label{fig:et-depth-encoding-l1}
\end{figure*}

\begin{figure*}[t]
\centering

\begin{subfigure}{\linewidth}
\includegraphics[width=\linewidth]{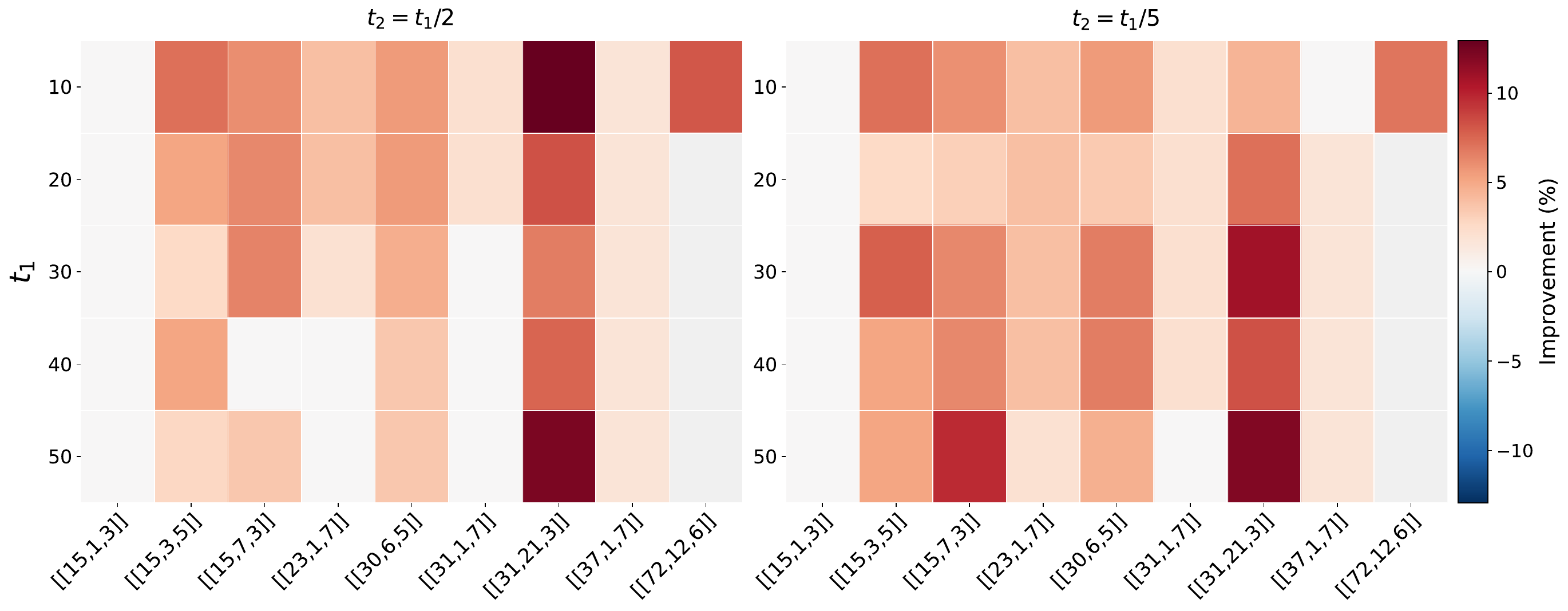}
    \caption{Relative improvement in two-qubit gate count.}
\end{subfigure}

\begin{subfigure}{\linewidth}
\includegraphics[width=\linewidth]{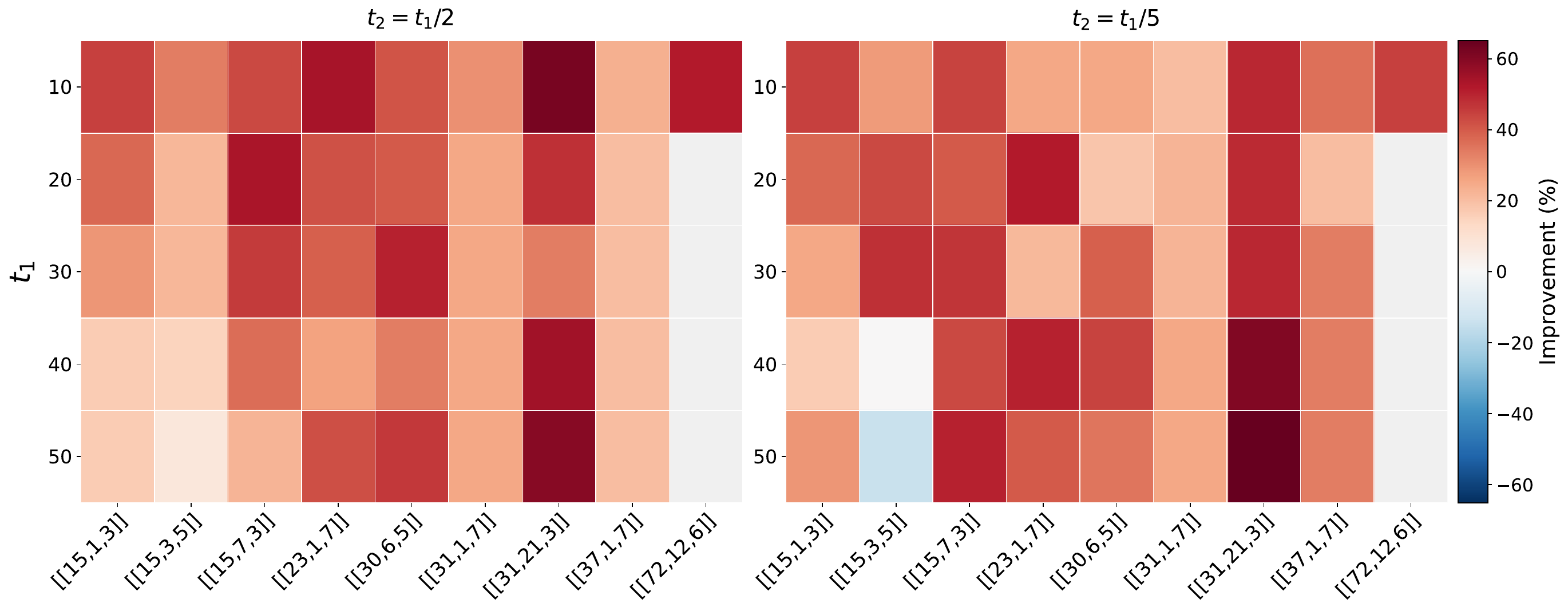}
    \caption{Relative improvement in circuit depth.}
\end{subfigure}

\caption{Impact of early termination on encoding circuits with rollout level $\ell=2$ when two-qubit gate count is used as the primary optimization objective. In each subfigure, the two panels correspond to the parameter settings $t_2=t_1/2$ and $t_2=t_1/5$, while the rows correspond to values of $t_1$. Colors indicate the relative improvement obtained by running without early termination; positive values denote an improvement. Subfigure~(a) reports the effect on two-qubit gate count, and subfigure~(b) reports the effect on circuit depth.}
\label{fig:et-gates-encoding-l2}
\end{figure*}

\begin{figure*}[t]
\centering

\begin{subfigure}{\linewidth}
    \centering
    \includegraphics[width=\linewidth]{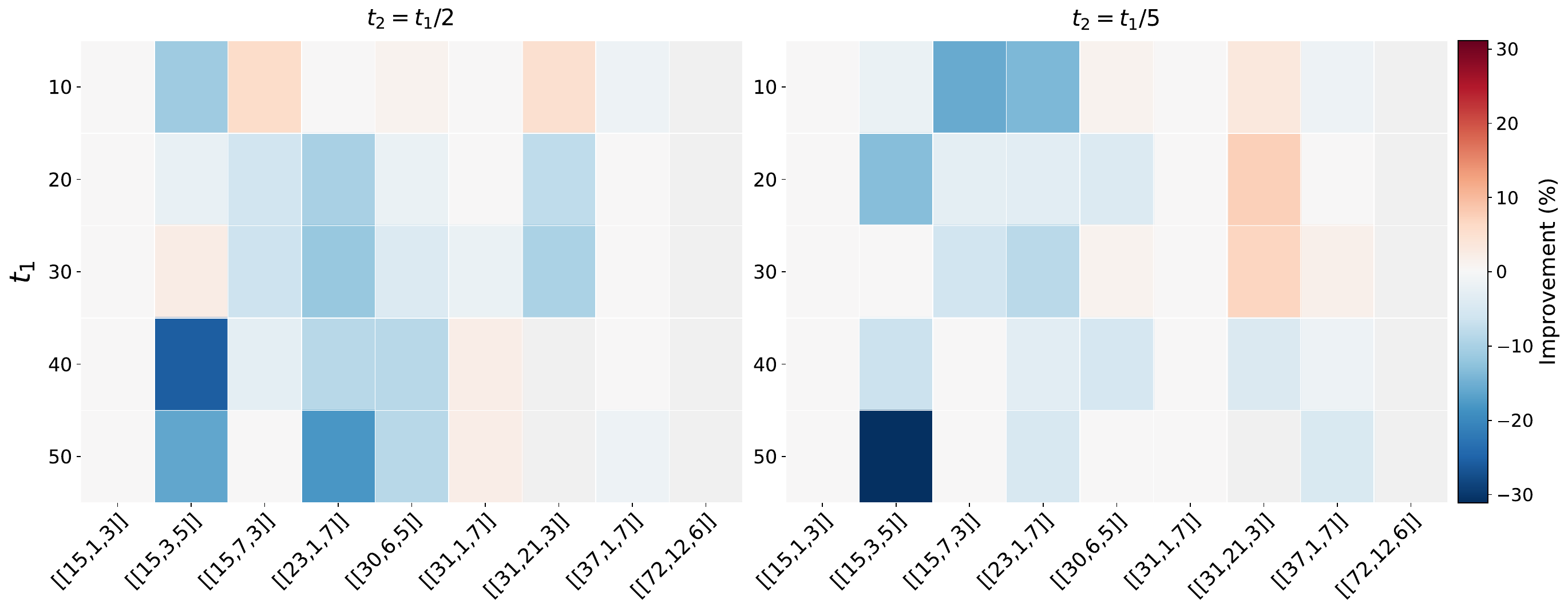}
    \caption{Relative improvement in two-qubit gate count.}
\end{subfigure}

\vspace{0.5em}

\begin{subfigure}{\linewidth}
    \centering
    \includegraphics[width=\linewidth]{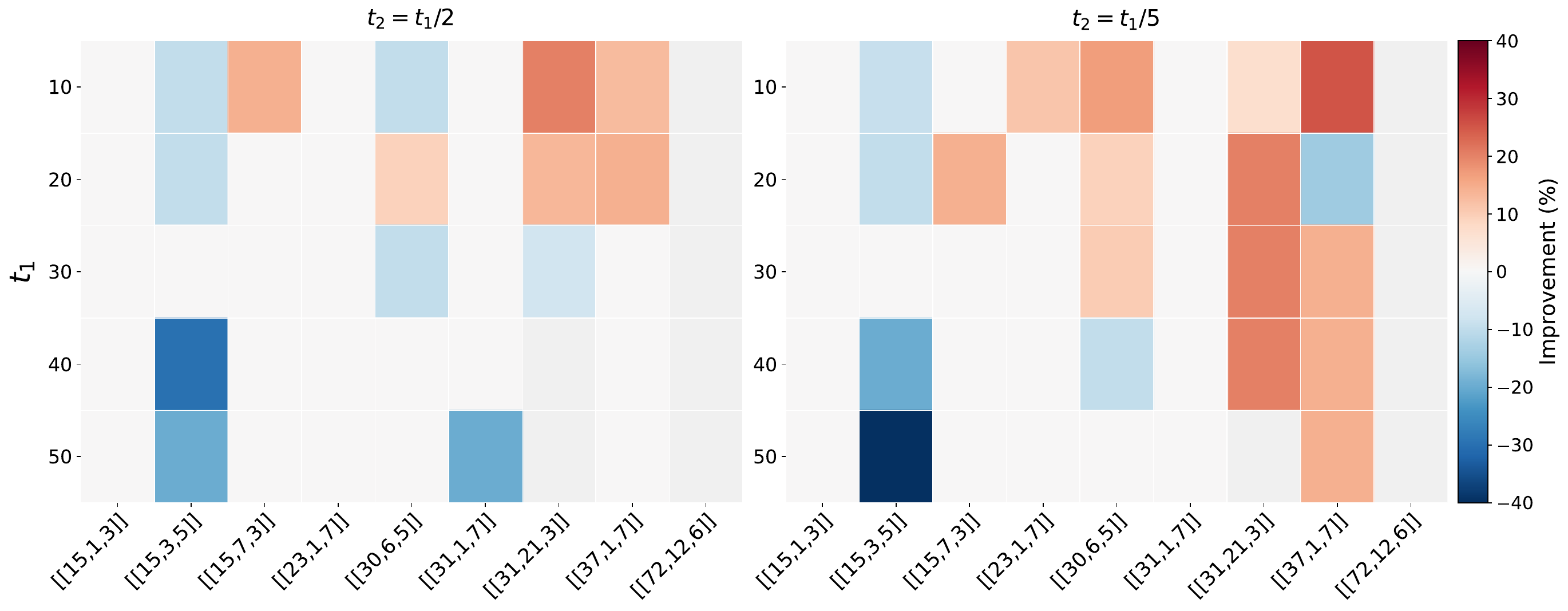}
    \caption{Relative improvement in circuit depth.}
\end{subfigure}

\caption{Impact of early termination on encoding circuits with rollout level $\ell=2$ when circuit depth is used as the primary optimization objective. In each subfigure, the two panels correspond to the parameter settings $t_2=t_1/2$ and $t_2=t_1/5$, while the rows correspond to values of $t_1$. Colors indicate the relative improvement obtained by running without early termination; positive values denote an improvement. Subfigure~(a) reports the effect on two-qubit gate count, and subfigure~(b) reports the effect on circuit depth.}
\label{fig:et-depth-encoding-l2}
\end{figure*}

As described in~\Cref{sec:rollout-synthesis}, terminating the search early if no single step in the search space suggests an improvement can speed up the synthesis significantly.
Even though no improvement is evident locally, continuing the search might reveal potential improvements later.
To investigate the impact of early termination, we compare the quality of synthesized circuits obtained with and without this setting across different values of $\ell$ and $t$.
The results for encoding circuit synthesis can be seen  in~\Cref{fig:et-gates-encoding-l1,fig:et-gates-encoding-l2} when targeting two-qubit gate count as the primary optimization objective and~\Cref{fig:et-depth-encoding-l1,fig:et-depth-encoding-l2} when targeting depth as the primary optimization objective.

From~\Cref{fig:et-gates-encoding-l1,fig:et-gates-encoding-l2}, we observe that disabling early termination often improves the synthesized circuits when two-qubit gate count is used as the primary optimization objective.
The effect is particularly pronounced for circuit depth, i.e., the secondary optimization objective in this setting, whereas the improvement in the required number of two-qubit gates is typically smaller.
This finding matches the expected structure of the optimization landscape.
A locally suboptimal two-qubit gate may be reflected immediately in the gate count, while its effect on circuit depth often becomes apparent only after an entire layer has been completed.
Since early termination stops the search as soon as no single step yields an immediate improvement in the rollout cost, these delayed improvements in depth are easily missed.

The behavior is different when circuit depth is the primary optimization objective.
In this case, disabling early termination does not uniformly improve the result, and for some instances it can even lead to worse outcomes, including with respect to the primary objective; see in particular~\Cref{fig:et-depth-encoding-l2}.
At first sight, this is surprising, since disabling early termination allows each rollout to inspect a superset of continuations.
However, the rollout score is itself only an approximate evaluation, and for $\ell=2$ it depends on lower-level searches that may terminate early.
Consequently, a more extensive lower-level search can alter the ranking of candidate moves, ultimately leading to a worse top-level decision.
Notably, this detrimental effect is observed mainly for the two-level rollout in~\Cref{fig:et-depth-encoding-l2}.
Since early termination is applied at every rollout level, there are cases in which a level-one rollout that is locally suboptimal due to early termination nevertheless leads to a better final decision at level two.

Overall, these findings suggest that further improvements to the synthesis procedure, particularly in how the search explores and evaluates longer-term continuations, may yield additional gains in circuit quality.

\end{document}